\documentclass{aa}
\usepackage{graphicx}
\usepackage{txfonts}
\usepackage{natbib}
\bibpunct{(}{)}{;}{a}{}{,}
\usepackage{color}
\usepackage{longtable}
\usepackage{epstopdf}
\begin{document}

   \title{ATLASGAL - Ammonia observations towards the southern Galactic Plane}

   \author{M. Wienen\inst{1}\thanks{Member of the International Max Planck Research School (IMPRS) for Astronomy and Astrophysics at the Universities of Bonn and Cologne.}, F. Wyrowski\inst{1}, K. M. Menten\inst{1}, J. S. Urquhart\inst{2}, C. M. Walmsley\inst{3,}\inst{4}, T. Csengeri\inst{1}, B. S. Koribalski\inst{5}
          \and F. Schuller\inst{1,}\inst{6}
          }

   \institute{\inst{1}Max-Planck-Institut f\"ur Radioastronomie, Auf dem H\"ugel 69, 53121 Bonn, Germany\\ \email{mwienen@mpifr-bonn.mpg.de}\\
             \inst{2}School of Physical Sciences, University of Kent, Ingram Building, Canterbury, Kent CT2 7NH, UK\\
             \inst{3}Osservatorio Astrofisico di Arcetri, Largo E. Fermi, 5, I-50125 Firenze, Italy\\
             \inst{4}Dublin Institute of Advanced Studies, Fitzwilliam Place 31, Dublin 2, Ireland\\
             \inst{5}Australia Telescope National Facility, CSIRO, P.O. Box 76, Epping, NSW 1710, Australia\\
             \inst{6}Alonso de Cordova 3107, Casilla 19001, Santiago 19, Chile}

   \date{Received }

  \abstract
   {The initial conditions of molecular clumps in which high-mass stars form are poorly understood. In particular, a more detailed study of the earliest evolutionary phases is needed. 
   The APEX Telescope Large Area Survey of the whole inner Galactic disk at 870 $\mu$m, ATLASGAL, has therefore been conducted to discover high-mass star-forming regions at different evolutionary phases.}
   {We derive properties such as velocities, rotational temperatures, column densities, and abundances of a large sample of southern ATLASGAL clumps in the fourth quadrant.}
   {Using the Parkes telescope, we observed the NH$_3$ (1,1) to (3,3) inversion transitions towards 354 dust clumps detected by ATLASGAL within a Galactic longitude range between 300$^{\circ}$ and 359$^{\circ}$ and a latitude within $\pm 1.5^{\circ}$. For a subsample of 289 sources, the N$_2$H$^+$ $(1-0)$ line was measured with the Mopra telescope.}
   {We measured a median NH$_3$ (1,1) line width
    of $\sim 2$ km~s$^{-1}$, rotational temperatures from 12 to 28 K with a mean of 18 K, and source-averaged NH$_3$ abundances from $1.6 \times 10^{-6}$ to $10^{-8}$. 
    For a subsample with detected NH$_3$ (2,2) hyperfine components, we found that the commonly used method to compute the (2,2) optical depth from the (1,1) optical depth and the (2,2) to (1,1) main beam brightness temperature ratio leads to an underestimation of the rotational temperature and column density. 
    A larger median virial parameter of $\sim 1$ is determined using the broader N$_2$H$^+$ line width than is estimated from the NH$_3$ line width of $\sim 0.5$ with a general trend of a decreasing virial parameter with increasing gas mass. We obtain a rising NH$_3$ (1,1)/N$_2$H$^+$ line-width ratio with increasing rotational temperature.
   }
   {A comparison of NH$_3$ line parameters of ATLASGAL clumps to cores in nearby molecular clouds reveals smaller velocity dispersions in low-mass than high-mass star-forming regions and a warmer surrounding of ATLASGAL clumps than the surrounding
of low-mass cores. The NH$_3$ (1,1) inversion transition of 49\% of the sources shows hyperfine structure anomalies. The intensity ratio of the outer hyperfine structure lines with a median of $1.27 \pm 0.03$ and a standard deviation of 0.45 is significantly higher than 1, while the intensity ratios of the inner satellites with a median of $0.9 \pm 0.02$ and standard deviation of 0.3 and the sum of the inner and outer hyperfine components with a median of $1.06 \pm 0.02$ and standard deviation of 0.37 are closer to 1.
}

   \keywords{Submillimeter --- Surveys --- ISM: molecules --- 
             ISM: kinematics and dynamics --- Stars: formation --- Stars: massive}               
\titlerunning{ATLASGAL - ammonia observations towards the southern Galactic Plane}
\authorrunning{M. Wienen et al.}
   \maketitle
%

\section{Introduction}
The earliest phases of high-mass star formation occur in the densest regions of giant molecular clouds.
Several surveys \citep{1975ApJ...202...30B,1977ApJ...217L.155C,1988ApJ...324..248B} have established the large-scale distribution of molecular clouds in the Galaxy. Observations have also revealed the physical properties of these objects, such as their mass, which ranges from $10^4$ to $10^6$ M$_{\odot}$, sizes of between 50 to 200 pc, and temperatures of $\sim 10$ K \citep{2001ApJ...547..792D}. Clumps consisting of molecular gas that are embedded within these molecular clouds have high masses ($> 500$ M$_{\odot}$), high densities ($> 10^5$ cm$^{-3}$), and low temperatures \citep[$< 20$ K,][]{2007ARA&A..45..339B}, and thus exhibit the initial conditions expected for high-mass star or cluster formation. While it is known that massive stars form in clusters, the detailed physical processes are little understood. In addition, because massive young stellar objects (YSOs) are rarer than low-mass stars according to the initial mass function \citep[only $\sim 500$ are identified throughout the Galactic plane,][]{2011arXiv1112.3340K,2013ApJS..208...11L}, they are more distant and heavily affected by extinction. Especially the earliest stages of high-mass star formation are not well known compared to an accepted evolutionary scenario explaining isolated low-mass star formation \citep{2000prpl.conf...59A}.\\
While high-mass protostars evolve in massive dense cores within the clumps \citep{2010A&A...524A..18B,2011ApJ...740L...5C}, their intense ultraviolet radiation heats, ionizes, and disrupts their natal molecular cloud. Embedded YSOs that have already formed an HII region have been studied in more detail because they can be observed in far-infrared and radio continuum surveys \citep{1989ApJS...69..831W,1990ApJ...358..485B,1994ApJS...91..659K,1998MNRAS.301..640W,2012PASP..124..939H,2013MNRAS.435..400U}. In addition to these regions that are in a more evolved phase of high-mass star formation, further progress required the observation of earlier stages. Surveys searched for the progenitors of ultracompact HII regions (UCHIIRs), thus very luminous ($> 10^3$ L$_{\odot}$) infrared protostars, that are embedded in a high-mass envelope and harbour hot gas, but are not detected at centimetre wavelengths. Association of high-density gas around these sources reveals an envelope and the detection of a hot core, and water or methanol masers indicate hot gas. These high-mass protostellar objects have been analysed in several studies \citep{1996A&AS..115...81B,2000AJ....119.2711H,2001A&A...370..230B,2002ApJ...566..931S,2013ApJS..208...11L,2014MNRAS.437.1791U}. Recently, massive dense cores and high-mass protostars at an early stage have been directly observed \citep{2010A&A...524A..18B,2013A&A...558A.125D,2014A&A...570A...1D}. Surveys for water and 6.7 GHz methanol maser emission detected numerous new high-mass star-forming regions \citep{1997ApJ...476..730P,2002A&A...392..277S,2002ApJ...566..931S,2011MNRAS.417.1964C,2009MNRAS.392..783G}. Other large-scale surveys conducted within the Galactic Plane are the Spitzer GLIMPSE \citep[Galactic Legacy Infrared Mid-Plane Survey Extraordinaire,][]{2003PASP..115..953B}  between 3 to 8 $\mu$m and the MIPSGAL survey \citep[MIPS Galactic Plane Survey,][]{2009PASP..121...76C} at 24 and 70 $\mu$m. Longer wavelengths are observed by the BGPS \citep[Bolocam Galactic Plane Survey,][]{2011ApJS..192....4A} at 1.1 mm and Hi-GAL \citep[Herschel Infrared Galactic Plane Survey,][]{2010PASP..122..314M} from 70 to 500 $\mu$m.\\
Even younger stellar objects are still deeply embedded in an envelope and thus so cold that they cannot be detected at mid-infrared wavelengths. They can be found in infrared dark clouds (IRDCs), which were discovered by two infrared satellites, ISO and MSX, as extinction features against the bright mid-infrared background \citep{1998ApJ...494L.199E,1996A&A...315L.165P}. High column densities ($\sim 10^{23}-10^{25}$ cm$^{-2}$) and low temperatures ($< 25$ K) of IRDCs \citep{1998ApJ...508..721C,2000ApJ...543L.157C,2006A&A...450..569P} indicate that they are in the earliest phases of star formation.\\ 
Surveys conducted toward massive star-forming regions have only traced one particular evolutionary phase, such as UCHIIRs or IRDCs \citep{1989ApJS...69..831W,2009A&A...499..149V}, or the targeted samples have been limited in number. To enlarge these samples and to trace the early, cold stages as well as later evolutionary phases, the first unbiased submillimeter-continuum survey of the whole inner Galactic disk, ATLASGAL (\textit{The APEX Telescope Large Area Survey of the Galaxy at 870 $\mu$m}) \citep{2009A&A...504..415S} was conducted. It observed the Galactic longitude range of $\pm 60^{\circ}$ and latitude of $\pm 1.5^{\circ}$ using the Large APEX Bolometer Camera \citep[LABOCA,][]{2007Msngr.129....2S,2008SPIE.7020E..03S} with a beam width of 19.2$\arcsec$ FWHM at the wavelength of 870 $\mu$m. ATLASGAL aims to derive physical properties of a statistically representative sample of objects at different evolutionary phases of high-mass star formation. To extract sources from the submillimeter (submm) maps, two algorithms are exploited: SExtractor \citep{1996A&AS..117..393B} provides global properties for clumps that are presented in the ATLASGAL Compact Source Catalogue \citep[CSC;][]{2013A&A...549A..45C,2014A&A...568A..41U}, while the Gaussclumps algorithm \citep{1990ApJ...356..513S,1998A&A...329..249K} separates compact emission from the more diffuse envelope \citep{2014A&A...565A..75C}. In addition to the identification of massive clumps, information from molecular line observations is also required. To derive the three-dimensional distribution of the dense material, for
instance, distances and therefore velocities are necessary \citep{2015A&A...579A..91W}. NH$_3$ is a useful tool for this task because it is still present in the gas phase of the very dense \citep[$\sim$ 10$^5$ cm$^{-3}$][]{2002ApJ...566..945B} and cold molecular cores \citep{2002ApJ...569..815T}, where other molecules such as CS and CO are partly frozen onto dust grains. In addition, NH$_3$ is an important probe to measure the temperature of molecular clouds because they exhibit low temperatures ($\sim 20$ K), at which the NH$_3$ inversion transitions in the lowest metastable $(J, K)$ rotational energy levels are excited \citep{1983ARA&A..21..239H}. The splitting of the inversion transitions into different hyperfine structure components allows deriving the optical depth and column density. The rotational temperature of the gas within molecular clumps can then be determined from the line ratios between different inversion transitions and the optical depth. The line width can also be used to determine the stability of the clumps from an analysis of their virial masses.\\
Previous studies have used ammonia to analyse the physical properties of nearby molecular clouds \citep{2008ApJS..175..509R,2009ApJ...697.1457F,2010ApJ...711..655J}. Various observations of ammonia have also been conducted toward high-mass star-forming regions \citep{1992ApJ...388..467M,1993ApJ...402..230W,1994A&A...288..903C,2004A&A...414..299F,2011ApJ...736..163R}. These studies tend to focus on a single stage of high-mass star formation such as IRDCs, YSOs, or HII regions \citep{2009A&A...505..405P,2013A&A...552A..40C,2014A&A...566A.150M,2014ApJ...784..107Z}, contain a small number of sources \citep{2006A&A...450..569P,2010ApJ...720..392W,2013A&A...556A..16G,2014ApJ...790...84L}, or are confined to the first quadrant \citep{2011ApJ...741..110D,2011MNRAS.418.1689U,2015MNRAS.452.4029U}. A shallow low-resolution ($2 \arcmin$) survey of NH$_3$ (1,1) and (2,2) has been part of the H$_2$O southern Galactic Plane Survey \citep[HOPS,][]{2011MNRAS.416.1764W,2012MNRAS.426.1972P} in the fourth quadrant. These measurements cover the Galactic longitude range of the ATLASGAL survey, but are confined to a Galactic latitude of $\pm 0.5^{\circ}$.

We observed ammonia towards a large sample of ATLASGAL sources in different evolutionary phases of high-mass star formation in the fourth quadrant, which allows a statistical analysis of derived line parameters. First, the NH$_3$ (1,1) to (3,3) line observations were made towards 862 northern ATLASGAL clumps in the first quadrant, within $|b| \leq 1.5^{\circ}$ and $l = 5^{\circ} - 60^{\circ}$; the results are presented in \cite{2012A&A...544A.146W}. We extended this survey to the fourth quadrant and observed 354 southern ATLASGAL sources within $|b| \leq 1.5^{\circ}$ and $l = 300^{\circ} - 359^{\circ}$. This article focuses on physical properties that could not be studied for the northern sources.
The structure of the paper is as follows.\\
Section \ref{s:obs} presents the NH$_3$ observations and data reduction. We analyse properties derived from the NH$_3$ inversion lines such as the velocity, line width, rotational temperature, and column density and show their variation with galactocentric radius in Section \ref{results}. 
We investigate the dependence of these line parameters on the environment within molecular clouds at different distances in Section \ref{discussion}. In addition, we compare virial parameters derived from the NH$_3$ line width with those calculated from the line width of a higher density tracer, N$_2$H$^+$. Furthermore, we analyse anomalies in the relative hyperfine satellite intensities of the NH$_3$ (1,1) inversion transition. Section \ref{summary} presents a summary of our work.

\section{Parkes observations and data reduction}
\label{s:obs}
A flux-limited subsample was selected from a preliminary ATLASGAL point-source catalogue similar to the observations made with the Effelsberg telescope described in \cite{2012A&A...544A.146W}. We selected clumps with peak fluxes above a threshold of 1.2 Jy/beam in the fourth quadrant, while our northern NH$_3$ sample exhibits peak flux densities above 0.4 Jy/beam. In 2009, ammonia observations of a total of 354 dust clumps located in the Galactic longitude range between $300^\circ$ and $359^\circ$ and latitude within $\pm$1.5$^\circ$ were conducted. 

We observed the NH$_3$ (1,1), (2,2), and (3,3) inversion transitions using the Parkes 64 m telescope from 20 to 27 June, 2009. The frontend was a 13mm receiver that covers the frequency range from 16 to 26 GHz. The beam width (FWHM) at the NH$_3$ (1,1) to (3,3) line frequencies at $\sim$ 24 GHz is 61$\arcsec$. The spectrometer was a Digital Filter Bank (DFB3) with a bandwidth of 256 MHz, which results in a spectral resolution of $\sim 0.4$ km~s$^{-1}$. Two polarizations of each of the three NH$_3$ transitions were measured simultaneously. The observations were conducted in position-switching mode with a constant offset of $\sim 15 \arcmin$ with a total integration time of about 8 min for each source, pointing and focus were measured approximately every hour. To calibrate the data, a reference source, G15.66$-$0.50, was measured in each observing session. To connect the calibration to the data observed with Effelsberg \citep{2012A&A...544A.146W}, we compared the northern and southern (3,3) line intensities of G15.66$-$0.50, which is assumed to be compact. This gives a calibration factor of 3.47 that also accounts for the atmospheric opacity, which is multiplied to the data. These were then corrected for the gain-elevation curve. The variation in main-beam brightness temperatures between the different observing days gives a calibration error of $\sim 7$\%. Typical system temperatures were about 50 K. We measured an rms noise level between 10 and 130 mK at a velocity resolution of 0.4 km~s$^{-1}$.

The NH$_3$ spectra were read into the CLASS software\footnote{available at http://www.iram.fr/IRAMFR/GILDAS}, which we used to reduce
the spectra. The ammonia lines of each source can be observed in two polarizations, which are averaged together. 
Since position-switching was performed as observing mode, the baseline is more stable than it was for the northern NH$_3$ measurements,
for which we used frequency-switching \citep{2012A&A...544A.146W}. As there are still some fluctuations in the baseline, they were corrected by subtracting a polynomial baseline of order 3 to 7 over a velocity range of 100 km~s$^{-1}$ for the NH$_3$ (1,1) line and over 70 km~s$^{-1}$ for the NH$_3$ (2,2) line. We used the method described by \cite{2012A&A...544A.146W} to place windows around the hyperfine structure components for the northern NH$_3$ lines. We followed the procedure described in \cite{2012A&A...544A.146W} to derive line parameters with fits in CLASS  of the ammonia (1,1) lines that include 18 hyperfine structure components. In contrast to the northern sources, a small southern subsample is detected in the (2,2) hyperfine structure. We did not use a single-Gaussian fitting to the main line of the (2,2) transition as for the northern sources, but fitted all the 22 hyperfine structure components of the whole southern sample (see Sects. \ref{linewidth} and \ref{rotational temperature} for details). The optical depth of the (3,3) line could not be derived because we did not detect the hyperfine structure, and we fitted a single Gaussian to the main line of the (3,3) transition. The derived positions of the sources, the optical depths of the NH$_3$ (1,1) lines, $\tau$(1,1), their LSR velocities, $\rm v$(1,1), FWHM line widths, $\Delta \rm v$(1,1), and main-beam brightness temperatures, $T_{\mbox{\tiny MB}}$(1,1), with their formal fit errors are given in Table \ref{parline11-atlasgal}. The LSR velocities, FWHM line widths, and main-beam brightness temperatures of the (2,2) and (3,3) lines together with their formal fit errors can be found in Table \ref{parline22_line33-atlasgal}. Using the standard formulation for NH$_3$ spectra \citep{1983ARA&A..21..239H,1986A&A...157..207U}, we determined physical parameters such as the rotational temperature ($T_{\mbox{\tiny rot}}$), kinetic temperature ($T_{\mbox{\tiny kin}}$), and ammonia column density ($N_{\mbox{NH}_3}$), see Sect. \ref{results}. Table \ref{parabgel-atlasgal} contains the parameters with the errors (1$\sigma$) calculated from Gaussian error propagation.

\subsection{Mopra observations}
We selected a flux-limited subsample of clumps from a preliminary ATLASGAL compact source catalogue. It ensured high enough column densities for the Mopra line detections and a coverage of all phases of high-mass star formation. The sample consists of sources with an MSX infrared association within a distance of 30$\arcsec$ and peak fluxes above a limit of 1.75 Jy/beam, and cold clumps without MSX association and peak fluxes above 1.2 Jy/beam. We observed a total of 700 ATLASGAL clumps within $l = 300^{\circ} - 359^{\circ}$ and $|b| \leq 1.5^{\circ}$ in 2009 and 2010.

The N$_2$H$^+$ ($1-0$) line was observed using the Mopra 22m Radiotelescope\footnote{Mopra is part of the Australia Telescope National Facility.}. The frontend was a 3 mm HEMT receiver with a frequency range of between 76 and 117 GHz. The UNSW Mopra spectrometer (MOPS) consists of four 2.2 GHz bands that overlap slightly and result in a total of $\sim 8$ GHz continuous bandwidth. We centred the 3 mm band on 89.3 GHz, covering the frequency range from $\sim 85.2$ GHz to $\sim 93.4$ GHz. The FWHM at the frequency of the N$_2$H$^+$ ($1-0$) line at $\sim 90$ GHz is 38$\arcsec$. MOPS was used in the broadband mode with a velocity resolution of 0.9 km~s$^{-1}$ of each 2.2 GHz band.

MOPS was able to measure two polarizations of the N$_2$H$^+$ line simultaneously. We conducted pointed observations in position-switching mode. Using ATLASGAL and MSX maps, we searched around each source for an offset position that was free of emission, and used a constant offset of $\pm 5^{\circ}$ in longitude or latitude. Each source was observed with a total integration time of $\sim 15$ min, which gives an rms noise level of 24 mK on average at a velocity resolution of 0.9 km~s$^{-1}$. Line pointings on SiO masers were measured every hour, a reference spectrum of G327 and M17 was taken each day.

The ASAP package\footnote{http://www.atnf.csiro.au/news/newsletter/jun06/ASAP.htm} was used for an initial processing of the data. This included processing of the on-off observing mode, the time and polarization averaging, and baseline subtraction. The data were calibrated to the T$_{\rm A}^{*}$ temperature scale that is transformed into $T_{\rm MB}$ by correcting for the beam efficiency of 0.49 \citep{2005PASA...22...62L}. For the subsequent analysis the data were exported to the CLASS software from the GILDAS package.

\begin{figure*}
\centering
\includegraphics[angle=0,width=13.0cm]{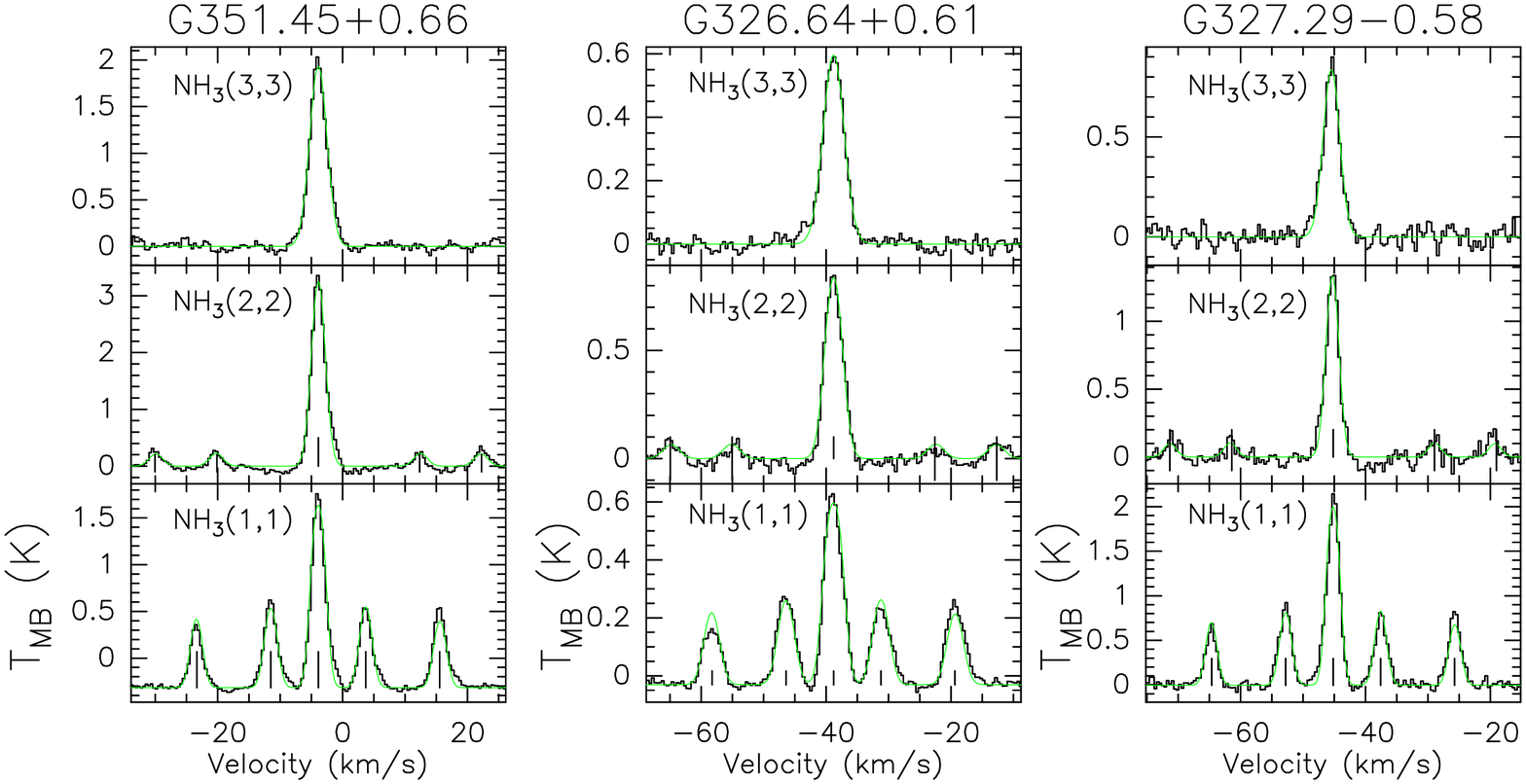}\vspace*{0.5cm}
\includegraphics[angle=0,width=13.0cm]{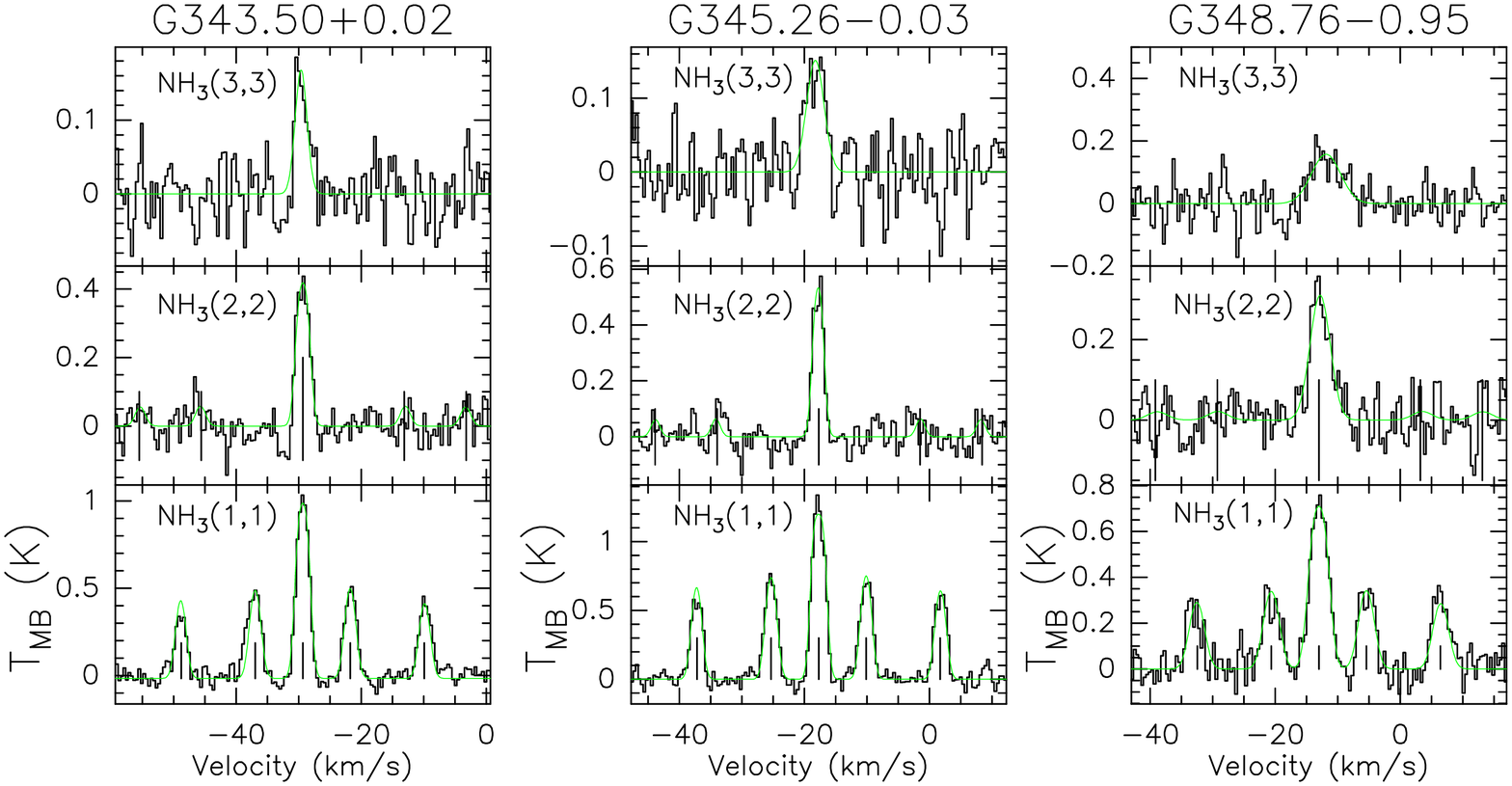}
\caption[spectra of observed sources]{Reduced and calibrated spectra of observed NH$_3$(1,1), (2,2), and (3,3) inversion transitions; the fit is shown in green. The frequencies of the (1,1) and (2,2) main line and satellites are indicated by straight lines.}\label{nh3lines}
\end{figure*}

\section{Results for and analysis of the ammonia sample in the fourth quadrant}
\label{results}
\subsection{Ammonia line parameters}
Of the 354 clumps observed, 315 were detected in NH$_3$ (1,1) (89\%), 262 sources (74\%) in NH$_3$ (2,2), and 187 clumps (53\%) in the (3,3) line with a signal-to-noise ratio (S/N)  $>$ 3. Most NH$_3$ (1,1) lines clearly show hyperfine structure, although it can be detected only in a few (2,2) lines and is too weak to be observed in the (3,3) lines. Examples of some ammonia spectra are displayed in Fig. \ref{nh3lines}. A few sources, such as G351.45+0.66, G326.64+0.61, and G327.29$-$0.58 in the upper panel, exhibit strong NH$_3$ lines, about 10\% have an S/N of the (1,1) line greater than 30. On the other hand, Fig. \ref{nh3lines} also shows example spectra with their fit for clumps with weak (3,3) lines, which are G343.50+0.02, G345.26$-$0.03, and G348.76$-$0.95 in the lower panel, their S/Ns of the (3,3) line are $\sim$ 4. A comparison of the peak intensities of the different lines given in Tables \ref{parline11-atlasgal} and \ref{parline22_line33-atlasgal} shows that the average main-beam brightness temperature of the (2,2) line is 67\% of the average (1,1) peak intensity, while the average (3,3) main-beam brightness temperature is 40\% of the average (1,1) peak intensity. This reveals that we are probing a range of excitation conditions and thus of star formation activity, which indicates that our sample consists of clumps in various evolutionary phases of high-mass star formation and in different environments.

In addition, we detect two different velocity components in nine NH$_3$ spectra, that is, in G328.2$-$0.55, G336.8+0.03, and G350.1$+$0.09. These result from different clouds located on the same line of sight at different distances. We fit a model including two velocity components to them and denote them with ``a'' and ``b'' after the source name in Tables 1 to 3.

\subsection{Ammonia velocities}
\begin{figure}[h]
\centering
\includegraphics[angle=0,width=10.0cm]{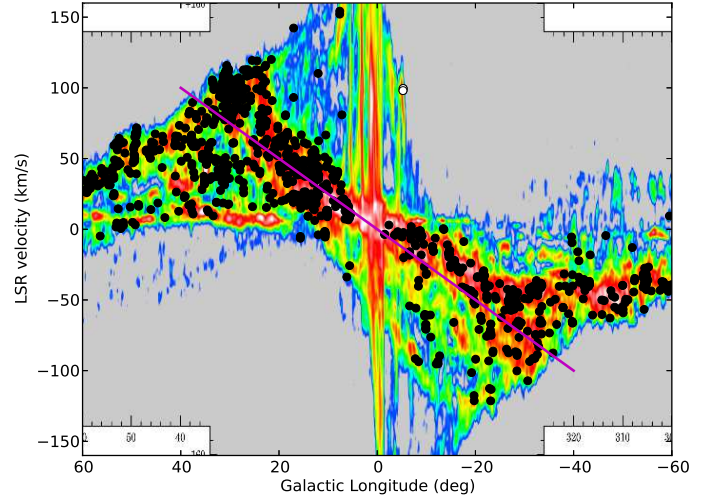}
\caption[dependence of the (1,1) LSR velocity from the Galactic longitude]{Galactic longitude and LSR velocities of detected northern and southern sources are shown with CO $(1-0)$ emission \citep{2001ApJ...547..792D} in the background. The pink straight line indicates the 5 kpc molecular ring \citep{2006ApJ...653.1325S}. Two sources associated with Bania's Clump 1 are illustrated as white points.}\label{coords-v11-atlasgal}
\end{figure}

We illustrate the NH$_3$ velocity and Galactic longitude range of the sources observed in the first quadrant \citep{2012A&A...544A.146W} and in the fourth quadrant in Fig. \ref{coords-v11-atlasgal}. The plot shows that more observations were conducted toward the northern than the southern ATLASGAL clumps. Most southern NH$_3$ velocities range from 5 km~s$^{-1}$ to $-120$ km~s$^{-1}$ with a smaller number of extreme velocities than in the north. Only two clumps, G354.72+0.30 and G354.75+0.37, have high velocities of about 99 km~s$^{-1}$. These are located close to Bania's Clump 1 \citep{1977ApJ...216..381B} and are indicated as white points in Fig. \ref{coords-v11-atlasgal}; it is suggested that they are likely in the Galactic bar because of their high velocities \citep{2010MNRAS.404.1029C}. The background in Fig. \ref{coords-v11-atlasgal} shows CO $(1-0)$ emission reported by \cite{2001ApJ...547..792D} and indicates a good correlation of most clumps with the CO intensity, which traces giant molecular clouds. The pink straight line denotes the 5 kpc molecular ring that is the most massive and active star-forming molecular gas accumulation in our Galaxy \citep{2006ApJ...653.1325S}. The NH$_3$ radial velocities are combined with the Galactic rotation model \citep{1993A&A...275...67B} to determine kinematic distances. We distinguish between near and far distances by examining HI self-absorption and HI absorption towards a large sample of ATLASGAL sources, which gives an unbiased three-dimensional view of high-mass star-forming regions in the first and fourth quadrant \citep{2015A&A...579A..91W}.

\subsection{Line width}
\label{linewidth}
The NH$_3$ (1,1) FWHM line widths lie between 0.8 km~s$^{-1}$ and $\sim 5$ km~s$^{-1}$, while the (2,2) inversion transitions have widths of up to 8 km~s$^{-1}$. Although the backend spectral resolution of the southern observations of 0.4 km~s$^{-1}$ is better than that of the northern NH$_3$ sample of 0.7 km~s$^{-1}$, the narrowest (1,1) FWHM line widths of the southern and northern clumps are similar, $\sim$ 0.8 km~s$^{-1}$. The average of measured southern (1,1) line widths is $\sim$ 2 km~s$^{-1}$, which agrees with the average obtained in the first quadrant and has mainly non-thermal contributions.
The range of the (2,2) and (3,3) line widths is similar, the (2,2) lines exhibit widths from 0.7 km~s$^{-1}$ to 8 km~s$^{-1}$ and the (3,3) lines from 0.8 km~s$^{-1}$ to 7.6 km~s$^{-1}$. However, a comparison shows that most ATLASGAL sources have a larger width of the (3,3) inversion transition than of the (2,2) line. Clumps emitting a NH$_3$ (3,3) line must be warmer as a high temperature is needed to excite the (3,3) inversion transition. This indicates that star formation activity heats these sources in the innermost parts of the clump where the density and turbulence are highest. 

The NH$_3$ (2,2) hyperfine components are clearly detected towards 27 sources in the fourth quadrant. We used a hyperfine structure fit for this subsample, which allowed us to measure the (2,2) optical depth. We give the optical depth, $\tau$(2,2), and the
line
width obtained from the hyperfine structure fit of the NH$_3$ (2,2) line, $\Delta \rm v_{\rm hfs}$(2,2), for the subsample of 31 sources in Table \ref{22-hyperfine}.

\begin{figure}[h]
\centering
\includegraphics[angle=0,width=9.0cm]{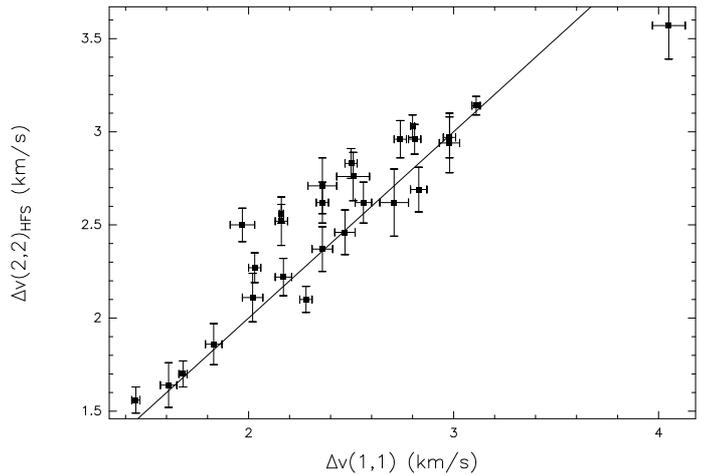}\vspace*{0.5cm}
\caption[comparison of (1,1) linewidths and (2,2) linewidths with measured optical depth]{Correlation plot of (1,1) and (2,2) line widths for sources for which the hyperfine structure of the NH$_3$ (2,2) line is detected. The solid line presents equal line widths.}\label{dv11-dv22-atlasgal}
\end{figure}

The (1,1) line widths are plotted against the (2,2) line widths for the subsample with detected hyperfine structure of the (2,2) line in Fig. \ref{dv11-dv22-atlasgal}. For most sources, the two line widths agree, while a few clumps exhibit broader (2,2) than (1,1) line widths. Some of them with the largest line broadening show signs of star formation activity, but a few are not detected at 24 $\mu$m. Among these ``24 $\mu$m dark'' clumps are G326.64+0.61, for which \cite{2009A&A...505..405P} did not find any hint of star formation activity within this IRDC, and G354.62+0.47, which is associated with 6.7 GHz methanol maser emission \citep{2013MNRAS.431.1752U}. Among the infrared bright sources are G333.02+0.77, which is located within an IRDC and associated with a 24 $\mu$m source \citep{2009A&A...505..405P}, G351.45+0.66, a bright hot core within the NGC 6334 molecular cloud complex \citep{1982ApJ...255..103R}, and G327.29$-$0.58, a bright hot core within a high-mass star-forming complex \citep{2009A&A...501L...1M,2012ApJ...752..146L}.

\begin{figure}[h]
\centering
\includegraphics[angle=0,width=9.0cm]{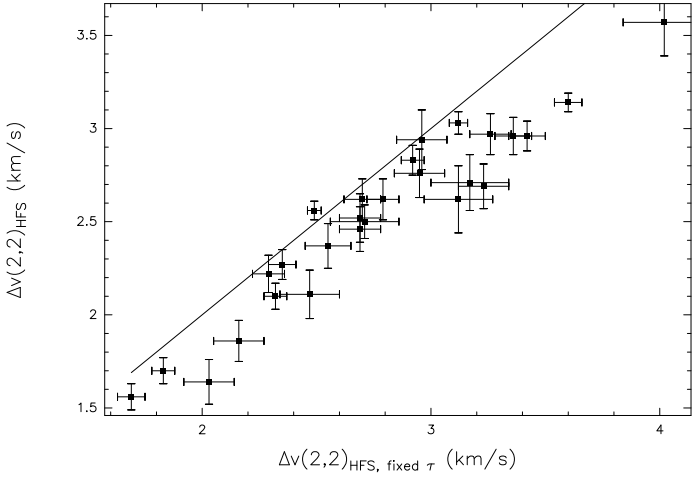}\vspace*{0.5cm}
\caption[comparison of (2,2) linewidths with measured and fixed optical depth]{Comparison of (2,2) line widths determined from a hyperfine structure fit with measured (2,2) optical depth against those with fixed (2,2) optical depth. The solid line displays equal line widths.}\label{dv22-tau22-atlasgal}
\end{figure}

Guided by the method described by \cite{2008ApJS..175..509R}, we calculated the (2,2) optical depth, $\tau$ (2,2), from the (1,1) and (2,2) main-beam brightness temperatures, $T_{\mbox{\tiny MB}}$, and the (1,1) optical depth, $\tau$ (1,1):
\begin{eqnarray}\label{tau22calc}
\tau (2,2) = \rm \ln\left(1- \frac{T_{\mbox{\tiny MB}}(2,2)}{T_{\mbox{\tiny MB}}(1,1)} (1- \rm exp(-\tau (1,1)))\right).
\end{eqnarray}
This was used as a fixed parameter in the hyperfine structure fit of the (2,2) transition of the whole ATLASGAL sample in the fourth quadrant. The derived (2,2) line widths are given in Table \ref{parline22_line33-atlasgal}. The intrinsic (2,2) line widths resulting from a hyperfine structure fit with the measured (2,2) optical depth against those derived from the hyperfine structure fit with fixed (2,2) optical depth are compared in Fig. \ref{dv22-tau22-atlasgal}. It reveals a constant offset from equal line widths, explained in Sect. \ref{rotational temperature}, with broader (2,2) line
widths resulting from the hypefine structure fit with fixed (2,2) optical depth. 

\begin{table*}
\begin{minipage}{\textwidth}
\caption{NH$_3$(1,1) line parameters. Errors are given in parentheses. The full table is available at the CDS.}              
\label{parline11-atlasgal}      
\centering                                     
\begin{tabular}{l l l l l l l}         
\hline\hline                      
 & RA\tablefootmark{1} & Dec\tablefootmark{1} & $\tau$(1,1) & $\rm v$(1,1) & $\Delta \rm v$(1,1) & T$_{\mbox{\tiny MB}}$(1,1) \\ 
Name  & (J2000) &  (J2000) &  & (km~s$^{-1}$) &  (km~s$^{-1}$) & (K) \\               
\hline                                 
 G300.72+1.20 & 12 32 50.23 & -61 35 27.4 & 1.98 $(\pm$0.89) & -43.96 $(\pm$0.19) & 2.89 $(\pm$0.38) & 0.18 $(\pm$0.05) \\
G300.82+1.15 & 12 33 40.83 & -61 38 52.9 & 2.06 $(\pm$0.39) & -42.72 $(\pm$0.05) & 1.78 $(\pm$0.11) & 0.41 $(\pm$0.04) \\
G300.91+0.88 & 12 34 14.23 & -61 55 22.7 & 1.37 $(\pm$0.23) & -40.87 $(\pm$0.04) & 2.52 $(\pm$0.09) & 0.82 $(\pm$0.06) \\
G300.97+1.15 & 12 34 52.75 & -61 39 47.9 & 0.71 $(\pm$0.37) & -42.51 $(\pm$0.07) & 2.16 $(\pm$0.16) & 0.28 $(\pm$0.03) \\
G301.01+1.11 & 12 35 14.97 & -61 41 55.4 & 1.36 $(\pm$0.37) & -42.02 $(\pm$0.06) & 2.22 $(\pm$0.13) & 0.56 $(\pm$0.06) \\
G301.14-0.23 & 12 35 34.87 & -63 02 31.1 & 0.10 $(\pm$0.12) & -39.58 $(\pm$0.16) & 3.42 $(\pm$0.31) & 0.14 $(\pm$0.03) \\
G301.12+0.96 & 12 36 01.82 & -61 51 27.8 & 1.41 $(\pm$0.19) & -40.13 $(\pm$0.02) & 1.56 $(\pm$0.05) & 1.10 $(\pm$0.05) \\
G301.12+0.98 & 12 36 02.85 & -61 50 26.0 & 1.68 $(\pm$0.14) & -40.14 $(\pm$0.01) & 1.33 $(\pm$0.03) & 1.49 $(\pm$0.05) \\
G301.14+1.01 & 12 36 14.07 & -61 48 39.3 & 1.72 $(\pm$0.17) & -39.89 $(\pm$0.03) & 2.33 $(\pm$0.05) & 0.95 $(\pm$0.05) \\
G301.68+0.25 & 12 40 32.82 & -62 35 54.8 & 3.08 $(\pm$0.45) & -37.67 $(\pm$0.04) & 1.38 $(\pm$0.07) & 0.64 $(\pm$0.05) \\
G301.74+1.10 & 12 41 19.90 & -61 44 41.1 & 1.72 $(\pm$0.14) & -39.49 $(\pm$0.02) & 1.90 $(\pm$0.04) & 0.47 $(\pm$0.02) \\
G301.81+0.78 & 12 41 53.36 & -62 04 12.3 & 0.42 $(\pm$0.63) & -37.13 $(\pm$0.08) & 1.57 $(\pm$0.21) & 0.28 $(\pm$0.04) \\
G302.39+0.28 & 12 46 44.26 & -62 35 11.9 & 1.34 $(\pm$0.28) & -43.19 $(\pm$0.04) & 1.87 $(\pm$0.09) & 0.44 $(\pm$0.03) \\
G304.20+1.34 & 13 02 06.13 & -61 30 25.3 & 1.54 $(\pm$0.51) & -43.63 $(\pm$0.05) & 1.34 $(\pm$0.12) & 0.42 $(\pm$0.05) \\
G304.76+1.34 & 13 06 45.70 & -61 28 39.9 & 0.67 $(\pm$0.21) & -26.80 $(\pm$0.02) & 1.43 $(\pm$0.06) & 0.90 $(\pm$0.04) \\

\hline                                          
\end{tabular}
\tablefoot{
\tablefoottext{1}{Units of right ascension are hours, minutes, and seconds, and units of declination are degrees, arcminutes, and arcseconds.}
}
\end{minipage}
\end{table*}

\begin{table*}
\begin{minipage}{\textwidth}
\caption{NH$_3$(2,2) and (3,3) line parameters. Errors are given in parentheses. The full table is available at the CDS.}              
\label{parline22_line33-atlasgal}     
\centering   
\begin{tabular}{l l l l l l l}   
\hline\hline                       
 & $\rm v$(2,2) & $\Delta \rm v$(2,2) & T$_{\mbox{\tiny MB}}$(2,2) & $\rm v$(3,3) & $\Delta \rm v$(3,3) & T$_{\mbox{\tiny MB}}$(3,3) \\ 
Name  & (km~s$^{-1}$) &  (km~s$^{-1}$) &  (K) & (km~s$^{-1}$) &  (km~s$^{-1}$) & (K) \\                    
\hline                                  
 G300.72+1.20 & - & - & $<$ 0.05	& - & - & $<$ 0.05 \\
G300.82+1.15 & -42.54 $(\pm$0.12) & 1.83 $(\pm$0.28) & 0.18 $(\pm$0.03)	& - & - & $<$ 0.04 \\
G300.91+0.88 & -41.20 $(\pm$0.14) & 2.77 $(\pm$0.35) & 0.33 $(\pm$0.06)	& - & - & $<$ 0.07 \\
G300.97+1.15 & -42.57 $(\pm$0.13) & 2.49 $(\pm$0.24) & 0.19 $(\pm$0.03)	& -42.76 $(\pm$0.18) & 4.09 $(\pm$0.45) & 0.14 $(\pm$0.03) \\
G301.01+1.11 & -41.84 $(\pm$0.19) & 3.40 $(\pm$0.41) & 0.18 $(\pm$0.04)	& -41.44 $(\pm$0.23) & 2.61 $(\pm$0.57) & 0.12 $(\pm$0.04) \\
G301.14-0.23 & -38.49 $(\pm$0.44) & 5.38 $(\pm$1.42) & 0.07 $(\pm$0.03)	& - & - & $<$ 0.03 \\
G301.12+0.96 & -40.28 $(\pm$0.06) & 1.77 $(\pm$0.14) & 0.48 $(\pm$0.05)	& -40.81 $(\pm$0.14) & 1.43 $(\pm$0.34) & 0.16 $(\pm$0.04) \\
G301.12+0.98 & -40.19 $(\pm$0.04) & 1.39 $(\pm$0.09) & 0.66 $(\pm$0.05)	& -40.29 $(\pm$0.43) & 3.44 $(\pm$1.31) & 0.10 $(\pm$0.05) \\
G301.14+1.01 & -40.02 $(\pm$0.09) & 2.65 $(\pm$0.17) & 0.38 $(\pm$0.05)	& -40.21 $(\pm$0.14) & 2.86 $(\pm$0.41) & 0.27 $(\pm$0.05) \\
G301.68+0.25 & -37.64 $(\pm$0.14) & 1.70 $(\pm$0.34) & 0.22 $(\pm$0.05)	& - & - & $<$ 0.05 \\
G301.74+1.10 & -39.57 $(\pm$0.07) & 2.18 $(\pm$0.13) & 0.17 $(\pm$0.02)	& -39.72 $(\pm$0.17) & 1.66 $(\pm$0.45) & 0.06 $(\pm$0.02) \\
G301.81+0.78 & -37.79 $(\pm$0.16) & 0.95 $(\pm$0.26) & 0.14 $(\pm$0.05)	& - & - & $<$ 0.05 \\
G302.39+0.28 & -43.20 $(\pm$0.10) & 2.15 $(\pm$0.24) & 0.19 $(\pm$0.03)	& - & - & $<$ 0.03 \\
G304.20+1.34 & -43.71 $(\pm$0.21) & 2.21 $(\pm$0.50) & 0.16 $(\pm$0.05)	& - & - & $<$ 0.04 \\
G304.76+1.34 & -26.37 $(\pm$0.11) & 1.37 $(\pm$0.26) & 0.21 $(\pm$0.05)	& - & - & $<$ 0.06 \\

\hline                   
\end{tabular}                              
\end{minipage}
\end{table*}

\begin{table*}
\begin{minipage}{\textwidth}
\caption{Parameters derived from the NH$_3$(1,1) to (3,3) inversion transitions. Errors are given in parentheses. The full table is available at the CDS.}             
\label{parabgel-atlasgal}      
\centering                                     
\begin{tabular}{l l l l }          
\hline\hline                        
 & T$_{{\mbox{\tiny rot}}}$ & T$_{\mbox{\tiny kin}}$ & log$_{10}$(N$_{\mbox{NH}_3}$)   \\ 
Name  & (K) &  (K) &   (cm$^{-2}$)  \\                  
\hline                                  
 G300.72+1.20 & - & - & - \\
G300.82+1.15 & 15.3 $(\pm$1.7) & 17.3 $(\pm$2.4) & 15.3 $(\pm$0.09) \\
G300.91+0.88 & 15.8 $(\pm$1.5) & 18.1 $(\pm$2.2) & 15.28 $(\pm$0.08) \\
G300.97+1.15 & 22.8 $(\pm$3.5) & 29.9 $(\pm$6.8) & 15.03 $(\pm$0.24) \\
G301.01+1.11 & 14.4 $(\pm$1.7) & 16.1 $(\pm$2.3) & 15.21 $(\pm$0.12) \\
G301.14-0.23 & 21.3 $(\pm$5.6) & 27.0 $(\pm$10.3) & 14.35 $(\pm$0.54) \\
G301.12+0.96 & 16.5 $(\pm$1.0) & 19.1 $(\pm$1.5) & 15.09 $(\pm$0.06) \\
G301.12+0.98 & 16.0 $(\pm$0.7) & 18.4 $(\pm$1.0) & 15.09 $(\pm$0.04) \\
G301.14+1.01 & 15.2 $(\pm$1.0) & 17.2 $(\pm$1.4) & 15.34 $(\pm$0.04) \\
G301.68+0.25 & 12.5 $(\pm$1.2) & 13.6 $(\pm$1.6) & 15.36 $(\pm$0.07) \\
G301.74+1.10 & 14.4 $(\pm$0.7) & 16.0 $(\pm$1.0) & 15.25 $(\pm$0.04) \\
G301.81+0.78 & 20.6 $(\pm$4.9) & 25.8 $(\pm$8.7) & 14.62 $(\pm$0.66) \\
G302.39+0.28 & 16.4 $(\pm$1.5) & 19.0 $(\pm$2.2) & 15.15 $(\pm$0.09) \\
G304.20+1.34 & 15.2 $(\pm$2.3) & 17.3 $(\pm$3.2) & 15.05 $(\pm$0.15) \\
G304.76+1.34 & 14.0 $(\pm$1.2) & 15.5 $(\pm$1.6) & 14.72 $(\pm$0.14) \\
G305.10+0.09 & 22.5 $(\pm$5.8) & 29.3 $(\pm$11.2) & 14.99 $(\pm$0.47) \\

\hline                                            
\end{tabular}
\end{minipage}
\end{table*}

\begin{figure}[h]
\centering
\includegraphics[angle=0,width=9.0cm]{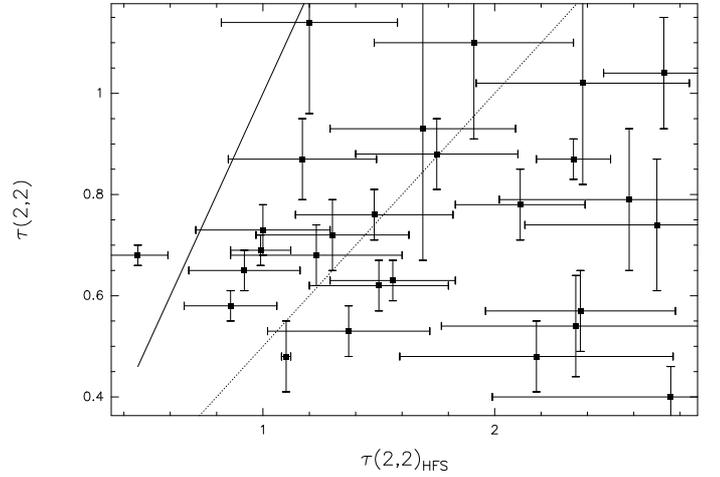}\vspace*{0.5cm}
\caption[comparison of calculated and fitted (2,2) optical depth]{(2,2) optical depth calculated from the main-beam brightness temperatures of the (1,1) and (2,2) lines, and the (1,1) optical depth is plotted against the depth derived from the hyperfine structure. The straight line indicates equal values and the dotted line $\tau (2,2) = \tau (2,2)_{\mbox{ \tiny HFS}}/2$.}\label{tau22}
\end{figure}

\begin{figure}[h]
\centering
\includegraphics[angle=0,width=9.0cm]{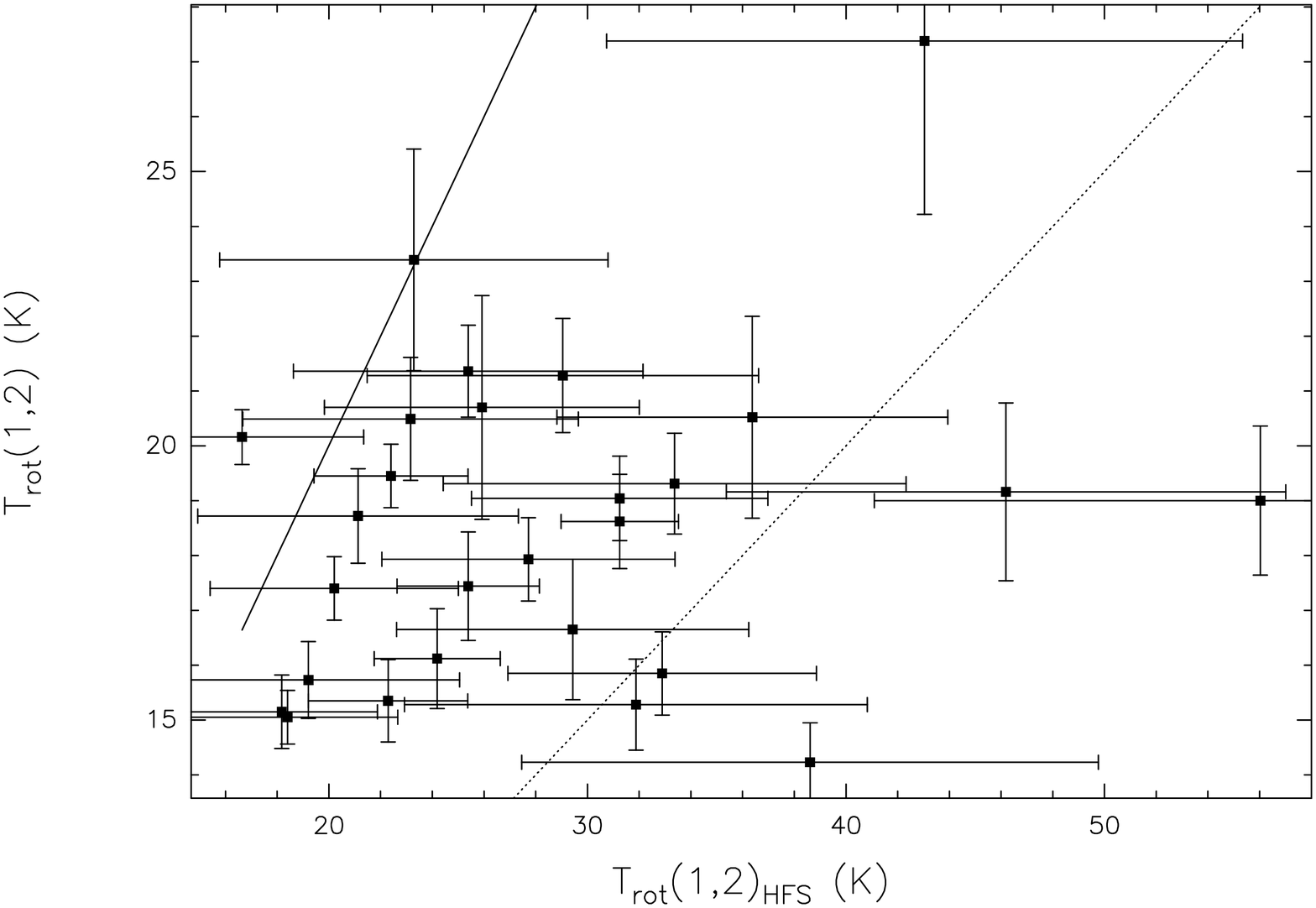}
\caption[comparison of (1,1) and (2,2) linewidth]{Comparison of the rotational temperature between the (1,1) and (2,2) inversion transition using the (2,2) optical depth derived from hyperfine structure fits, $T_{\mbox{\tiny rot}}(1,2)_{\mbox{\tiny HFS}}$, and the temperature obtained from the calculated (2,2) optical depth, T$_{\mbox{\tiny rot}}$ (1,2). The straight line shows equal values, while the dotted line illustrates T$_{\mbox{\tiny rot}}$ (1,2) = 1/2$T_{\mbox{\tiny rot}}(1,2)_{\mbox{\tiny HFS}}$.}\label{trot}
\end{figure}

\subsection{Rotational temperature and source-averaged NH$_3$ column density}
\label{rotational temperature}
Assuming equal beam-filling factors, we calculated the rotational temperature between the (1,1) and (2,2) inversion transition using \citep{1983ARA&A..21..239H}
\begin{eqnarray}\label{Trot}
T_{\mbox{\tiny rot}}=\frac{-41.5}{\ln\left( -0.282 \frac{\tau (2,2)}{\tau (1,1)} \right)}
,\end{eqnarray}
where $\tau (1,1)$ and $\tau (2,2)$ are the optical depths of the (1,1) and (2,2) main lines. Because we did not detect the hyperfine structure of most (2,2) lines, we computed the (2,2) optical depth from Equation \ref{tau22calc}. The same excitation temperature was assumed for the (1,1) and (2,2) lines, $T_{\mbox{\tiny ex}}$, in Equations \ref{tau22calc} and \ref{Trot}. We tested this assumption for the subsample given in Sect. \ref{linewidth}, for which the optical depth could be derived from the detected (1,1) and (2,2) hyperfine components. The excitation temperature can be computed from
\begin{eqnarray}
 T_{\rm ex} = \frac{T_{\rm MB}(1,1)}{1- {\rm exp}( -\tau (1,1))} + T_{\rm bg}
\end{eqnarray}
with T$_{\rm bg} = 2.73$ K and assuming a beam-filling factor of 1. The excitation temperature of the NH$_3$ (2,2) line is on average lower but still within 10-20\% of the (1,1) excitation temperature value. It is unclear whether this small difference is due to an actual change in the excitation temperature or a change in the filling factor.

The rotational temperatures range between 12 K and 28 K (see Table \ref{parabgel-atlasgal}), and the mean value is 18.6 K with an uncertainty of 15\%. For the subsample described in Sect. \ref{linewidth} with detected hyperfine satellites of the (2,2) line, we compared the (2,2) optical depth, $\tau(2,2)_{\mbox{\tiny HFS}}$, to the values computed with Equation \ref{tau22calc}. The solid line in Fig. \ref{tau22} indicates $\tau (2,2) = \tau(2,2)_{\mbox{\tiny HFS}}$. Figure \ref{tau22} illustrates that the (2,2) optical depths from hyperfine structure fits are larger than the calculated values, except for three sources, which might explain the offset seen in Fig. \ref{dv11-dv22-atlasgal}. The clump with the lowest (2,2) optical depths of these is G331.51$-$0.10, for which \cite{1997A&AS..124..509S} detected OH-maser emission at 1612 MHz. The other two sources are G351.45+0.66, the bright hot core described in Sect. \ref{linewidth} that is also a known methanol maser emitting at 6.6 GHz and 12 GHz \citep{1995MNRAS.272...96C}, and G305.37+0.17, a compact HII region in the rich high-mass star-forming complex G305 \citep{2004A&A...427..839C,2013MNRAS.435.2003H}. The dotted line indicates $\tau (2,2) = \tau(2,2)_{\mbox{\tiny HFS}}/2$, many sources lie between the two relations. Clumps with large fitted (2,2) optical depths lie below the dotted line, they have the largest deviations between calculated and fitted (2,2) optical depth. These are weak sources with the lowest S/Ns and larger errors of the (2,2) optical depth. 

Because hyperfine anomalies that affect mostly the outer satellites are found in the NH$_3$ sample in the fourth quadrant (see Sect. \ref{anomalies measurement}), the excitation temperatures of the main line and satellites are likely to be different, which might influence the estimation of the optical depth from the hyperfine structure. To test this, we compared the (1,1) optical depth derived from the main line and inner satellite lines, $\tau_{\rm inner \ HFS}$, with the (1,1) optical depth determined from all hyperfine structure lines, $\tau_{\rm all \ HFS}$, including the outer satellites associated with the largest hyperfine anomalies. A fit to the two optical depths yields $\tau_{\rm inner \ HFS} = 0.92 \tau_{\rm all \ HFS} +0.17$ and a Spearman correlation test gives a coefficient of 0.86 with a p-value $< 0.0013$. This shows that the correlation between the optical depth obtained from the main line and the inner satellites and the optical depth derived from all hyperfine structure lines is significant over 3$\sigma$. Most sources in the fourth quadrant therefore exhibit similar $\tau_{\rm inner \ HFS}$ and $\tau_{\rm all \ HFS}$ ,
and the ratio $\tau_{\rm inner \ HFS}/\tau_{\rm all \ HFS}$ also remains constant as a function of the hyperfine anomalies.

The rotational temperature, T$_{\mbox{\tiny rot}}$, determined from Equations \ref{Trot} and \ref{tau22calc}, is compared with that using the (2,2) optical depth derived from the hyperfine structure fit, $T_{\mbox{\tiny rot}}(1,2)_{\mbox{\tiny HFS}}$. Figure \ref{trot} shows that the rotational temperature with calculated $\tau (2,2)$ is mostly lower than $T_{\mbox{\tiny rot}}(1,2)_{\mbox{\tiny HFS}}$. This trend is similar to that of the (2,2) optical depth in Fig. \ref{tau22} and results from the difference between calculated and fitted (2,2) optical depths. The relation T$_{\mbox{\tiny rot}}$ = 1/2$T_{\mbox{\tiny rot}}(1,2)_{\mbox{\tiny HFS}}$ is a lower boundary to the rotational temperature distribution of most clumps with detected (2,2) hyperfine structure fits.

To test how representative clumps with (2,2) hyperfine structure are for the whole southern sample, we compared their rotational temperature and (1,1) optical depth. A Kolmogorov-Smirnov test does not contradict the assumption that the subsample and all sources in the fourth quadrant have the same rotational temperature distribution. A comparison of the (1,1) optical depth shows that small optical depths are missing in the subsample with (2,2) hyperfine structure, but it still covers a wide range from $\sim 1$ to 5 and is not strongly skewed to sources exhibiting high optical depths.

We present the rotational temperature and column density obtained from the measured (2,2) optical depth in Table \ref{22-hyperfine}. We estimate a ratio of T$_{\rm rot}(1,2)_{\rm HFS}$ to T$_{\rm rot}$ of $1.56 \pm 0.5$ and a ratio of the NH$_3$ column density derived using the (2,2) optical depth from the hyperfine structure fit to that determined from the (1,1) optical depth and the line temperature ratio of $1.56 \pm 0.64$.

These factors are not used to correct the rotational temperature and column density of the whole sample in the fourth quadrant, to be consistent with our method in the first quadrant \citep{2012A&A...544A.146W}. Moreover, the (2,2) hyperfine structure could only be detected for a small subsample of 27 sources, resulting in a large scatter of the data. The hyperfine structure fit with the (2,2) optical depth derived from the (1,1) optical depth and line temperature ratio leads to an underestimation of the rotational temperature and column density.

\begin{table*}
\begin{minipage}{\textwidth}
\caption{NH$_3$ line parameters derived from hyperfine structure fits of the (2,2) inversion transition for the subsample with detected (2,2) satellites.}             
\label{22-hyperfine}     
\centering 
\renewcommand{\arraystretch}{1.4}
                          
\begin{tabular}{l c c c c}          
\hline\hline                       
Name & $\tau $(2,2) & $\Delta \rm v_{\rm hfs}$(2,2) & T$_{\rm rot, hfs}$ & N$_{\rm NH_3,hfs}$  \\ 
     &   & (km~s$^{-1}$) & (K) & (10$^{14}$ cm$^{-2}$)   \\    
\hline                                 
 G316.81$-$0.06 & 1.91 $(\pm$0.43) & 2.37 $(\pm$0.12) & 25.9 $(\pm$6.1) & 50.3 $(\pm$13.4) \\
G337.92$-$0.48 & 1.69 $(\pm$0.40) & 2.76 $(\pm$0.13) & 43.0 $(\pm$12.3) & 49.4 $(\pm$22.8) \\
G326.64+0.61 & 0.99 $(\pm$0.13) & 3.03 $(\pm$0.06) & 22.4 $(\pm$3.0) & 34.4 $(\pm$4.2) \\
G351.45+0.66 & 0.46 $(\pm$0.13) & 2.56 $(\pm$0.05) & 16.6 $(\pm$4.7) & 19.2 $(\pm$2.3) \\
G351.58$-$0.35 & 2.73 $(\pm$0.26) & 2.96 $(\pm$0.08) & 24.2 $(\pm$2.4) & 89.4 $(\pm$9.5) \\
G351.77$-$0.54 & 2.34 $(\pm$0.16) & 3.14 $(\pm$0.05) & 31.3 $(\pm$2.3) & 77.4 $(\pm$7.5) \\
G331.88+0.06 & 2.37 $(\pm$0.41) & 2.69 $(\pm$0.12) & 32.9 $(\pm$6.0) & 71.4 $(\pm$17.8) \\
G332.09$-$0.42 & 1.17 $(\pm$0.32) & 2.22 $(\pm$0.10) & 23.2 $(\pm$6.5) & 30.6 $(\pm$8.3) \\
G305.37+0.17 & 1.20 $(\pm$0.38) & 2.94 $(\pm$0.16) & 23.3 $(\pm$7.5) & 42.9 $(\pm$13.4) \\
G328.26$-$0.53 & 1.50 $(\pm$0.30) & 2.97 $(\pm$0.11) & 27.7 $(\pm$5.7) & 48.5 $(\pm$11.8) \\
G327.29$-$0.58 & 0.92 $(\pm$0.24) & 2.27 $(\pm$0.08) & 25.4 $(\pm$6.8) & 21.1 $(\pm$6.1) \\
G329.03$-$0.20 & 1.23 $(\pm$0.37) & 2.52 $(\pm$0.13) & 19.2 $(\pm$5.9) & 39.8 $(\pm$8.3) \\
G333.02+0.77 & 0.86 $(\pm$0.20) & 2.83 $(\pm$0.08) & 20.2 $(\pm$4.8) & 30.0 $(\pm$5.5) \\
G333.13$-$0.56 & 2.11 $(\pm$0.28) & 2.96 $(\pm$0.10) & 22.3 $(\pm$3.1) & 72.2 $(\pm$9.4) \\
G336.02$-$0.83 & 2.70 $(\pm$0.57) & 2.71 $(\pm$0.15) & 46.2 $(\pm$10.8) & 77.4 $(\pm$29.1) \\
G316.77$-$0.02 & 1.00 $(\pm$0.29) & 2.62 $(\pm$0.11) & 21.1 $(\pm$6.2) & 31.2 $(\pm$7.6) \\
G337.92$-$0.46 & 1.30 $(\pm$0.33) & 2.62 $(\pm$0.11) & 29.1 $(\pm$7.6) & 35.7 $(\pm$11.6) \\
G342.71+0.13 & 1.56 $(\pm$0.27) & 2.10 $(\pm$0.07) & 31.3 $(\pm$5.7) & 37.8 $(\pm$9.3) \\
G343.50+0.02 & 2.18 $(\pm$0.59) & 1.86 $(\pm$0.11) & 31.9 $(\pm$9.0) & 42.4 $(\pm$16.0) \\
G305.24+0.26 & 1.37 $(\pm$0.35) & 2.46 $(\pm$0.12) & 33.4 $(\pm$9.0) & 36.1 $(\pm$13.4) \\
G351.78$-$0.52 & 1.48 $(\pm$0.34) & 1.70 $(\pm$0.07) & 18.4 $(\pm$4.3) & 39.8 $(\pm$5.6) \\
G331.55$-$0.07 & 2.35 $(\pm$0.58) & 2.62 $(\pm$0.18) & 56.0 $(\pm$14.9) & 89.3 $(\pm$38.6) \\
G354.62+0.47 & 1.10 $(\pm$0.02) & 2.50 $(\pm$0.09) & 25.4 $(\pm$2.8) & 24.5 $(\pm$3.9) \\
G347.63+0.15 & 2.38 $(\pm$0.46) & 3.57 $(\pm$0.18) & 36.4 $(\pm$7.6) & 104.5 $(\pm$31.9) \\
G351.74$-$0.58 & 1.75 $(\pm$0.35) & 1.56 $(\pm$0.07) & 18.2 $(\pm$3.7) & 41.5 $(\pm$5.2) \\
G345.72+0.82 & 2.76 $(\pm$0.77) & 1.64 $(\pm$0.12) & 38.6 $(\pm$11.2) & 49.1 $(\pm$21.1) \\
G347.97$-$0.43 & 2.58 $(\pm$0.56) & 2.11 $(\pm$0.13) & 29.4 $(\pm$6.8) & 55.7 $(\pm$16.6) \\

\hline                                            
\end{tabular}
\end{minipage}
\end{table*}

\subsection{Comparison of NH$_3$ line parameters in the first and fourth quadrant}
\begin{figure}[h]
\centering
\includegraphics[angle=-90,width=9.0cm]{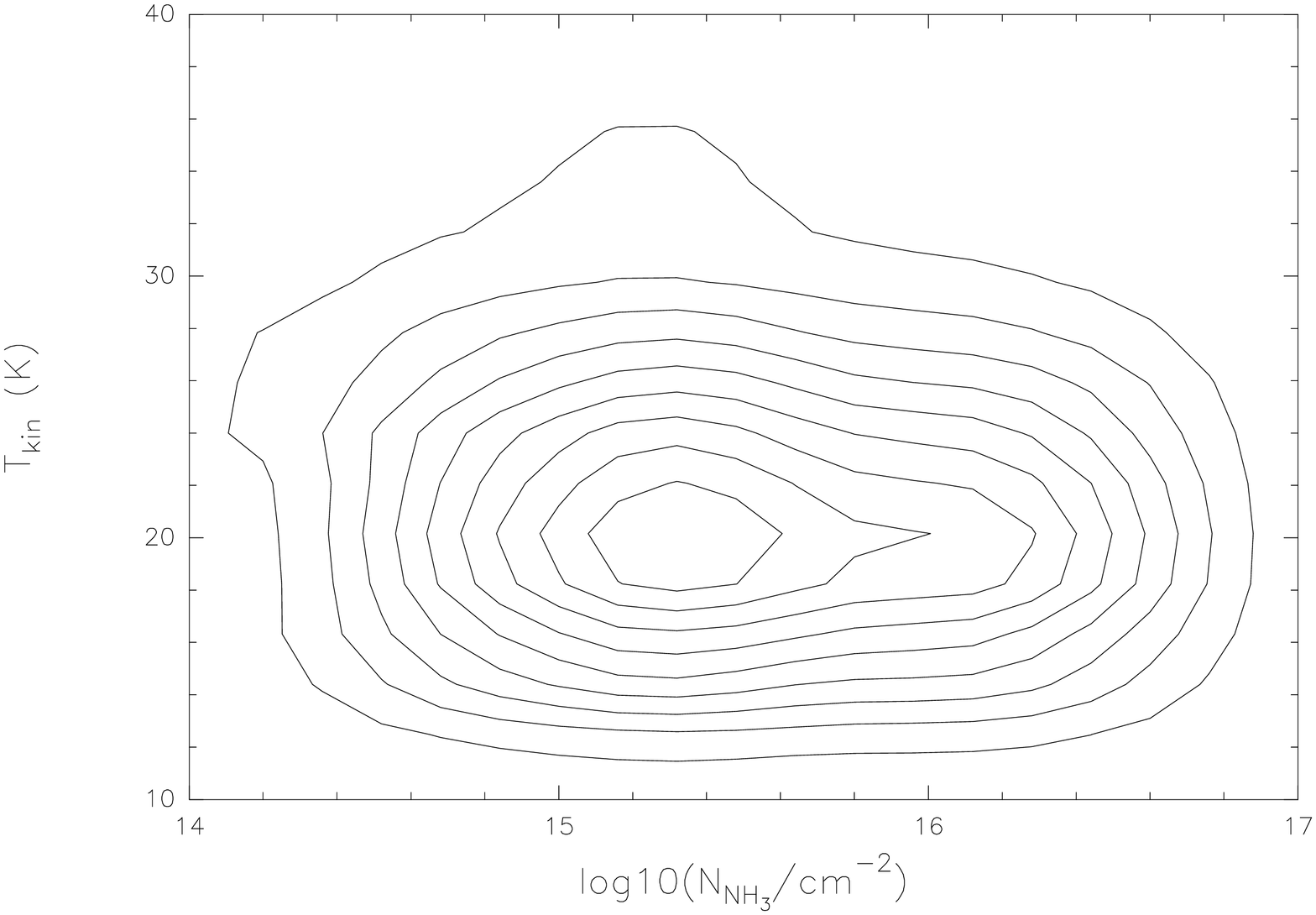}\vspace*{0.5cm}
\includegraphics[angle=0,width=9.0cm]{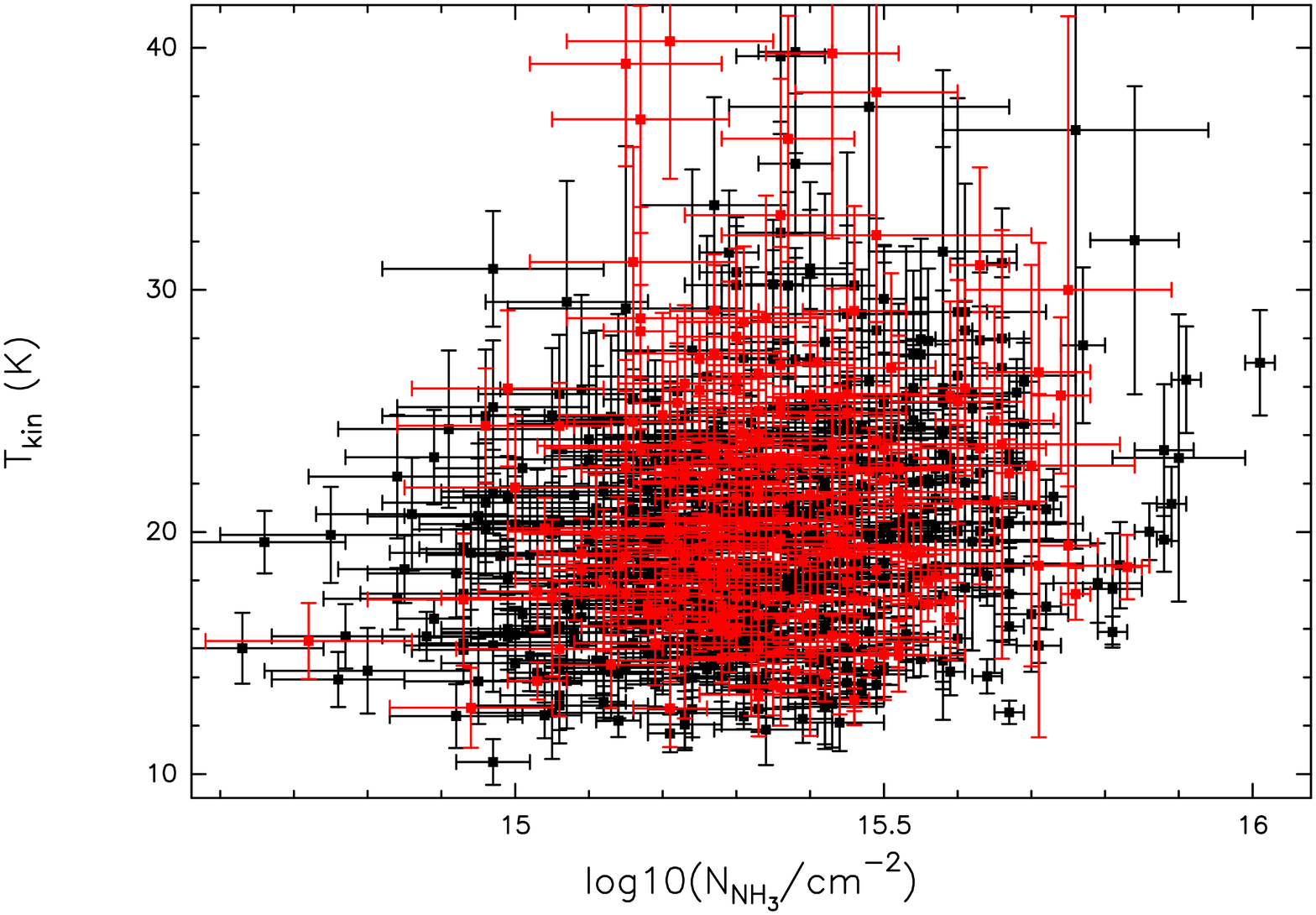}
\caption[comparison of logarithm of column density and kinetic temperature]{Kinetic temperature plotted against the logarithm of the NH$_3$ column density. Northern ATLASGAL sources are shown in black and the southern sample in red. For the contour plot in the upper panel, we counted the number of sources in each column density bin of 1 cm$^{-2}$ and in each kinetic temperature bin of 3 K. The contours illustrate 10\%\ to 90\% in steps of 10\% of the peak source number per bin. These levels are shown in all contour plots in this article.}\label{ncol}
\end{figure}

\begin{figure}[h]
\centering
\includegraphics[angle=0,width=9.0cm]{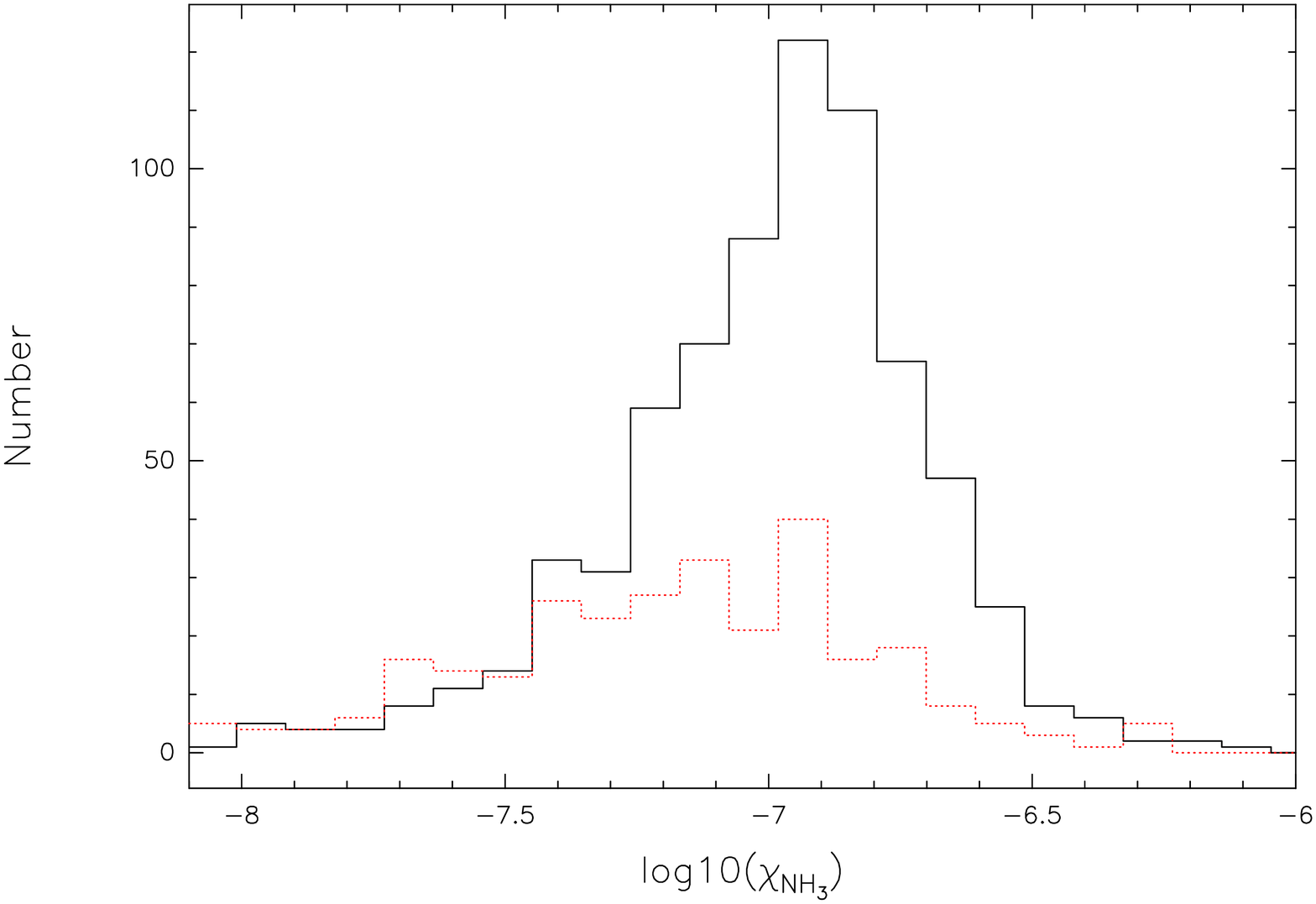}
\caption[comparison of logarithm of abundance]{Histogram of the logarithm of the NH$_3$ abundance for sources in the first quadrant are plotted as a solid black curve and for clumps in the fourth quadrant as a dashed red curve.}\label{abund}
\end{figure}

We used the same assumptions and equations to calculate the source-averaged NH$_3$ column density and the kinetic temperature from the rotational temperature of the ATLASGAL sources in the fourth quadrant as in the first quadrant, which are given in \cite{2012A&A...544A.146W}. Figure \ref{ncol} displays the kinetic temperature plotted against the logarithm of the NH$_3$ column density for the southern sources as red points and the northern sample in black. We obtained kinetic temperatures of the southern clumps in a similar range as those of the northern NH$_3$ sample from 13 K to 40 K with an uncertainty of 24\%. The median kinetic temperature of clumps in the first quadrant is 19 K, which is slightly lower than the peak of 23 K of the southern sources in Fig. \ref{ncol}. However, the sample in the first quadrant contains sources with lower peak fluxes, decreasing to 0.4 Jy/beam, than that in the fourth quadrant. The median kinetic temperature of northern clumps with a peak flux density above 1.2 Jy/beam is $\sim 22$ K and agrees with that of the sources in the fourth quadrant. Values of the NH$_3$ column density lie between $5\times 10^{14}$ cm$^{-2}$ and $7\times 10^{15}$ cm$^{-2}$, the peak at $2\times 10^{15}$ cm$^{-2}$ is the same as obtained for the northern clumps. Figure \ref{abund} presents the distribution of the logarithm of the NH$_3$ abundance for the sample in the first quadrant as a solid black curve and for sources in the fourth quadrant as a dashed red curve. The comparison shows a consistent range of the abundance for the two subsamples. The slightly higher abundance peak of the northern sample observed with the Effelsberg telescope might result from a smaller beam width tracing higher abundances towards the central parts of the clumps.

\subsection{Variation of NH$_3$ line parameters}
\label{NH3 variation}
The longitude and radial velocity of a source determine a unique galactocentric radius using a model of the Galactic rotation curve \citep[e.g.][]{1993A&A...275...67B}. We analysed trends of physical parameters obtained from NH$_3$ inversion transitions within the inner Galaxy. We divided galactocentric radii of the NH$_3$ subsamples in the first and fourth quadrant of between 3 kpc and 8 kpc into bins of width 1 kpc and investigated the distribution of the means of each bin. The rotational temperature, calculated with Equation \ref{Trot}, shows a flat distribution for the northern and southern subsamples with mean values of
between 16 K and 23 K, which is consistent with kinetic temperatures of sources from the BGPS \citep[Bolocam Galactic Plane Survey,][]{2011ApJS..192....4A} within the inner Galaxy observed in NH$_3$ \citep{2011ApJ...741..110D}. We also inspected the influence of the heliocentric distance on the distribution of the rotational temperature and NH$_3$ (1,1) line width (see Sect. \ref{linewidth}), but no dependence was found. Furthermore, we computed the H$_2$ column density via Equation 11 in \cite{2012A&A...544A.146W} using the beam width of 61$\arcsec$ of the Parkes telescope. We smoothed the ATLASGAL maps to the resolution of the Parkes telescope and extracted the peak flux of the clumps. The mean values of the H$_2$ column density in the first and fourth quadrants lie between $10^{22}$ and $4 \times 10^{22}$ cm$^{-2}$ within galactocentric radii from 3 to 8 kpc, which is similar to the mean H$_2$ column density of the BGPS sample. Moreover, the (1,1) line widths, exhibiting mean values between $\sim 2$ and 4 km~s$^{-1}$, are also approximately constant within a $3 \sigma$ uncertainty for the northern and southern subsamples. This shows that clumps in the spiral arms are not hotter or more turbulent, which indicates that the ratio of star-forming to quiescent clumps is the same in the spiral arms and the interarm regions. This is consistent with the investigations of \cite{2013MNRAS.431.1587E} and \cite{2014ApJ...780..173B}, who found no changes in clump formation efficiency with different environments located in the Galactic plane. 

\begin{figure}[h]
\centering
\includegraphics[angle=0,width=9.0cm]{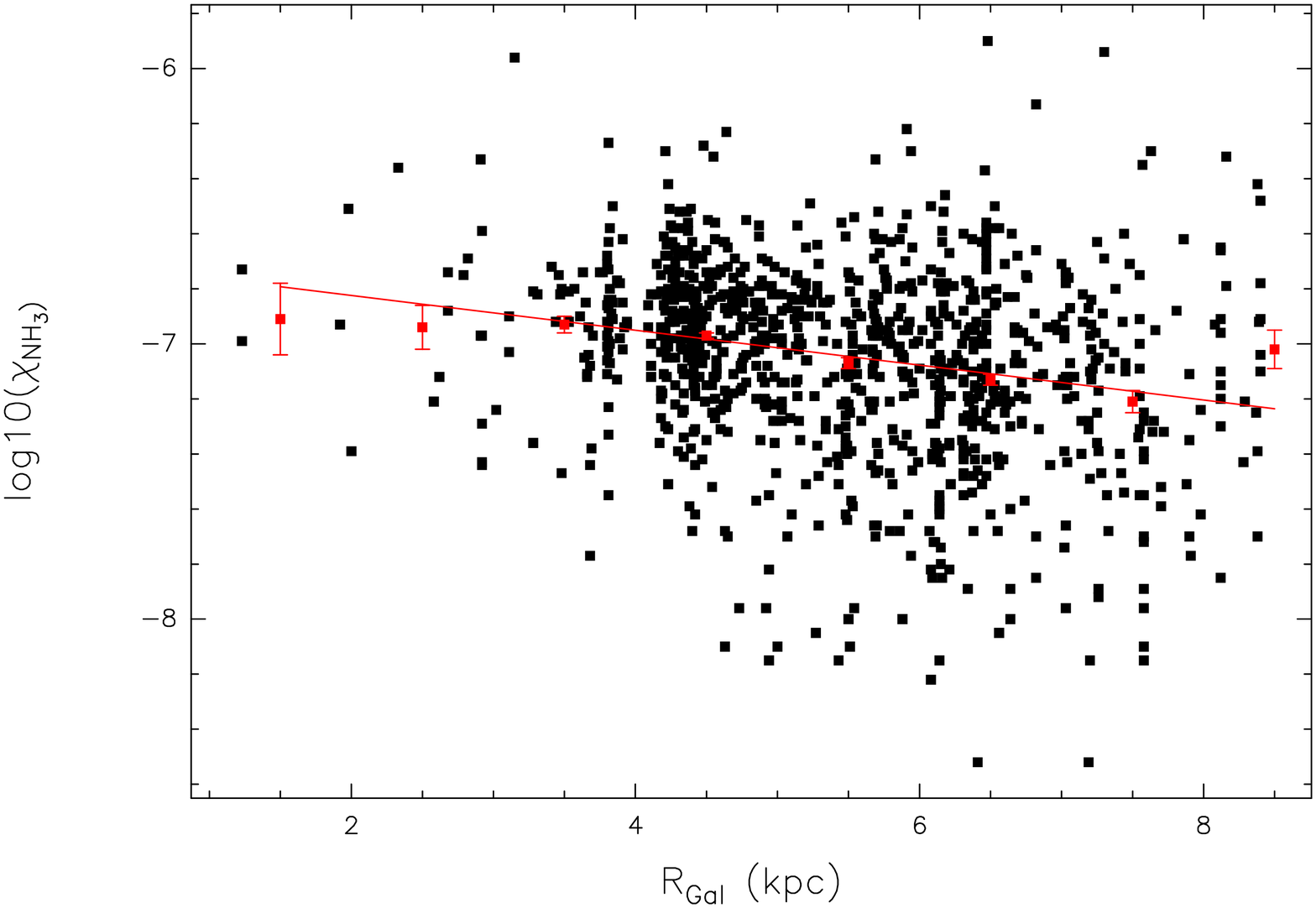}\vspace*{0.5cm}
\includegraphics[angle=0,width=9.0cm]{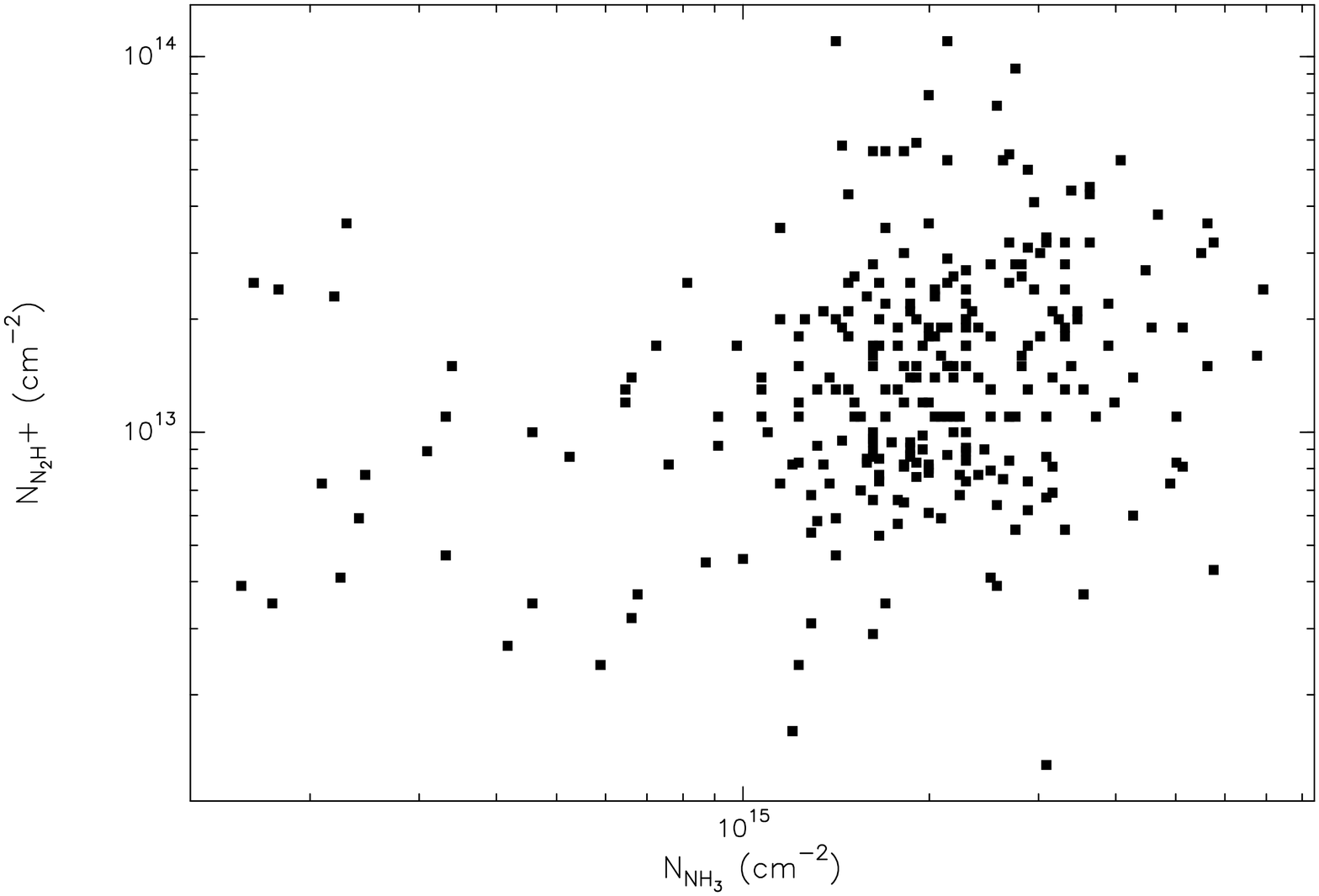}
\caption[NH$_3$ and N$_2$H$^+$ abundances]{Dependence of the logarithm of the NH$_3$ abundance on the galactocentric radius (upper panel). The whole sample is displayed in black, while the mean of each bin in galactocentric radius is indicated in red. The straight line shows a fit to the binned data. The correlation plot of the NH$_3$ and N$_2$H$^+$ column density is illustrated in the lower panel.}\label{nh3-n2hp}
\end{figure}

\begin{figure}[h]
\centering
\includegraphics[angle=0,width=9.0cm]{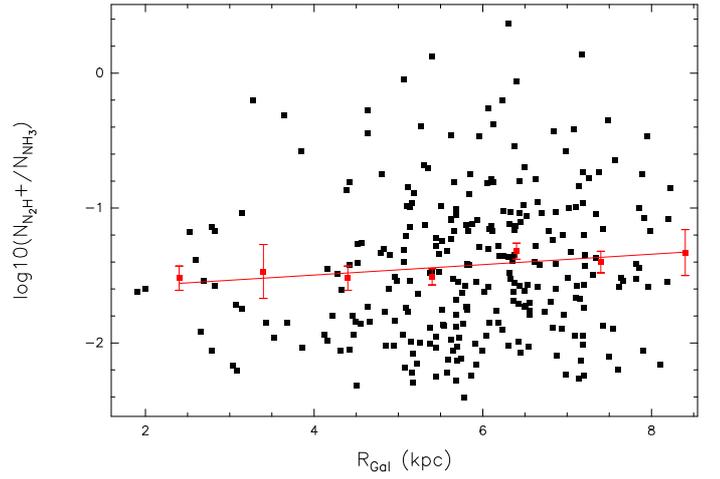}
\caption[N$_2$H$^+$ to NH$_3$ ratio]{Distribution of the N$_2$H$^+$/NH$_3$ column density ratio within the inner Galaxy. The whole sample is shown as black points and the mean of each bin in galactocentric radius as red points.}\label{n2hp-nh3-ratio}
\end{figure}

The analysis of the NH$_3$ abundance as a function of galactocentric radius towards BGPS sources by \cite{2011ApJ...741..110D} shows a decreasing trend. We also determined the NH$_3$ abundance of ATLASGAL sources in the first and fourth quadrant as described in \cite{2012A&A...544A.146W}. While we used the same relation to calculate the H$_2$ column density as \cite{2010ApJ...717.1157D}, we derived a source-averaged NH$_3$ column density similar as in \cite{2012A&A...544A.146W}, and \cite{2010ApJ...717.1157D} reported a beam-averaged NH$_3$ column density. Because our NH$_3$ abundances are affected by the unknown filling factor of the sources as described in \cite{2012A&A...544A.146W}, we obtain higher values from $\sim 10^{-6}$ to 10$^{-8}$ than the range determined by \cite{2011ApJ...741..110D} between $10^{-7}$ and $10^{-9}$. The upper panel of Fig. \ref{nh3-n2hp} shows the logarithm of the NH$_3$ abundance plotted against the galactocentric radius. The whole sample is illustrated as black points and the mean of each bin in galactocentric radius as red points. The distribution is fitted by the relation
\begin{eqnarray}
 \chi_{\mbox{\tiny NH}_3} = 10^{-6.70 \pm 0.04}10^{-0.063 \pm 0.007R_{\mbox{\tiny gal}}}
.\end{eqnarray}
This reveals a decrease in NH$_3$ abundance of $-0.063 \pm 0.007$ dex/kpc. Using the NH$_3$ abundance of BGPS sources \citep{2011ApJ...741..110D} only within galactocentric radii between $\sim 1$ and 8 kpc, we obtain a gradient of $-0.067 \pm 0.016$ dex/kpc, which is slightly steeper than our value, but agrees within 3$\sigma$. While the ATLASGAL sample lies within galactocentric radii from 1.2 to 8.4 kpc, \cite{2011ApJ...741..110D} added the Gemini OB1 molecular cloud from the outer Galaxy, which lowers their gradient in the NH$_3$ abundance to $-0.096$ dex/kpc. They considered varying dust properties such as a changing gas-to-dust ratio, different dust temperatures, or different metallicities in the Galaxy as explanations for this trend. In addition, they suggested that the decreasing nitrogen abundance with rising galactocentric radii \citep[e.g.][]{1983MNRAS.204...53S,2000A&A...363..537R} is likely to reduce the NH$_3$ formation. \cite{1983MNRAS.204...53S} determined a gradient in the nitrogen abundance of $-0.09 \pm 0.015$ dex/kpc mainly in the outer Galaxy, but it is unclear if the decrease still exists in the inner Galaxy. We only obtain a gradient for the ATLASGAL sample when we include NH$_3$ abundances at the innermost galactocentric radii between 1 and 4 kpc. If weaker NH$_3$ formation resulted from a reduced amount of nitrogen, we would expect an even steeper decrease in N$_2$H$^+$ abundance, consisting of two nitrogen atoms, than of NH$_3$. This would result in a decreasing trend of the N$_2$H$^+$/NH$_3$ column density ratio as a function of galactocentric radius. We measured the N$_2$H$^+$ $(1-0)$ line using the Mopra telescope for a southern ATLASGAL subsample of 293 sources observed in NH$_3$. The N$_2$H$^+$ $(1-0)$ column density was calculated using \citep{1998ApJ...506..743B}
\begin{eqnarray}\label{column density}
 N_{\rm N_2H^+} = \frac{3.3 \times 10^{11} \tau \Delta \rm v_{\rm N_2H^+} T_{\rm kin}}{1 - \rm exp(-4.47/T_{\rm kin})}
,\end{eqnarray}
with the optical depth of the N$_2$H$^+$ $(1-0)$ line, $\tau$, the N$_2$H$^+$ line width, $\Delta \rm v_{\rm N_2H^+}$, and the kinetic temperature derived from NH$_3$ observations, T$_{\rm kin}$. While \cite{1998ApJ...506..743B} used in this equation the excitation temperature, it can be low due to an additional small beam-filling factor or sub-thermal excitation. To be consistent with our NH$_3$ analysis, we used instead the kinetic temperature in Equation \ref{column density} to estimate the N$_2$H$^+$ column density. Systematic errors are introduced by replacing the excitation temperature with the kinetic temperature. To estimate these, we computed with RADEX \citep{2007A&A...468..627V} the expected excitation temperature for a range of H$_2$ densities using typical column densities and kinetic temperatures of our sample. For a density of 10$^5$ cm$^{-3}$ similar to the median of the ATLASGAL sample \citep{2015A&A...579A..91W}, the NH$_3$ column density would be underestimated by 37\%, while for higher densities of 10$^6$ cm$^{-3}$, more appropriate to regions dominated by the more compact N$_2$H$^+$ emission, we would overestimate the column density by a factor of 1.45. The comparison of the N$_2$H$^+$ and NH$_3$ column density is presented in the lower panel of Fig. \ref{nh3-n2hp}, which shows higher NH$_3$ column densities ranging from $1.5 \times 10^{14}$ to $7 \times 10^{15}$ cm$^{-2}$ than N$_2$H$^+$ column densities between $6 \times 10^{12}$ and $1.7 \times 10^{15}$ cm$^{-2}$ and a trend of increasing N$_2$H$^+$ column density with rising NH$_3$ column density. The logarithm of the N$_2$H$^+$/NH$_3$ column density ratio is illustrated in Fig. \ref{n2hp-nh3-ratio}, which indicates an approximately constant distribution fitted by 
\begin{eqnarray}
 N_{\mbox{\tiny N$_2$H}^+}/N_{\mbox{\tiny NH}_3} = 10^{-1.65 \pm 0.12}10^{0.039 \pm 0.02R_{\mbox{\tiny gal}}}.
\end{eqnarray}
We performed a t-test to investigate whether the slope of the distribution is equal to 0. This is rejected if the p-value is lower than the assumed significance level of 0.05. Our t-test result gives a p-value of 0.11 and therefore does not contradict the assumption that the distribution can be fitted by a function with a slope of 0. This is consistent with the formation of N$_2$H$^+$ and NH$_3$ from N$_2$ in most gas-phase models. For the production of N$_2$H$^+$ , the reaction of CH with N forms CN, which reacts with N and gives N$_2$. This reacts with H$_3^+$ and produces N$_2$H$^+$ \citep{1990MNRAS.246..183N,1992ApJ...387..417W}. NH$_3$ can be formed through the reaction of He$^+$ with N$_2$, which gives N$^+$. Subsequent hydrogenation reactions with H$_2$ result in the formation of NH$^+$, NH$_2^+$, NH$_3^+$ , and NH$_4^+$. The recombination of NH$_4^+$ with an electron produces NH$_3$ \citep{1991A&A...242..235L}. Consequently, the NH$_3$ and N$_2$H$^+$ abundances are found to be approximately proportional to N$_2$ in models of prestellar cores in dark clouds \citep{2010A&A...513A..41H}, which would lead us to expect a similar trend of the N$_2$H$^+$ and NH$_3$ column density with galactocentric radius. Moreover, NH$_3$ can also be formed on dust grains and return to the gas through evaporation or sputtering in shocks when the temperature rises \citep[e.g.][]{2012ApJ...760...40A}. This formation process is important close to protostars and in hot cores and is therefore likely to occur in ATLASGAL sources, which are in a more evolved phase of high-mass star formation and exhibit high rotational temperatures and broad line widths.

The decrease in the NH$_3$ abundance across the Galaxy as shown in \cite{2011ApJ...741..110D} and Fig. \ref{nh3-n2hp} might result from a lowered nitrogen abundance, but a larger number of nitrogen atoms within a certain molecule might not be related to an enhanced decrease of its abundance.

\subsection{NH$_3$ (3,3) masers}
The inversion of the NH$_3$ (3,3) level is explained with a pumping process of the (3,3) maser by \cite{1983A&A...124..322G} and \cite{1983A&A...122..164W}. Depending on the nuclear spin configuration of the hydrogen atoms, we distinguish between two species of the ammonia molecule: all hydrogen spins are parallel in ortho-NH$_3$ with $K = 3n,$ and one pair is anti-parallel in para-NH$_3$ with $K \neq 3n$. Radiative transitions follow the selection rule $\Delta K = 0$ and collisional transitions $\Delta K = 3n$, which cannot change the spin orientation and does therefore not transfer between ortho- and para-NH$_3$. Collisions between levels of different parity are favoured. As the lower (3,3)$^-$ state has the same parity as the (0,0)$^-$ level, the upper state (3,3)$^+$ is collisionally connected with the (2,0)$^-$ and (0,0)$^-$ states. The lower (3,3)$^-$ level is depopulated through $\Delta K =3$ collisional transitions to the (1,0)$^+$ state, which decays through radiation to the (0,0)$^-$ level. The inversion of the (3,3) line results from the population of the upper level by collisions with the (0,0)$^-$ state. \cite{1983A&A...122..164W} estimated an H$_2$ density of 7000 cm$^{-3}$, above which this pumping process is expected. For densities higher than 10$^5$ cm$^{-3}$ , the K = 0 levels become populated, can thermalize, and the maser is quenched.

\begin{figure}[h]
\centering
\includegraphics[angle=0,width=9.0cm]{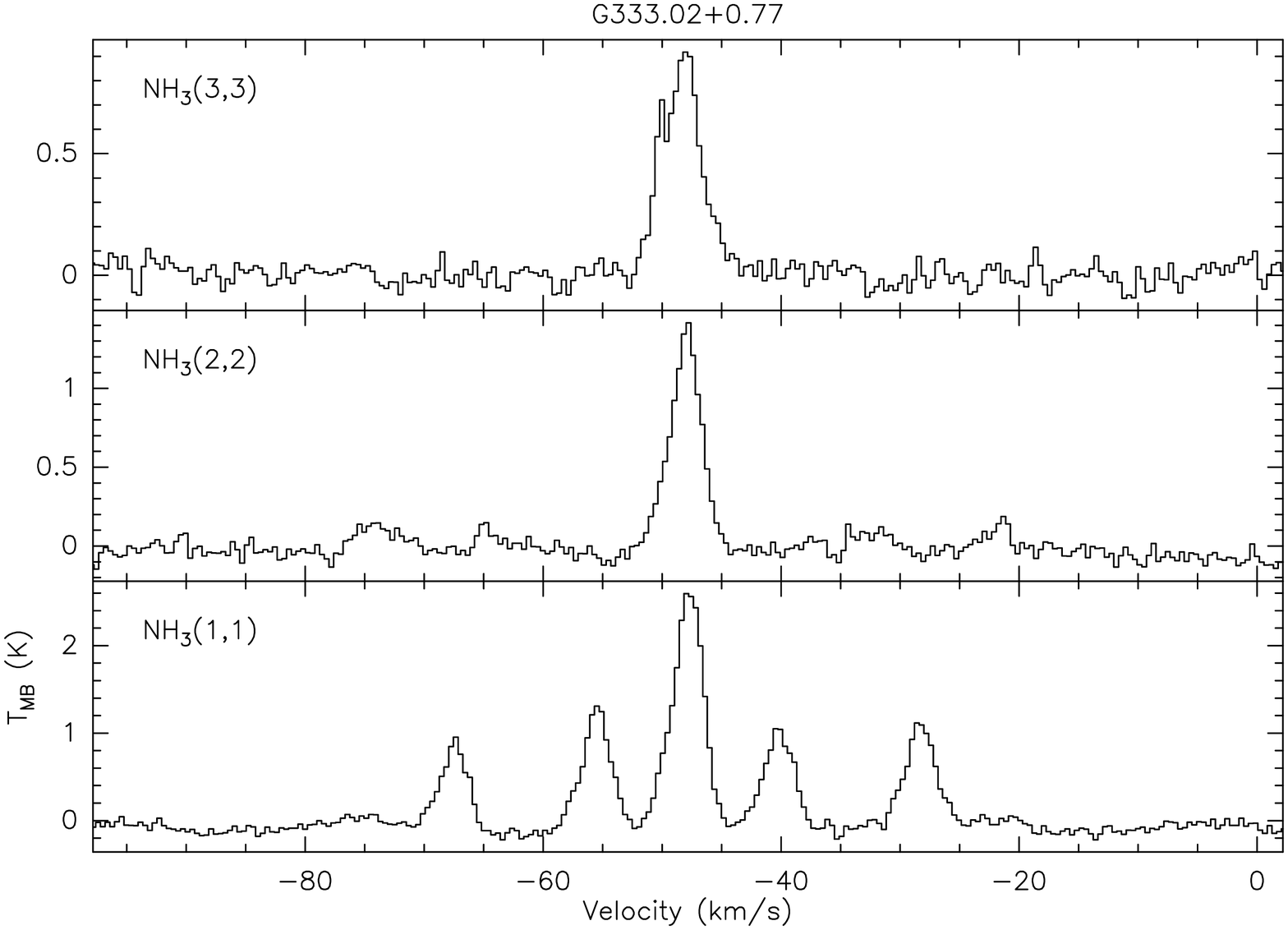}\vspace*{0.5cm}
\includegraphics[angle=0,width=9.0cm]{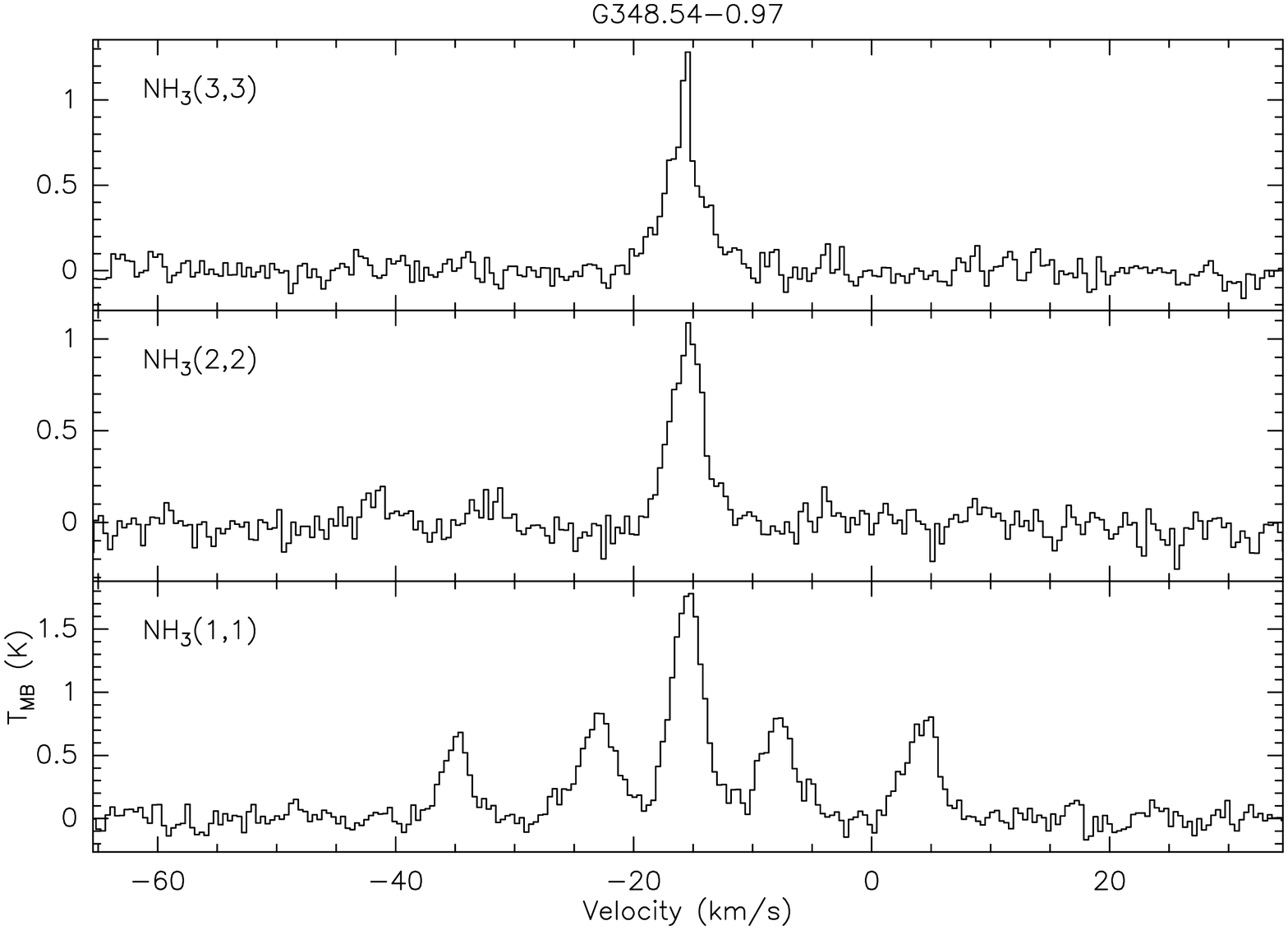}
\caption[NH$_3$ (3,3) maser]{NH$_3$ (1,1) to (3,3) inversion transitions of two ATLASGAL sources with a (3,3) line profile that deviates from a Gaussian fit and indicates a maser line.}\label{33-maser}
\end{figure}

NH$_3$ (3,3) masers have been detected in several high-mass star-forming regions such as DR 21(OH) \citep{1994ApJ...428L..33M}, NGC 6334I \citep{1995ApJ...439L...9K}, IRAS 20126+4104 \citep{1999ApJ...527L.117Z}, G5.89-0.39 \citep{2008ApJ...680.1271H}, G23.33$-$0.30 \citep{2011MNRAS.416.1764W}, G30.7206$-$00.0826 \citep{2011MNRAS.418.1689U}, and G35.03+0.35 \citep{2011ApJ...739L..16B}. We searched for NH$_3$ (3,3) masers in the ATLASGAL sample using the following criteria:
\begin{enumerate}
 \item We searched for sources in which the noise in the Gaussian fit of the (3,3) line is higher than the noise of the baseline. Such clumps have a (3,3) line profile that deviates from a Gaussian.
 \item The (3,3) line has a Gaussian profile, the noise of the baseline is therefore higher than or similar  to the error in the fitted (3,3) line, but the (3,3) line width is smaller than the thermal (1,1) and (2,2) line widths. To ensure that we selected spectra with high S/N, the intensity of the (3,3) line must be higher than 5$\sigma$, where $\sigma$ is the noise of the baseline.  
\end{enumerate}
Taking sources with the largest deviation of the error in the Gaussian fit from the noise of the baseline, $\sigma_{\rm Gauss}/\sigma_{\rm base} > 1.3$, into account, we find three NH$_3$ (3,3) masers. Examining the largest deviation of the noise of the baseline from the error in the fit, $\sigma_{\rm base}/\sigma_{\rm Gauss} > 1.6$, and the smallest line-width ratio, $\Delta v(3,3)/\Delta v(1,1) < 1.2$, results in an additional five masers with an S/N of the (3,3) line higher than 5$\sigma$. We give two examples with the largest $\sigma_{\rm Gauss}/\sigma_{\rm base}$ in Fig. \ref{33-maser}: The (3,3) line of G333.02+0.77 shows an additional peak at a sligthly lower velocity than that of the (3,3) thermal emission. This source is also known as IRAS $16115-4941$, where \cite{1989A&A...221..105S} observed H$_2$O maser emission. \cite{2007IAUS..237..482U} used high-resolution radio continuum observations at 3.6 and 6 cm to identify G333.02+0.77 as an irregular/multi-peaked UCHIIR. \cite{2011ApJS..194...32A} observed hydrogen radio recombination lines at 9 GHz towards G348.54$-$0.97, which is identified as an HII region based on its association with the radio continuum at 20 cm and 24 $\mu$m emission. Their investigation of GLIMPSE maps shows a surrounding layer of 8 $\mu$m emission, which characterizes the source as a bubble. Because G348.54$-$0.97 is associated with methanol masers \citep{1995MNRAS.272...96C,2005A&A...432..737P} and CS ($2 - 1$) emission \citep{1996A&AS..115...81B}, it is likely to be a young HII region.

\subsection{Measuring anomalies in the NH$_3$ (1,1) quadrupole hyperfine structure}
\label{anomalies measurement}
\begin{figure}[h]
\centering
\includegraphics[angle=0,width=9.0cm]{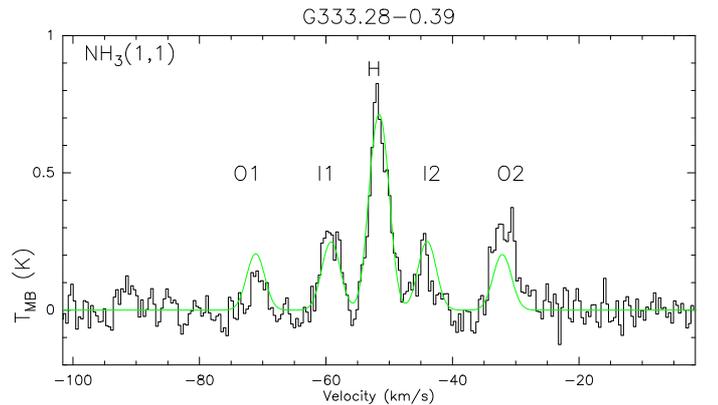}
\caption[hyperfine structure anomalies spectrum]{Example of observed NH$_3$ (1,1) hyperfine structure lines, the outer satellite lines are denoted by O1 and O2, the inner hyperfine lines by I1 and I2, and the main line by H. The NH$_3$ hyperfine structure fit assuming LTE conditions is overlaid in green, which displays equal intensities of O1 and O2 as well as of I1 and I2. The outer satellite lines of the observed spectrum deviate from this prediction.}\label{NH3-outer-satellites}
\end{figure}

It has been suggested that deviations in the intensity ratios of the NH$_3$ (1,1) inner and outer satellite lines to the main line from a symmetric distribution in local thermal equilibrium
(LTE) can be caused by two processes: a nonthermal excitation of the (1,1) hyperfine levels as described by \cite{1977ApJ...214L..67M}, or systematic motion as suggested by \cite{2001A&A...376..348P}. A large portion of the NH$_3$ sample shows such hyperfine anomalies; an example spectrum with an overlaid fit indicating the expected symmetric satellite ratios is given in Fig. \ref{NH3-outer-satellites}. The outer satellites are denoted by O1 and O2, the inner satellites by I1 and I2, and the main hyperfine line by H. The intensities of the outer satellite lines are decreased (O1) or enhanced (O2) with respect to the predicted spectrum.

\begin{figure}[h]
\centering
\includegraphics[angle=0,width=9.0cm]{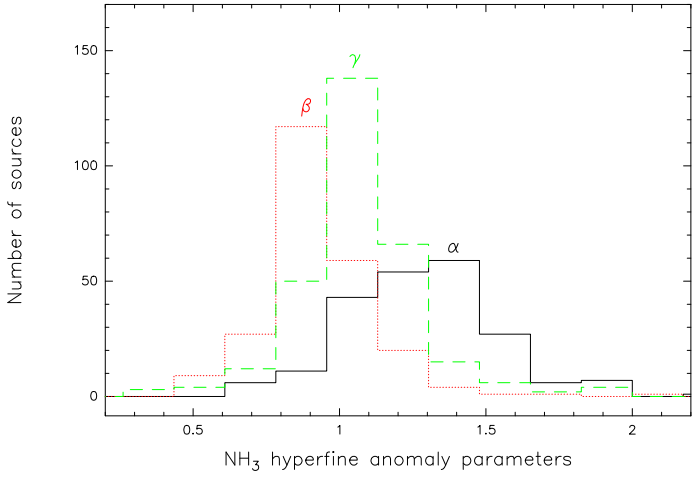}
\caption[alpha and beta distribution]{Number distribution of the NH$_3$ hyperfine anomaly parameters $\alpha$ as the solid black curve, $\beta$ as the dotted red curve, and $\gamma$ as the dashed green curve.}\label{hyperfine-parameters}
\end{figure}

\begin{figure}[h]
\centering
\includegraphics[angle=-90,width=9.0cm]{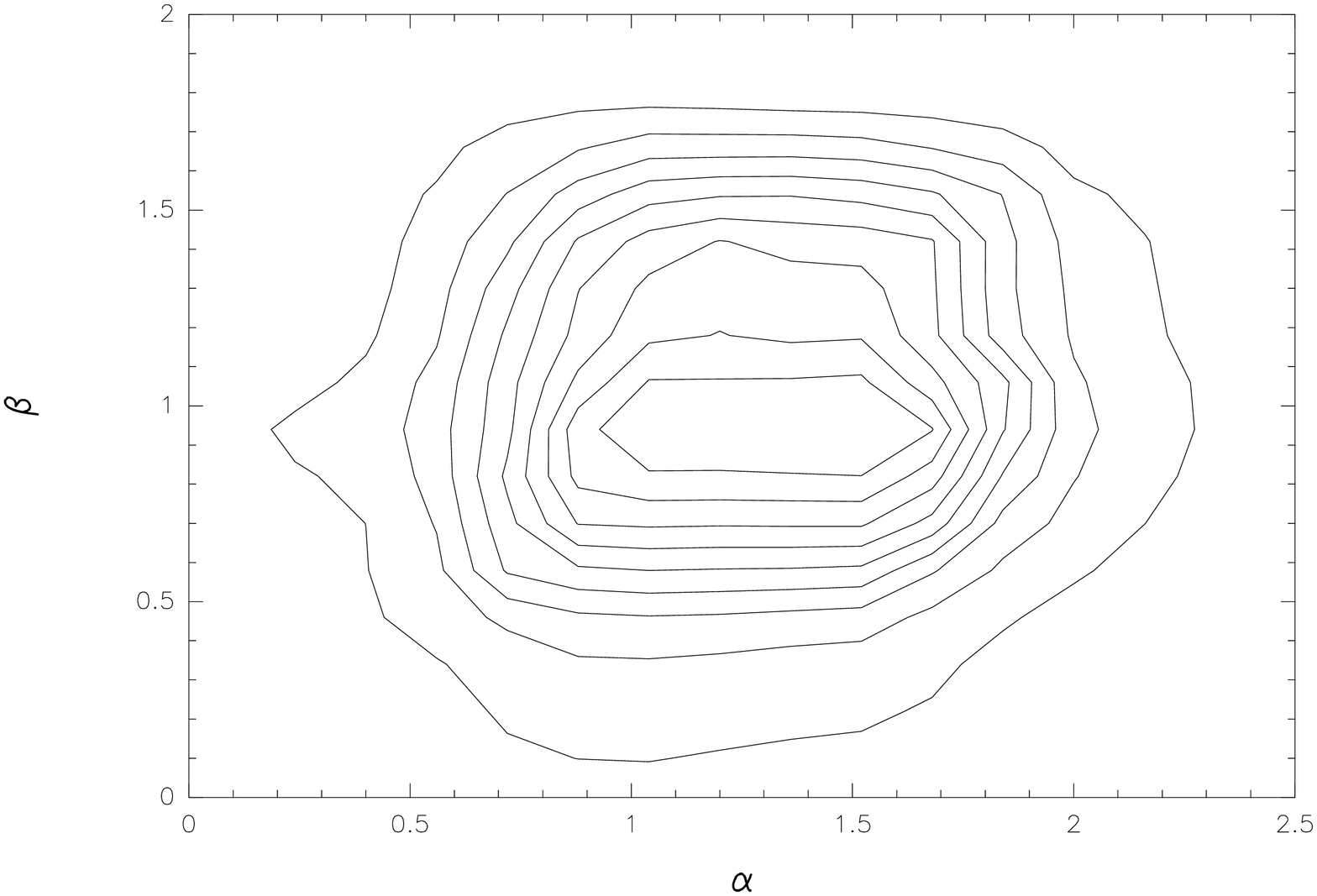}\vspace*{0.5cm}
\includegraphics[angle=-90,width=9.0cm]{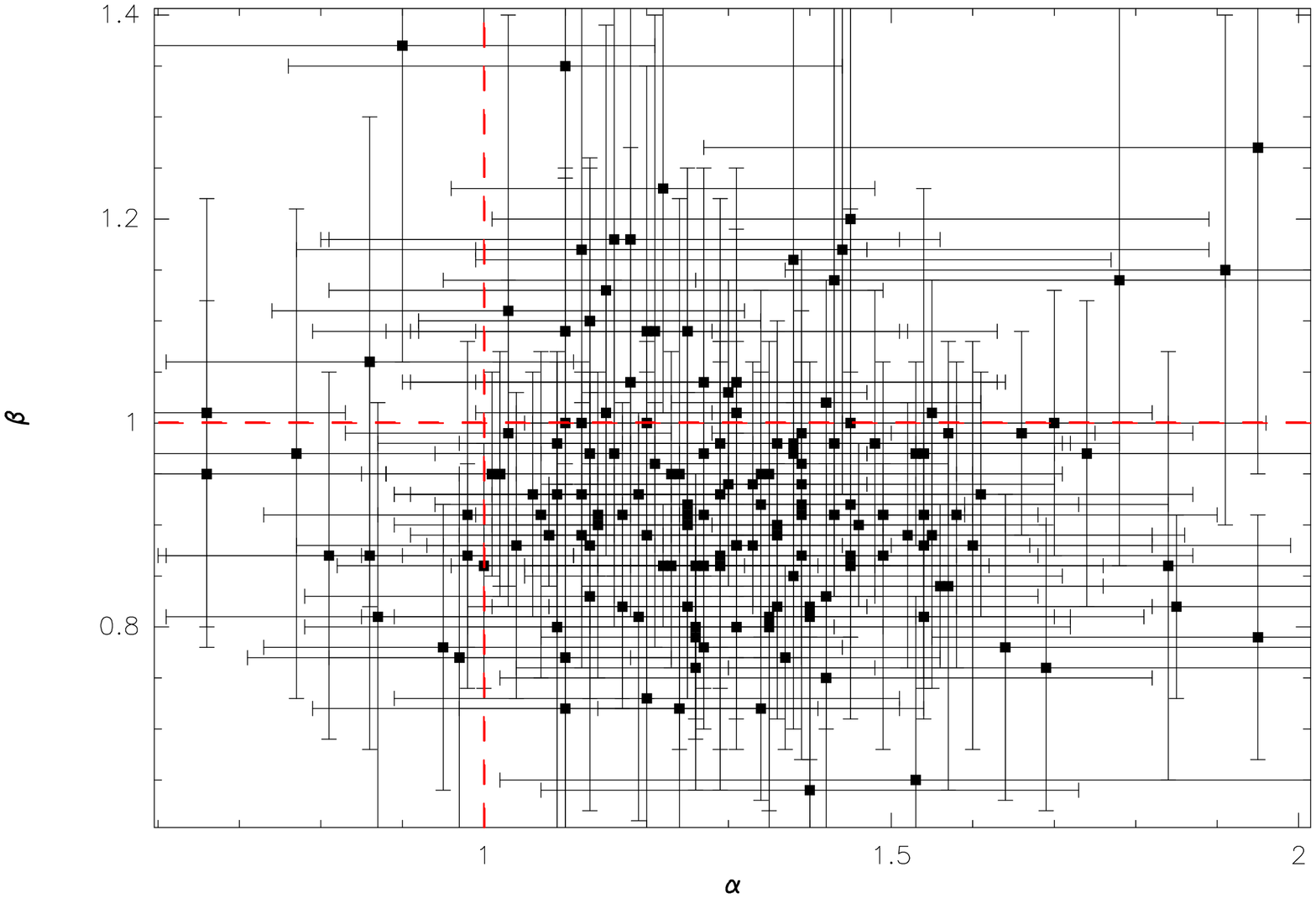}
\caption[comparison of alpha and beta]{Correlation plot of $\alpha$ and $\beta$. The contour plot in the upper panel shows no correlation between the two parameters. For this plot we counted the number of sources in each $\alpha$ and $\beta$ bin of 0.5. The scatter plot in the lower panel indicates that some $\alpha$ and $\beta$ values deviate from 1, as shown by the dashed red lines.}\label{alpha-beta}
\end{figure}

To analyse which of the two suggestions leads to our observed hyperfine anomalies, we examined the ratios of the inner and outer satellite lines as described by \cite{2007MNRAS.379..535L}. If the selective radiative trapping in non-LTE conditions creates the anomalies, the intensity of one outer satellite (O2 in Fig. \ref{NH3-outer-satellites}) will be higher than that of the other outer hyperfine line (O1 in Fig. \ref{NH3-outer-satellites}). In contrast, systematic motion can produce asymmetries in the outer and inner hyperfine structure lines. Hence, the satellites O1 and I1 are stronger than I2 and O2 for infall as a result
of core contraction, while I2 and O2 have increased intensities compared to O1 and I1 for outflow in expanding sources, as shown by \cite{2001A&A...376..348P}. We adapted the technique from \cite{2007MNRAS.379..535L} and fitted individual Gaussian profiles to the five hyperfine structure lines to compute their main-beam brightness temperatures. \cite{2007MNRAS.379..535L} measured the hyperfine anomalies by calculating the ratio of the temperatures of the different hyperfine components, where
\begin{eqnarray}
 \alpha = \frac{\rm O2}{\rm O1},
\end{eqnarray}
\begin{eqnarray}
 \beta = \frac{\rm I2}{\rm I1},
\end{eqnarray}
\begin{eqnarray}
 \gamma = \frac{\rm I2+O2}{\rm O1+I1}.
\end{eqnarray}
These parameters of our NH$_3$ observations are given in Table \ref{hyperfine} with the errors calculated from Gaussian error propagation. We also add the noise of the spectrum, which is used to exclude weak hyperfine components with brightness temperatures lower than the noise. The distributions of $\alpha$ and $\beta$ are shown in Fig. \ref{hyperfine-parameters}, the values of $\alpha$ range from 0.45 to 3.39, $\beta$ lies between 0.45 and 1.95, and $\gamma$ between 0.57 and 1.85. The median of $\alpha$ is $1.27 \pm 0.03$ with a standard deviation of 0.45, a one-sample t-test shows a significant deviation of the peak in the distribution of $\alpha$ from 1 with a p-value $< 0.0001$. While the median of $\beta$ of $0.9 \pm 0.02$ with a standard deviation of 0.3 and the median of $\gamma$ of $1.06 \pm 0.02$ with a standard deviation of 0.37 are closer to 1, a one-sample t-test yields that the peaks of $\beta$ and $\gamma$ differ significantly from 1 with a p-value $<0.0001$. Figure \ref{hyperfine-parameters} also indicates that the distribution of $\alpha$ exhibits a much larger scatter than that of $\beta$ and $\gamma,$ consistent with their standard deviations. Figure \ref{alpha-beta} presents our calculated estimates of $\beta$ plotted against our $\alpha$ values with straight lines indicating where the two parameters equal 1. Defining sources with hyperfine anomalies such as as those exhibiting $\alpha \pm \Delta \alpha$ or $\beta \pm \Delta \beta$ different from 1, $\sim 61$\% of the observed NH$_3$ (1,1) spectra in the fourth quadrant show hyperfine anomalies. We note that the hyperfine model of the NH$_3$ (1,1) line taking all 18 hyperfine components into account leads to a splitting of the outer satellite lines into two components, separated by 0.5 km~s$^{-1}$ for O1 and by 0.14 km~s$^{-1}$ for O2. This can cause a difference of the temperature ratio $\alpha$ from 1 for narrow lines even without hyperfine anomalies. A hyperfine model including a small NH$_3$ (1,1) line width of 1 km~s$^{-1}$ , for example,
leads to $\alpha$ of 1.18, while the deviation from 1 is even smaller for an average NH$_3$ (1,1) line width of 2 km~s$^{-1}$ in our sample, which leads to $\alpha$ of 1.03. Hence, this effect cannot account for the measured median of $\alpha$, $1.27 \pm 0.03$ in our sample, which therefore indicates true hyperfine anomalies.

To assess the influence of the baseline subtraction on the significance of the hyperfine anomalies, we reduced the NH$_3$ (1,1) spectra of the whole sample in the fourth quadrant again using different orders of the baseline. Two choices for an odd order of the baseline and also an even order lead to a variation of the hyperfine anomaly parameters by only 8 to 14\%, which demonstrates that the anomalies are real features.

\cite{2007MNRAS.379..535L} suggested that $\alpha > 1$, $\beta = 1$ for nonthermal excitation, while systematic motion are related to $\alpha, \beta < 1$ for infall and $\alpha, \beta > 1$ for outflow. The median values of $\alpha$ and $\beta$ and their comparison in Fig. \ref{alpha-beta} indicate that non-LTE conditions prevail in most clumps. This result is similar to that of \cite{2007MNRAS.379..535L}, who analysed NH$_3$ hyperfine anomalies in a much smaller sample of 21 southern hot molecular cores. Their comparison of $\alpha$, $\beta,$ and $\gamma$ values in correlation plots indicates that the nonthermal excitation of the cores are more likely than systematic motion. 

\begin{table*}
\begin{minipage}{\textwidth}
\caption{Main-beam brightness temperature ratios of the NH$_3$ (1,1) quadrupole hyperfine structure components. Errors are given in parentheses. $\sigma$ denotes the noise of the spectrum; weak hyperfine structure lines with brightness temperatures lower than $\sigma$ are not used. The full table is available at the
CDS.}              
\label{hyperfine}     
\centering                                     
\begin{tabular}{l l l l l }          
\hline\hline                        
Name & $\alpha$ & $\beta$ & $\gamma$ & $\sigma$   \\ 
         \hline                                 
 G300.72+1.20 & 1.26 $(\pm$1.47) & 1.10 $(\pm$0.80) & 1.16 $(\pm$0.74) & 0.05 \\
G300.82+1.15 & 0.87 $(\pm$0.33) & 0.77 $(\pm$0.23) & 0.82 $(\pm$0.19) & 0.04 \\
G300.91+0.88 & 1.29 $(\pm$0.42) & 0.98 $(\pm$0.24) & 1.10  $(\pm$0.21) & 0.06 \\
G300.97+1.15 & 1.48 $(\pm$0.89) & 1.06 $(\pm$0.51) & 1.23 $(\pm$0.46) & 0.03 \\
G301.01+1.11 & 1.31 $(\pm$0.51) & 0.88 $(\pm$0.26) & 1.05 $(\pm$0.25) & 0.05 \\
G301.12+0.96 & 1.48 $(\pm$0.30) & 0.98 $(\pm$0.15) & 1.17 $(\pm$0.14) & 0.05 \\
G301.12+0.98 & 1.01 $(\pm$0.13) & 0.95 $(\pm$0.10) & 0.98 $(\pm$0.08) & 0.05 \\
G301.14+1.01 & 1.29 $(\pm$0.28) & 0.93 $(\pm$0.15) & 1.07 $(\pm$0.14) & 0.05 \\
G301.68+0.25 & 0.66 $(\pm$0.17) & 1.01 $(\pm$0.21) & 0.83 $(\pm$0.14) & 0.06 \\
G301.74+1.10 & 1.15 $(\pm$0.16) & 1.01 $(\pm$0.12) & 1.07 $(\pm$0.09) & 0.02 \\
G301.81+0.78 & 0.06 $(\pm$0.63) & 0.56 $(\pm$0.45) & 0.37 $(\pm$0.36) & 0.05 \\
G302.39+0.28 & 1.95 $(\pm$0.68) & 1.27 $(\pm$0.32) & 1.54 $(\pm$0.31) & 0.03 \\
G304.20+1.34 & 1.63 $(\pm$0.72) & 0.79 $(\pm$0.28) & 1.10 $(\pm$0.29) & 0.04 \\
G304.76+1.34 & 1.27 $(\pm$0.37) & 1.04 $(\pm$0.21) & 1.13 $(\pm$0.19) & 0.05 \\

\hline                                             
\end{tabular}
\end{minipage}
\end{table*}

\section{Discussion}
\label{discussion}
\subsection{Comparison with nearby molecular clouds}
In this section we investigate how some ATLASGAL clumps compare with properties of nearby molecular clouds. To do this, we compared NH$_3$ properties of ATLASGAL clumps with cores within a region in different nearby molecular clouds that corresponds to the Parkes beam width of 2 pc at an average distance of the ATLASGAL sample of 5 kpc. \cite{2008ApJS..175..509R} observed NH$_3$ (1,1) and NH$_3$ (2,2) lines in the Perseus molecular cloud using the 100 m Robert F. Byrd Green Bank Telescope (GBT). Their beam width of 31$\arcsec$ corresponds to 0.04 pc at a distance of 260 pc. Because we wish to compare the NH$_3$ line parameters of the cores given in \cite{2008ApJS..175..509R} with those of our NH$_3$ sample, we moved the cores to the same distance as the ATLASGAL sources. A region of 26$\arcmin$ in the Perseus molecular cloud corresponds to the beam width of 2 pc used for the NH$_3$ observations of ATLASGAL clumps. Calculating the mean NH$_3$ properties of the cores within 26$\arcmin$, we obtain a velocity dispersion of 0.34 km~s$^{-1}$ for the subregion L1455/L1448, 0.41 km~s$^{-1}$ for NGC1333, and 0.54 km~s$^{-1}$ for IC348. These values as well as the line widths of cores in the Perseus molecular cloud are lower than the mean NH$_3$ (1,1) line width of ATLASGAL sources of 2 km~s$^{-1}$. Another still closer low-mass star-forming region is the Pipe Nebula at a distance of 130 pc. A survey of NH$_3$ (1,1) and (2,2) lines of starless cores in this molecular cloud complex was conducted by \cite{2008ApJS..174..396R} with the GBT. The beam width is $\sim 30\arcsec$ and corresponds to 0.019 pc at a distance of 130 pc. NH$_3$ line parameters within 53$\arcmin$ are analysed to compare with our NH$_3$ observations of ATLASGAL clumps. The velocity dispersion of the subregion located at $l \simeq 1^{\circ}$ and $b \simeq 4^{\circ}$ is 0.58 km~s$^{-1}$, while cores around $l \simeq -3^{\circ}$ and $b \simeq 7^{\circ}$ yield a lower velocity dispersion of 
0.16 km~s$^{-1}$. These values are similar to those calculated in Perseus. In contrast to the two low-mass star-forming regions, the ATLASGAL sample has much larger line widths, indicating that
different dynamics prevail in the high-mass star-forming clumps, which are located in a more turbulent environment. In addition, NH$_3$ (1,1) and (2,2) lines were observed by \cite{2009ApJ...697.1457F} in the Ophiuchus molecular cloud, which is a nearby region of active clustered star formation. \cite{2009ApJ...697.1457F} combined their interferometer and single-dish telescope measurements using the Australia Telescope Compact Array, the Very Large Array, and the GBT. They thus obtained a beam width of 15$\arcsec$, which corresponds to 0.009 pc at a distance of 120 pc. A region of 56$\arcmin$ in Ophiuchus, which is comparable to the beam
width of 2 pc used for ATLASGAL sources, includes all observed cores that are located in the central Ophiuchus region. These exhibit a velocity dispersion of 0.32 km~s$^{-1}$ with line widths
of up to 1 km~s$^{-1}$ that lie in the range of the smallest ATLASGAL line widths. Similar properties of cores in Ophiuchus and the ATLASGAL sample indicate that some ATLASGAL clumps could be low-mass cores. This also supports the process explaining NH$_3$ hyperfine anomalies (see Sect. \ref{hyperfine structure}) by assuming that the observed line width of a molecular cloud results from the relative motion and small line widths of individual clumps.

The mean NH$_3$ column density is similar in the three nearby molecular clouds and lower than that of the ATLASGAL sample. In Perseus we obtain a mean value of $5.3 \times 10^{14}$ cm$^{-2}$ for L1455/L1448, $3.5 \times 10^{14}$ cm$^{-2}$ for NGC1333 and $1.6 \times 10^{14}$ cm$^{-2}$ for IC348. The average ammonia column density in the Pipe Nebula is $\sim 3 \times 10^{14}$ cm$^{-2}$ , and Ophiuchus has a slightly higher mean value of 5.8 $\times 10^{14}$ cm$^{-2}$. It might be expected that the average column density measured within a larger region of the ATLASGAL clumps with a beam width of 40$\arcsec$ is lower than the values obtained for the low-mass star-forming clouds with a smaller beam width of $\sim 30\arcsec$ and 15$\arcsec$. However, the column density of the low-mass cores is still lower than that of the ATLASGAL sample with an average of $1.8 \times 10^{15}$ cm$^{-2}$, which might result from a lower peak column density of the low-mass sources.

The mean kinetic temperature of the local molecular clouds is lower than that of ATLASGAL sources. We calculate mean values of 12.2 K for L1455/L1448 in Perseus, 14.3 K for NGC1333 and 12.4 K for IC348. Cores in the Pipe Nebula have an average kinetic temperature of $\sim 12$ K and Ophiuchus has a mean of 14 K, while we obtain 20.8 K on average for the ATLASGAL sample. The mean kinetic temperature of ATLASGAL sources is calculated within a beam width of 2 pc, which is a more extended region around each clump compared to the smaller beam widths of the close molecular clouds. Hence, the lower kinetic temperatures of the low-mass cores are measured toward their peak, while the higher temperatures of ATLASGAL clumps indicate that they are in a warmer environment or that the more massive and luminous stars are better able to warm their surrounding. Moreover, the kinetic temperatures obtained for the ATLASGAL sample are similar to the temperatures measured for dust that is heated by the interstellar radiation field \citep{2010A&A...518L..88B}.

\subsection{Virial parameters}
To determine virial masses, it is important that the line tracer used to calculate the velocity dispersion traces the same material as seen in the dust emission with ATLASGAL. Ideally, we would use maps to investigate whether there is a correlation between the extent of the regions probed by the molecular line and dust emission, but such data are not available for NH$_3$. It is therefore useful to compare the velocity dispersion in different line tracers such as N$_2$H$^+$ and NH$_3$.
 
\begin{figure}[h]
\centering
\includegraphics[angle=0,width=9.0cm]{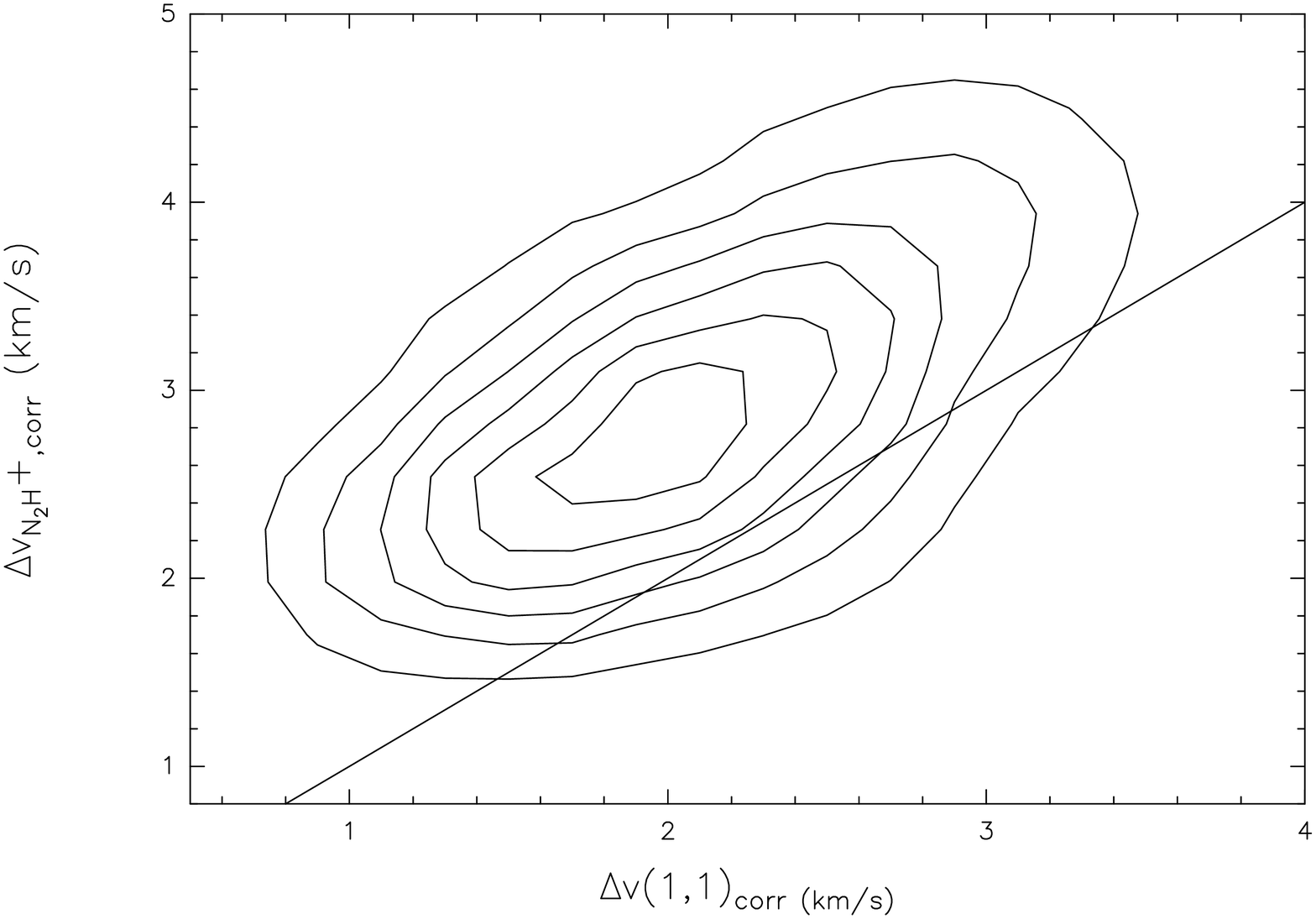}\vspace*{0.5cm}
\includegraphics[angle=0,width=9.0cm]{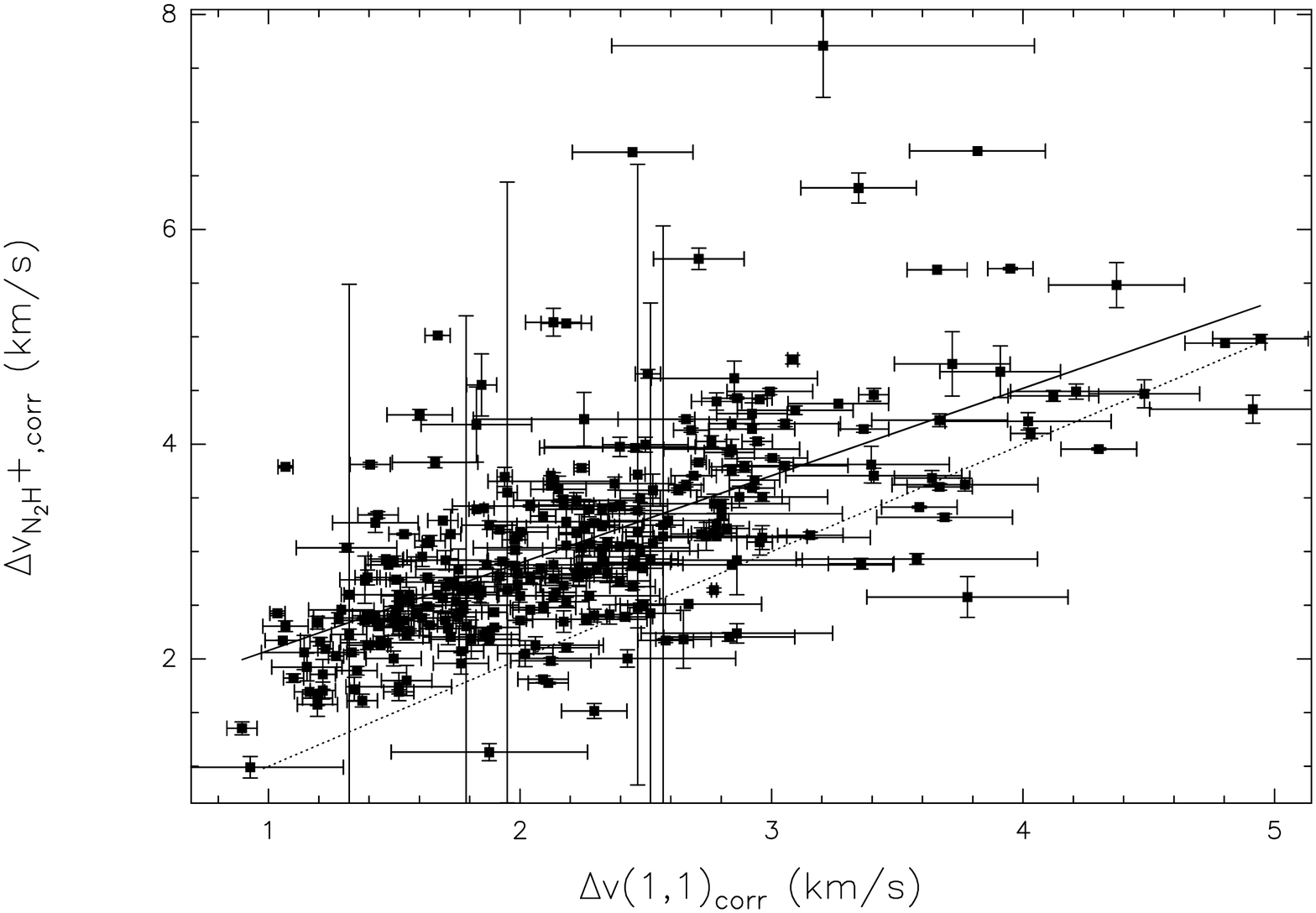}
\caption[NH$_3$ and N$_2$H$^+$ linewidths]{Dependence of N$_2$H$^+$ on NH$_3$ line widths is shown as contour plot in the top panel and as a scatter plot in the lower panel. The binning of the two line widths in the contour plot is 0.5 km~s$^{-1}$. The dotted straight line indicates equality, the solid line denotes a fit to the data.}\label{n2hp-nh3-dv}
\end{figure}

\begin{figure}[h]
\centering
\includegraphics[angle=0,width=9.0cm]{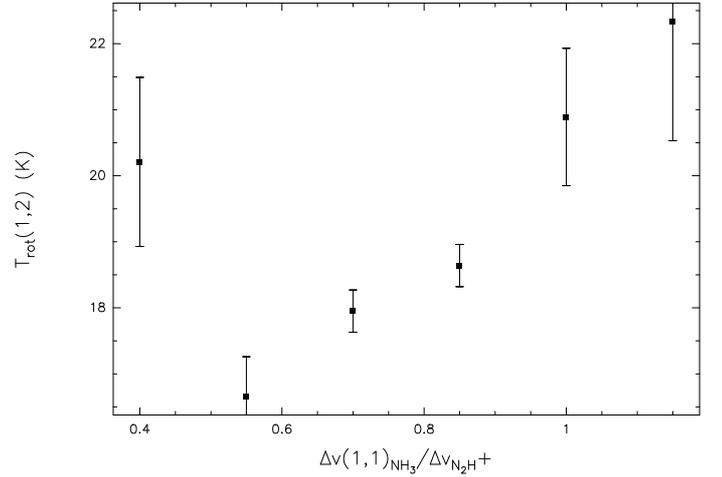}
\caption[NH$_3$ and N$_2$H$^+$ linewidths]{Mean of the rotational temperature calculated in each bin of NH$_3$ to N$_2$H$^+$ line-width ratios of 0.15 and plotted against the line-width ratio.}\label{dvratio-trot}
\end{figure}

Gas masses and virial masses were compared for an ATLASGAL sample of 348 clumps observed in NH$_3$, which have known distances from the GRS survey \citep{2009ApJ...699.1153R} or are located at the tangent points with similar near and far distances, in \cite{2012A&A...544A.146W}. This revealed smaller virial parameters with a mean of 0.21 for sources with narrow line widths, $\lesssim 1.8$ km~s$^{-1}$, than for clumps exhibiting broad line widths, $\gtrsim 1.8$ km~s$^{-1}$, with a mean virial parameter of 0.45. We continue this study here with a comparison of virial parameters by using NH$_3$ and the higher density tracer N$_2$H$^+$. We detect the N$_2$H$^+$ line in 610 ATLASGAL sources, which is 87\% of the observed southern sample. A subsample of 293 clumps in the fourth quadrant is detected in N$_2$H$^+$ and NH$_3$. Because the line widths of the two probes affect the virial mass estimates, we searched for trends of the line-width ratio. We corrected the line widths for the resolution of the spectrometer taking into account the channel width of 0.87 km~s$^{-1}$ of the Mopra telescope and of 0.4 km~s$^{-1}$ of the Parkes telescope with the correlation between the channels. This results in a N$_2$H$^+$ line width corrected for the velocity resolution, $\Delta \rm v_{\mbox{\tiny N$_2$H$^+$, corr}} = \sqrt{\Delta \rm v_{\mbox{\tiny N$_2$H$^+$}}^2-(1.044{\rm km~s^{-1}})^2}$, and a deconvolved NH$_3$ (1,1) line width, $\Delta \rm v(1,1)_{\mbox{\tiny corr}} = \sqrt{\Delta \rm v(1,1)^2-(0.4{\rm km~s^{-1}})^2}$. The lower panel of Fig. \ref{n2hp-nh3-dv} shows the N$_2$H$^+$ line widths corrected for the velocity resolution plotted against the deconvolved NH$_3$ (1,1) line widths, the dotted straight line indicates equality. The solid line presents a fit to the data, which is given by $\Delta \rm v_{\mbox{\tiny N$_2$H$^+$, corr}} = 0.81 \Delta \rm v(1,1)_{\mbox{\tiny corr}}+ 1.26$. Both are intrinsic line widths from hyperfine structure fits. We took the seven hyperfine components into account for the fit of the N$_2$H$^+$ lines. For the contour plot in the upper panel of Fig. \ref{n2hp-nh3-dv} we chose a binning of 0.5 km~s$^{-1}$ for the deconvolved NH$_3$ (1,1) and N$_2$H$^+$ line widths. The N$_2$H$^+$ lines have widths between 0.9 and 7.7 km~s$^{-1}$ with a peak at 3 km~s$^{-1}$ and are therefore broader than the NH$_3$ (1,1) lines with a peak at 2 km~s$^{-1}$. The difference in line width might indicate that the lines originate from different parts of the clumps. While NH$_3$ probes densities of $\sim$ 10$^4$ cm$^{-3}$ \citep{1986A&A...157..207U}, the critical density of N$_2$H$^+$ is $\sim 2 \times 10^5$ cm$^{-3}$\citep{2005ApJ...620..330A}. Assuming a density and temperature gradient over the extent of the sources, N$_2$H$^+$ traces the denser inner region of a source and NH$_3$ also probes the less dense envelope of the clump. In addition, the N$_2$H$^+$ lines were measured with a smaller beam width of 38$\arcsec$ compared to the 61$\arcsec$ beam width of the NH$_3$ observations. The narrower NH$_3$ line widths measured with a larger beam width of the Parkes telescope compared to the N$_2$H$^+$ line widths contradict the power-law relationship between the velocity dispersion and size of a cloud \citep{1981MNRAS.194..809L}. Such a trend was also found in a survey of massive clumps in the inner Galaxy by \cite{2010ApJS..188..313W}, who obtained larger line widths of the CS ($7-6$) transition than of the CS ($2-1$) line despite the much larger size of the CS ($2-1$) emission. However, the smaller inner part of a clump might be influenced by turbulence generated by heating or outflows, which causes broader line widths in the inner region probed by the higher density tracer. This is also presented in Fig. \ref{dvratio-trot}, which illustrates the NH$_3$ rotational temperature plotted against the NH$_3$ (1,1) to N$_2$H$^+$ line-width ratio, taken as the mean of each bin of 0.15. A low line-width ratio associated with low rotational temperatures is caused by a large N$_2$H$^+$ line
width probing the inner dense part of the source and a smaller width of the NH$_3$ line, which is emitted over the whole extent of the clump. This region also includes the less dense and cold envelope, which likely results in a decrease in line width and rotational temperature by averaging over the source size. A line-width ratio of $\sim 1$ and high rotational temperatures are associated with a higher density in the inner heated part of the clump, where turbulence is prevailing, NH$_3$ and N$_2$H$^+$ are excited and exhibit similar line widths.

\begin{figure}[h]
\centering
\includegraphics[angle=-90,width=9.0cm]{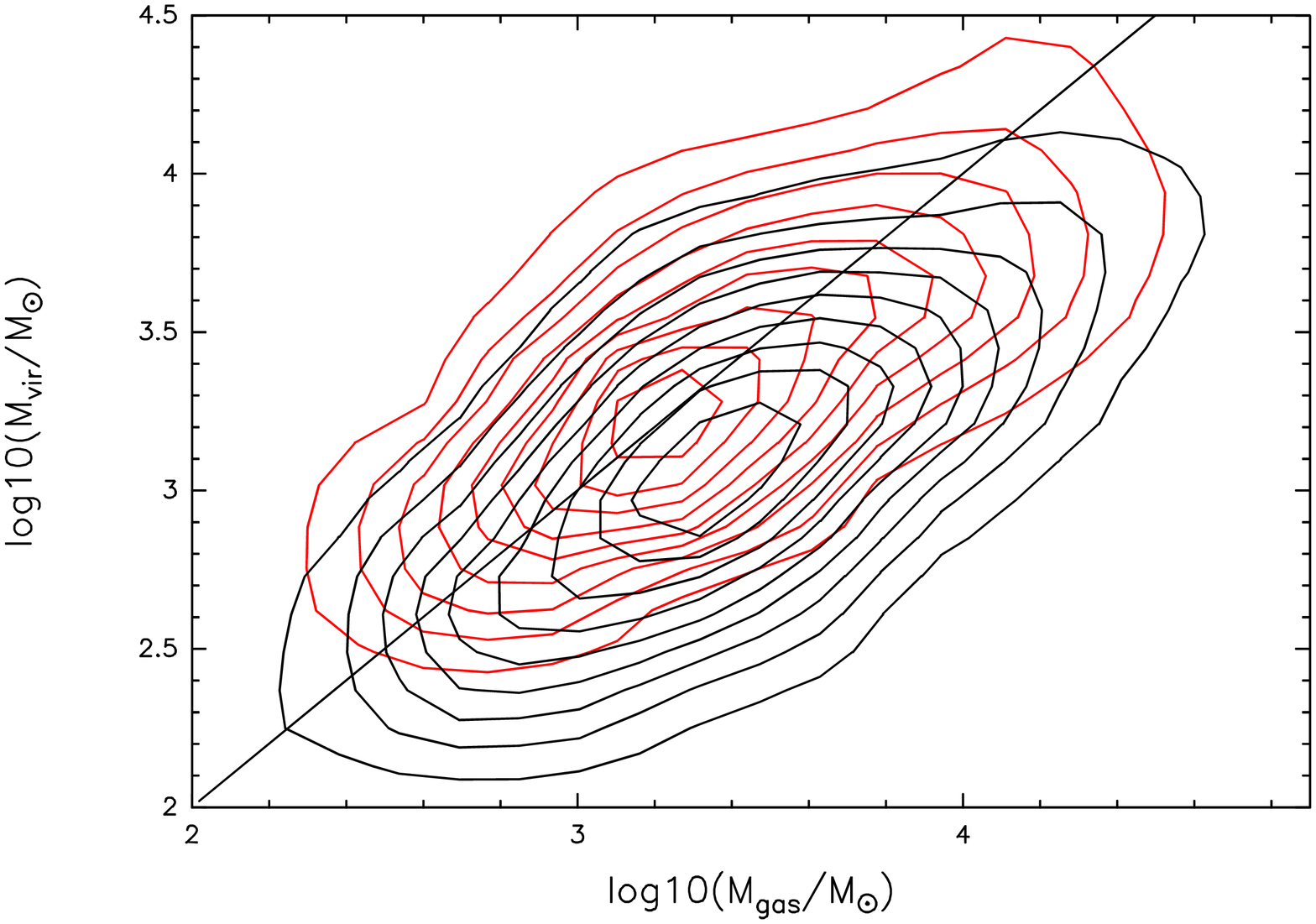}\vspace*{0.5cm}
\includegraphics[angle=0,width=9.0cm]{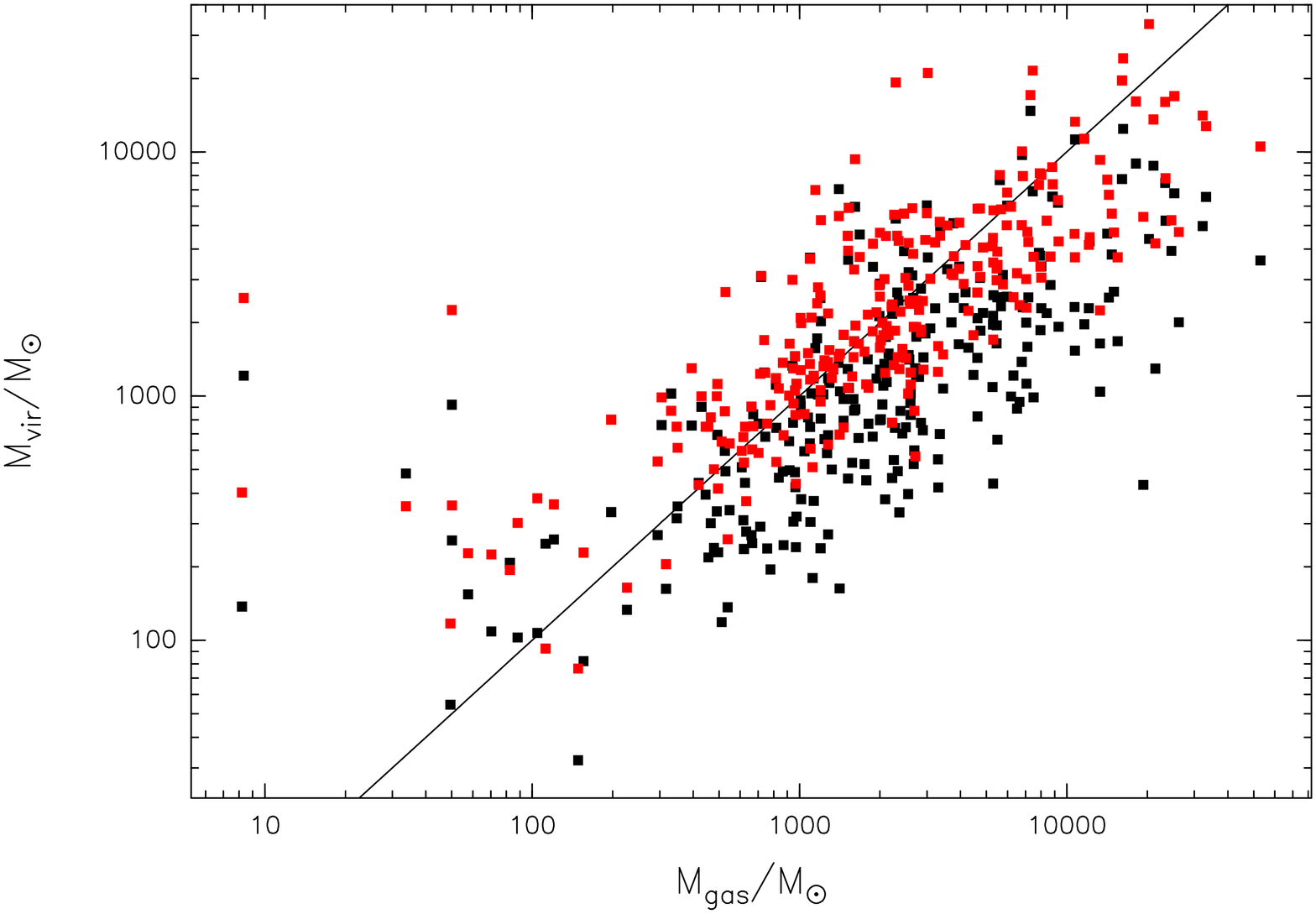}
\caption[gas mass and virial mass]{Comparison of the logarithm of the virial mass and gas mass for masses derived from the NH$_3$ line widths in black and from the N$_2$H$^+$ line widths in red. The contour plot of the two parameters is plotted in the upper panel with the range of the logarithm of the virial mass and gas mass divided into bins of 0.3. The straight line shows equal masses.}\label{mvir-mgas}
\end{figure}

\begin{figure}[h]
\centering
\includegraphics[angle=0,width=9.0cm]{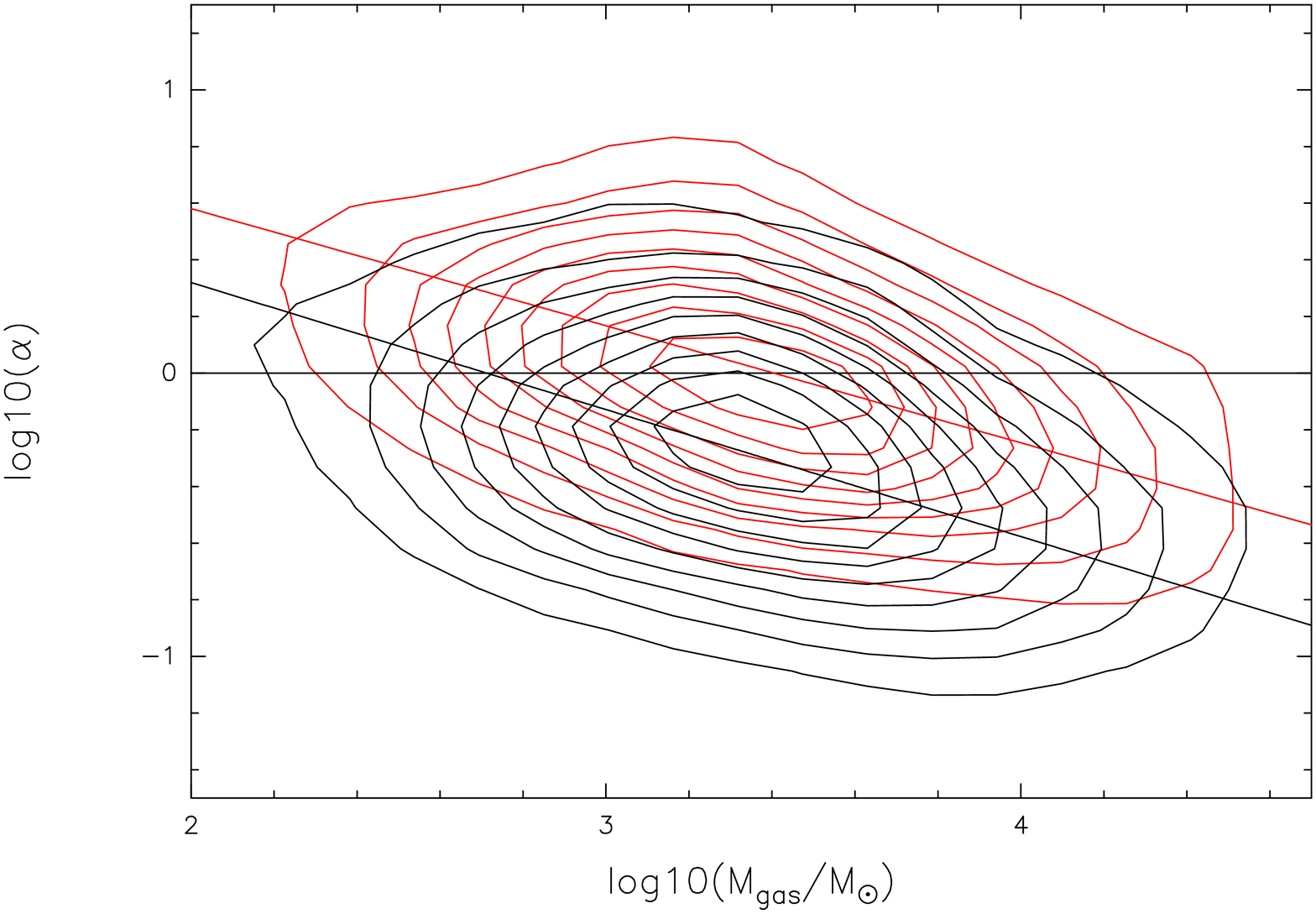}\vspace*{0.5cm}
\includegraphics[angle=0,width=9.0cm]{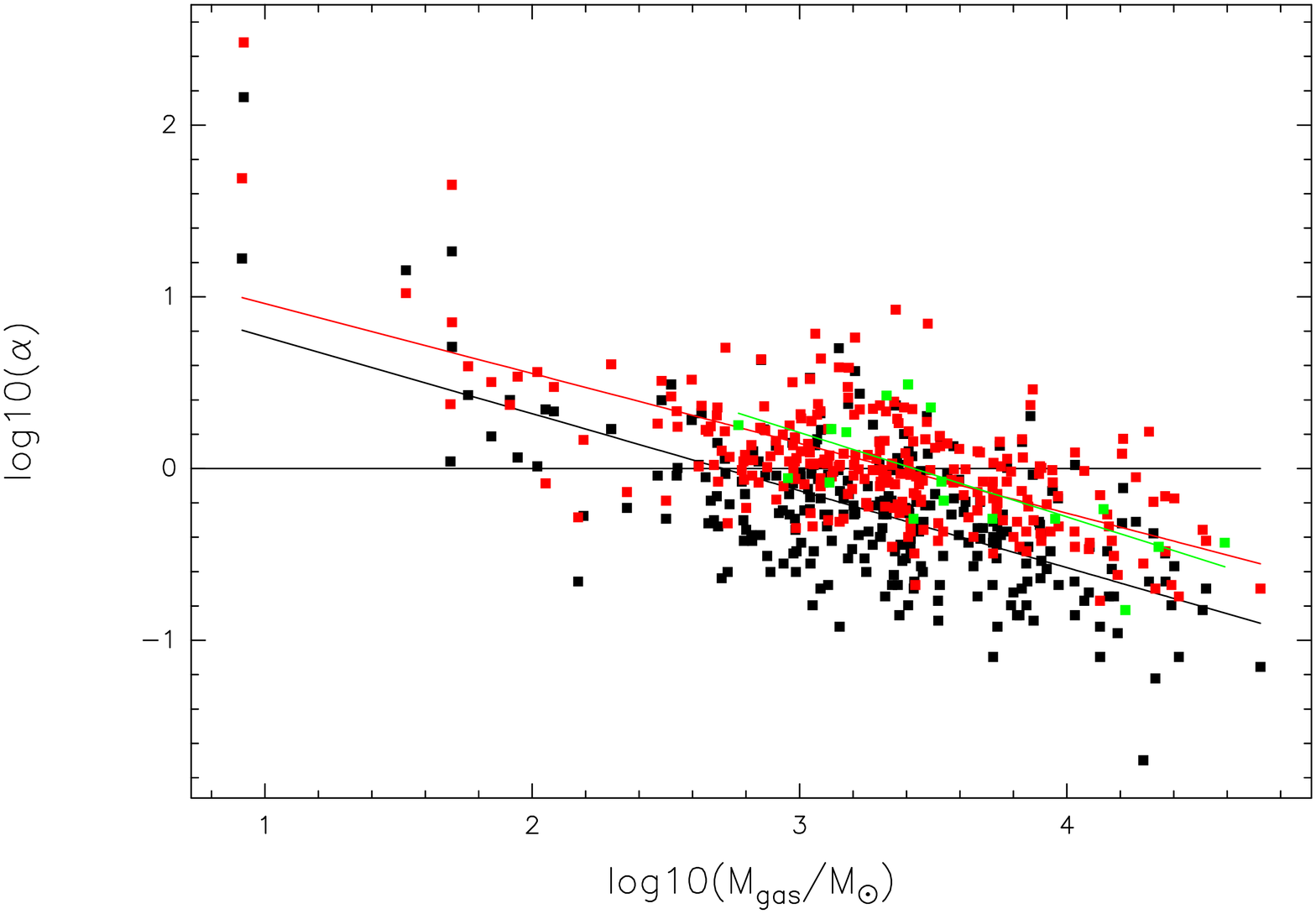}
\caption[gas mass and virial mass]{Correlation plot of the logarithm of the virial parameter and the logarithm of the gas mass as black points for masses calculated using the NH$_3$ line width, as red points using the N$_2$H$^+$ line width, and as green points for the C$^{17}$O line width. The horizontal black line indicates a virial parameter of 1. A fit to each data set is shown as straight lines. The upper panel illustrates the contour plot, for which we counted the number of sources in each logarithmic gas mass and virial parameter bin of 0.3 in black for masses computed from the NH$_3$ line width and in red for the N$_2$H$^+$ line
width.}\label{mgas-alpha}
\end{figure}

To calculate the gas mass of a southern NH$_3$ subsample of 280 sources with known distances, we used the same relation as for the northern clumps (see Equation 12 in \cite{2012A&A...544A.146W}) with the 870 $\mu$m flux density from the CSC, the kinematic distance to the complex, in which the source is located, given in \cite{2015A&A...579A..91W}, and the kinetic temperature listed in Table \ref{parabgel-atlasgal}. The virial masses are computed assuming virial equilibrium \citep{2004tra..book.....R},
\begin{eqnarray}\label{virial mass}
 M_{\mbox{\tiny vir}}(\rm M_{\odot})=250 \Delta \rm v(1,1)_{\mbox{\tiny corr}}^2 \times R
,\end{eqnarray}
with the NH$_3$ line width, $\Delta \rm v(1,1)_{\mbox{\tiny corr}}$, in km~s$^{-1}$ and the N$_2$H$^+$ line width, $\Delta \rm v_{\mbox{\tiny N$_2$H$^+$, corr}}$, in km~s$^{-1}$ corrected for the velocity resolution and the deconvolved effective radius in pc, as given in the CSC. We show the distribution of gas and virial masses calculated from the NH$_3$ line width in black and from the N$_2$H$^+$ line width in red in Fig. \ref{mvir-mgas}. The black solid line indicates equality between gas and virial masses. The gas mass ranges from $\sim 10$ to $5 \times 10^4$ M$_{\odot}$ with a peak at $\sim 2.5 \times 10^3$ M$_{\odot}$ , and the virial mass lies between $\sim 30$ and $3 \times 10^4$ M$_{\odot}$ with a peak at $\sim 1000$ M$_{\odot}$ using the NH$_3$ line width and a peak at $\sim 2000$ M$_{\odot}$ for values computed from the N$_2$H$^+$ line width. We obtain on average higher gas masses than virial masses calculated from the NH$_3$ line width, the same trend as found for the northern sample \citep{2012A&A...544A.146W}. When we use the N$_2$H$^+$ line width, the virial masses are higher than those resulting from the NH$_3$ line width, and the red sample is well described by the black line. Higher virial masses also result in larger virial parameters, which are computed through the relation \citep{1992ApJ...395..140B}
\begin{eqnarray}
 \alpha = \frac{M_{\mbox{\tiny vir}}}{M_{\mbox{\tiny gas}}}.
\end{eqnarray}
We plot the logarithm of virial parameters against the logarithm of gas masses in Fig. \ref{mgas-alpha}. The lack of clumps with low gas masses and high virial parameters results from the limiting sensitivity of the NH$_3$ observations. The straight line illustrates $\alpha = 1$ for sources, which are in hydrostatic equilibrium. Because the mean virial parameter is skewed to high values due to some sources that exhibit gas masses with large errors, we give the median $\alpha$ values. These are 0.54 for virial masses calculated using the NH$_3$ line width, similar to the virial parameters of northern ATLASGAL sources observed in NH$_3$ with a broad line width \citep{2012A&A...544A.146W} and 0.99 for estimates from the N$_2$H$^+$ line width. We compare these values to the virial parameter of a sample of ATLASGAL sources observed in C$^{17}$O \citep{2014A&A...570A..65G}. Although this molecule has a low critical density of $\sim 10^3$ cm$^{-3}$ \citep{1997ApJ...482..245U}, it also traces the dense part within a source because it is optically thin. The virial masses are determined using Equation \ref{virial mass}, the C$^{17}$O line width corrected for the velocity resolution, $\Delta$v$_{\rm C^{17}O, corr}$ = $\sqrt{(\Delta \rm v_{\rm C^{17}O})^2 - (0.64 \rm km~s^{-1})^2}$, and the deconvolved effective radius from the CSC. The C$^{17}$O line width leads to virial mass estimates
that are similar to those using the N$_2$H$^+$ line width, and to a median virial parameter of 0.8. Previous studies have shown that the distribution in Fig. \ref{mgas-alpha} can be fitted by a power law \citep{1992ApJ...395..140B,2008ApJ...672..410L}. A fit to our data gives $\alpha \sim M^{-0.45 \pm 0.03}$ using NH$_3$, $\alpha \sim M^{-0.41 \pm 0.03}$ for N$_2$H$^+$ and $\alpha \sim M^{-0.49 \pm 0.13}$ for C$^{17}$O. The power-law slopes agree within the errors and are also consistent with the exponent of the NH$_3$ sample in the first quadrant fitted by \cite{2013ApJ...779..185K}. Although the median virial parameters depend on the kind of tracer used, the slopes of the distributions are similar and in the narrow range between 0 and -1 as obtained for different star-forming samples by \cite{2013ApJ...779..185K}. ATLASGAL sources observed in NH$_3$ are close to be virialised with a trend of a decreasing virial parameter with rising gas mass.

\begin{figure}[h]
\centering
\includegraphics[angle=0,width=9.0cm]{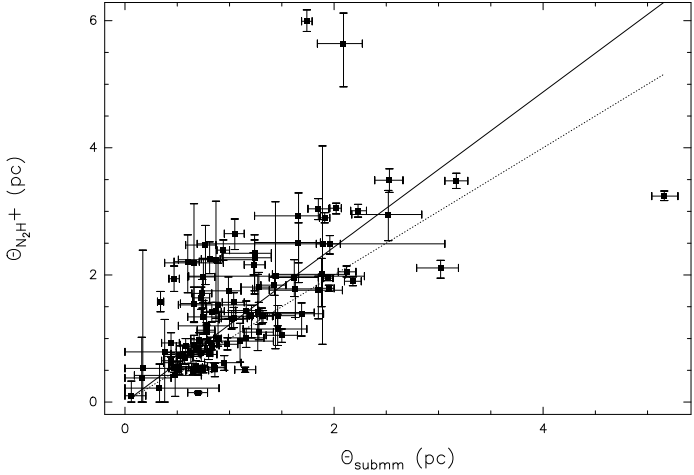}
\caption[gas mass and virial mass]{Deconvolved effective radius derived from Gaussian fits to N$_2$H$^+$ maps plotted against the radius determined from the 870 $\mu$m emission. The dotted straight line denotes equal effective radii, the solid line shows a fit to the data.}\label{radius-n2hp-nh3}
\end{figure}

To investigate whether NH$_3$ or N$_2$H$^+$ better represents the mass of a clump determined through the dust properties, we compared the H$_2$ volume density of ATLASGAL sources with the critical density of NH$_3$ and N$_2$H$^+$. The H$_2$ volume density of ATLASGAL sources with resolved kinematic distances \citep{2015A&A...579A..91W} lies between 500 and $2.8 \times 10^5$ cm$^{-3}$ with a mean of $2 \times 10^4$ cm$^{-3}$. N$_2$H$^+$ has a critical density of $\sim 2 \times 10^5$ cm$^{-3}$ \citep{2005ApJ...620..330A}. The study of the dust continuum at 1.2 mm and CS lines towards high-mass star-forming regions by \cite{2002ApJ...566..945B} showed that the densities probed by high-density tracers are typically ten times higher than the densities measured from dust emission, which might result from fragmentation. N$_2$H$^+$ could therefore trace several clumps of high densities within an ATLASGAL source. In contrast, the NH$_3$ critical density is only $\sim$ 10$^4$ cm$^{-3}$ \citep{1986A&A...157..207U} and it might probe a larger region than the dust emission. When we use the NH$_3$ line
width and the deconvolved effective radius derived from the 870 $\mu$m dust continuum to estimate the virial mass, we find lower limits of the virial parameter because the radius is too small compared to the region of NH$_3$ emission. However, N$_2$H$^+$ might also trace a smaller region than the dust. To investigate whether there is a correlation between the two, we derived deconvolved effective radii from Gaussian fits to N$_2$H$^+$ $(1-0)$ maps observed by the MALT90 survey \citep[The Millimetre Astronomy Legacy Team 90 GHz Survey;][]{2013PASA...30...57J}, $\theta_{\rm N_2H^+}$, and the deconvolved effective radii obtained from the 870 $\mu$m dust emission given in the CSC, $\theta_{\rm submm}$, for a sample of 105 ATLASGAL sources. Their comparison is displayed in Fig. \ref{radius-n2hp-nh3}, where the dotted straight line indicates equal effective radii. A fit to the data, shown by the solid line, yields $\theta_{\rm N_2H^+} = 1.22 \theta_{\rm submm} + 0.06$. While there is indication of larger N$_2$H$^+$ radii than dust radii, they agree in 90\% of the cases within 50\%. Statistically, N$_2$H$^+$ therefore represents the size of the dust emission envelope well. 

In summary, our analysis of virial parameters obtained from different high-density tracers reveals that care is required in interpreting absolute values of the virial parameter. These estimates can also vary depending on the gas masses, which can change when different dust properties are used. There is a variation in turbulence within different parts of a source, which is related to different widths of spectral lines. N$_2$H$^+$ exhibits broader line widths than NH$_3$, leading to twice larger virial parameters. In addition, the size of the dust continuum and of the N$_2$H$^+$ line emission agree well, which shows that the dust is better represented by N$_2$H$^+$ than by NH$_3$.

\subsection{Anomalies in the NH$_3$ (1,1) quadrupole hyperfine structure}\label{hyperfine structure}
It is expected that the inner and outer satellite intensity ratios of the NH$_3$ (1,1) transition are symmetric in LTE. For optically thin hyperfine lines, the intensity ratio of the inner satellite line to the main line is expected to be 0.28, while the ratio of the outer satellites to the main component is 0.22 \citep{1983ARA&A..21..239H}. However, $\sim$ 61\% of the observed NH$_3$ (1,1) spectra show deviations from this prediction. These hyperfine anomalies in the NH$_3$ (1,1) emission have previously been observed towards several star-forming regions \citep{1985A&A...144...13S,1977ApJ...214L..67M,1982A&A...111..201S}.

One process that explains the nonthermal excitation of the NH$_3$ (1,1) hyperfine levels is given by \cite{1977ApJ...214L..67M}. They assumed that the observed molecular cloud is clumped and the individual cores have small line widths ($\sim$0.3 km~s$^{-1}$) and high densities ($10^6-10^7$ cm$^{-3}$). The relative motion of the clumps result then in the observed line width, and the measured line is an average over the clump spectra. The NH$_3$(1,1) hyperfine levels are populated mainly by the far IR (J,K) = (2,1)$\rightarrow$(1,1) photons, which are affected by selective trapping. The inner hyperfine satellites of the IR transition overlap with the main line, while the outer satellites are separated from those that
are due to the small line widths. Consequently, it is more likely that photons in the outer satellite lines with a low optical depth can escape than those in the inner satellites with an increased optical depth. This leads to an overpopulation of the NH$_3$ (1,1) outer hyperfine levels, which is stronger for the level with total angular momentum $F_1 = 0$ than for $F_1 = 1$ state. Because photons of the $F_1 = 0 \rightarrow 1$ transition come from a strongly overpopulated to a weaker populated level, the intensity of this outer satellite line on the red side (O2 in Fig. \ref{NH3-outer-satellites}) is enhanced with regard to the intensity on the blue side (O1), $F_1 = 1 \rightarrow 0$, which goes to the strongly overpopulated $F_1 = 0$ state.
The nonthermal excitation of the (1,1) hyperfine levels explains anomalies in the outer satellite lines. However, some NH$_3$ observations also show unequal intensities of the inner hyperfine structure lines, which are explained by \cite{2001A&A...376..348P}. They used Monte Carlo radiative-transfer calculations to investigate the influence of the velocity field from systematic motion such as expansion and contraction on the line anomaly. For a uniform core collapse with a velocity field $v (r) \propto r,$ the photons emitted by an ammonia molecule and incident on another one are blueshifted. As a consequence, photons from a low-energy transition of the far IR (J,K) = (2,1) $\longrightarrow$ (1,1) decay can increase their energy and thus be reabsorbed, while higher energy photons leave the gas. The process works in the opposite way for systematic outflow. For different velocity fields such as inside-out collapse ($v (r) \propto r^{-0.5}$), the procedure is the same, although it is weaker because some photons are redshifted.

\begin{figure}[h]
\centering
\includegraphics[angle=-90,width=8.8cm]{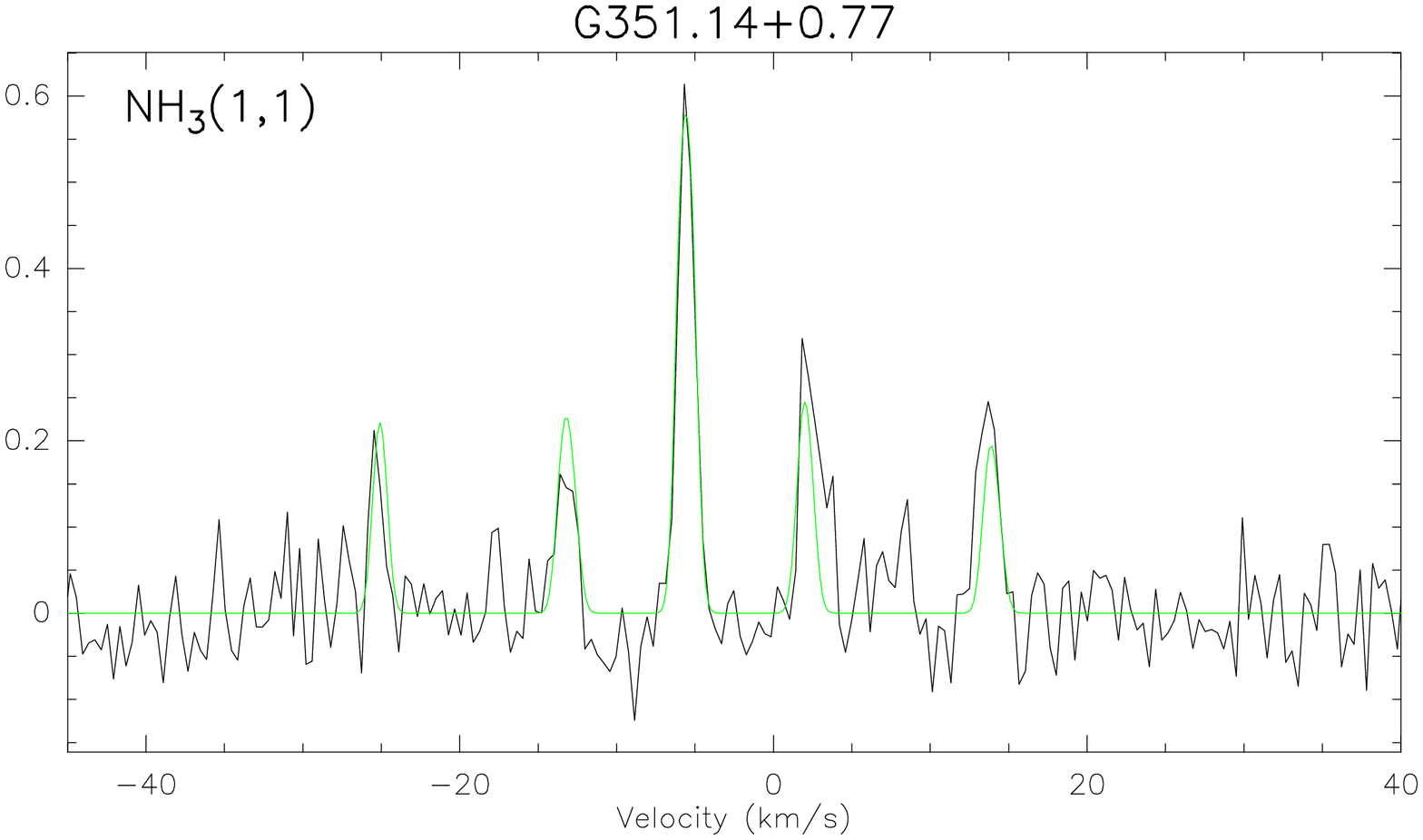}\vspace*{0.5cm}
\includegraphics[angle=-90,width=8.8cm]{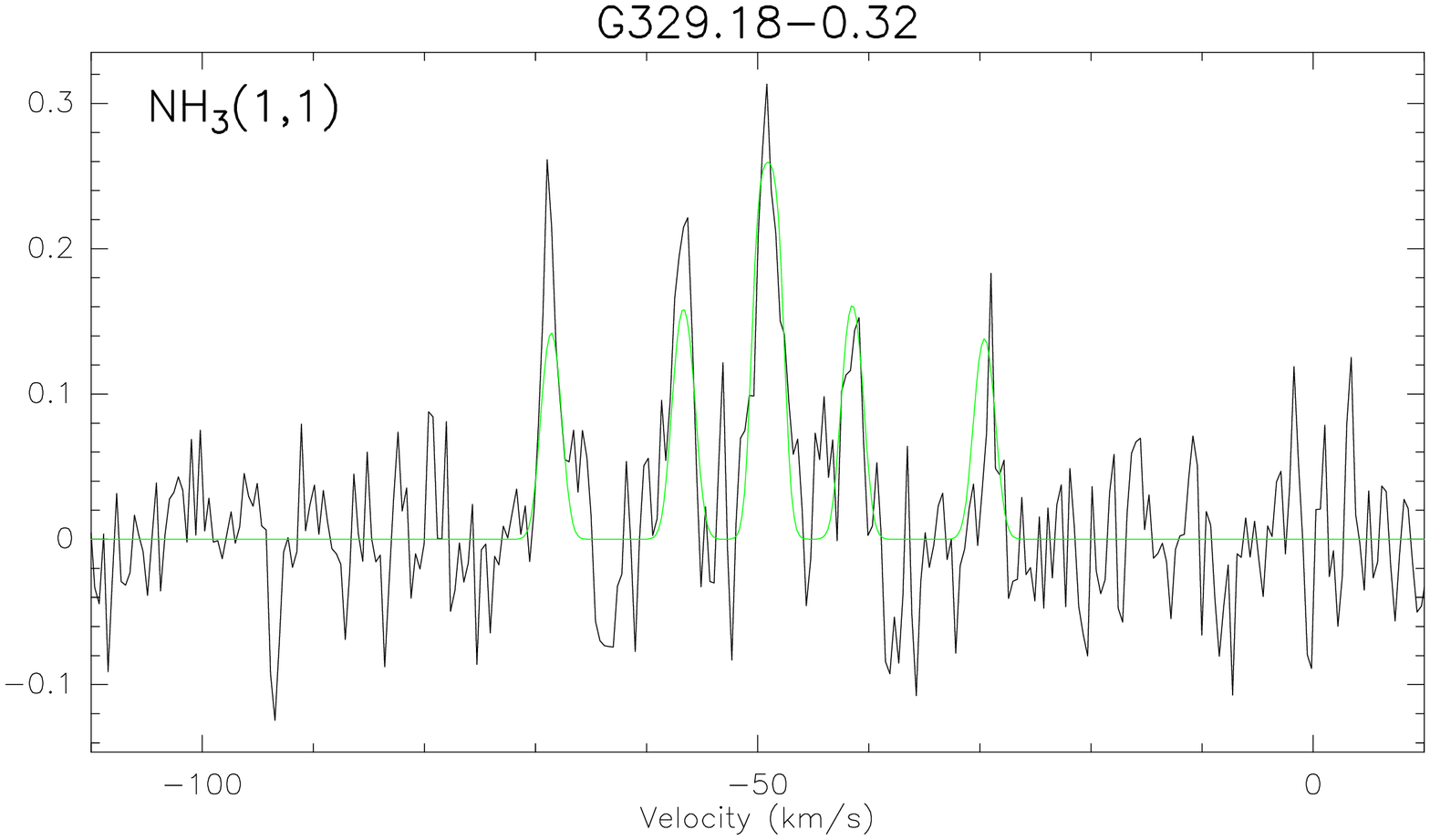}\\
\includegraphics[angle=-90,width=9.0cm]{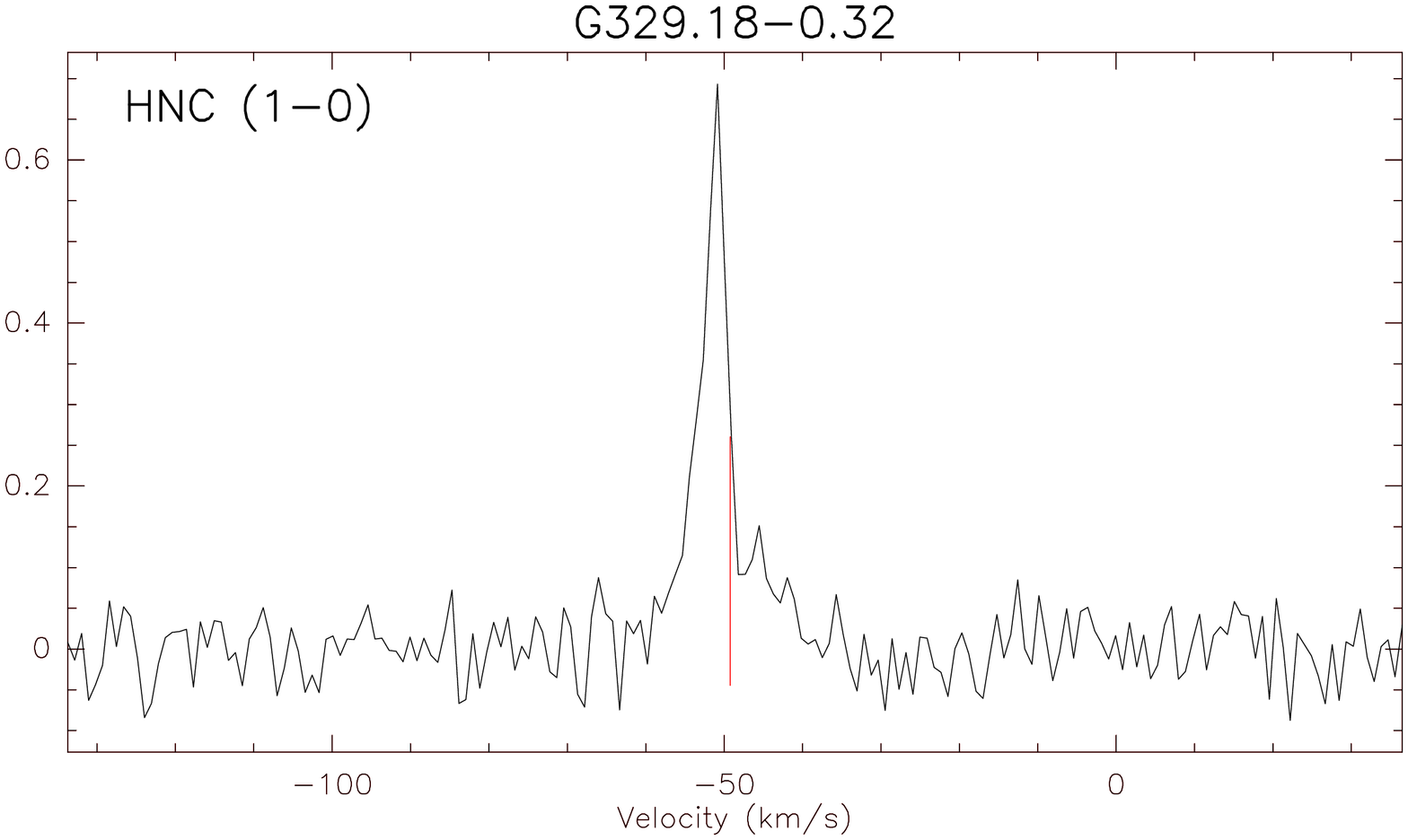}
\caption[hyperfine anomalies spectra]{NH$_3$ (1,1) hyperfine anomaly of G351.14+0.77, which is illustrated in the upper panel, indicates a systematic outflow in this source. The middle panel displays the NH$_3$ (1,1) line profile of G329.18-0.32, which indicates an infall. This is also supported by its HNC (1-0) line in the lowest panel with a stronger blue- than redshifted peak. The source velocity is shown by the straight red line.}\label{hyperfine-lines}
\end{figure}

Section \ref{anomalies measurement} describes our estimation of the hyperfine anomalies resulting in the three parameters $\alpha$, $\beta,$ and $\gamma$ following the work by \cite{2007MNRAS.379..535L}. The contour plot in Fig. \ref{alpha-beta} shows that there is no correlation between $\alpha$ and $\beta$, the scatter plot illustrates that $\alpha$ and $\beta$ of a few sources differ from 1. Taking the errors of the parameters into account, the NH$_3$ hyperfine anomaly might be produced by a systematic outflow in one source of our sample, G351.14+0.77, with $\alpha = 1.65 \pm 0.63$ and $\beta = 1.73 \pm 0.58$, while there is an indication of infall in one clump, G329.18$-$0.32, with $\alpha = 0.45 \pm 0.26$ and $\beta = 0.71 \pm 0.27$, which results in the asymmetry of its satellite lines (see Fig. \ref{hyperfine-lines}). Another indication of the infall in that source is the HNC (1-0) line profile shown in Fig. \ref{hyperfine-lines} in addition to the NH$_3$ (1,1) inversion transition. A blueshifted peak that is stronger than the redshifted peak is usually observed in an infalling envelope with a centrally peaked density and temperature distribution \citep{1999ARA&A..37..311E}. The blue peak originates from part of the cloud on the far side of the envelope, closer to the centre with a high excitation temperature, while the red peak is emitted from a point on the near side with a low excitation temperature. The central dip at the source LSR velocity is produced by self-absorption from low excited gas on the near side of the cloud.

We compared $\alpha$ and $\beta$ of the southern sources observed in NH$_3$ with a massive star-forming subsample of clumps associated with an RMS source \citep{2013ApJS..208...11L}. However, no trend is found, which shows that on the scale of a clump, star formation processes such as outflow or infall driven by the embedded source do not influence the properties of the clumps.

Because our investigation yields that the hyperfine structure anomalies result from the nonthermal excitation of the NH$_3$ (1,1) levels populated by the far IR transition for most sources, we would expect a correlation between the asymmetry of the outer satellite lines and the luminosity of the source. We therefore
searched for a correlation between $\alpha$ and the NH$_3$ kinetic temperature as well as the ammonia (1,1) line widths, which rise with increasing infrared luminosity of the source, but we did not find any correlation. 
We investigated whether the subsample with $\alpha, \beta < 1$, which might indicate infall motion, exhibits lower virial masses than gas masses and has therefore the smallest virial parameters. We plot the logarithm of the virial parameter against the logarithm of the gas mass as presented in Fig. \ref{mgas-alpha} for all ATLASGAL sources observed in NH$_3$ in the fourth quadrant. A fit to the sources with $\alpha, \beta < 1$ does not show whether the distribution is different from that of the whole sample because of the small number of clumps and the large uncertainties of the hyperfine component intensity ratios. Moreover, we analysed whether expanding sources associated with outflows are in a warmer and more turbulent environment. To do this, we searched for a trend of an increasing ratio $\alpha$ of the hyperfine components with rising line broadening $\Delta \rm v (3,3)/ \Delta \rm v(1,1)$, but we found no correlation.
  
To summarize, our mean value of $\alpha$ of 1.29, $\beta$ of 0.93 and $\gamma$ of 1.09 indicate that the nonthermal excitation of the NH$_3$ (1,1) levels likely produces hyperfine structure anomalies. The parameters have large uncertainties and the comparison of $\alpha$ and $\beta$ shows no correlation between the two. The errors of $\alpha$ and $\beta $ we considered show that one source of our sample with $\alpha, \beta > 1$ could be associated with an outflow and one clump with $\alpha, \beta < 1$ with an infall.

\section{Summary}
\label{summary}
We observed the NH$_3$ (1,1) to (3,3) inversion transitions of 354 dust clumps discovered by the ATLASGAL survey within a Galactic longitude from 300$^{\circ}$ to 359$^{\circ}$ and a latitude of $\pm 1.5^{\circ}$ using the Parkes telescope. Our main results can be summarized as follows:
\begin{enumerate}
 \item We derived LSR velocities from the NH$_3$ (1,1) lines mostly between 5 and -120 km~s$^{-1}$. They are coincident with CO emission \citep{2001ApJ...547..792D} and thus trace dense cores associated with large-scale molecular cloud structure. These velocities, together with those in the first quadrant \citep{2012A&A...544A.146W}, are required to calculate near and far kinematic distances using the \cite{1993A&A...275...67B} rotation curve \citep{2015A&A...579A..91W}.
 \item The NH$_3$ (1,1) line widths of the observed southern ATLASGAL sources, lying between 0.8 and 5 km~s$^{-1}$, are in a similar range as those of the northern clumps. We obtained mostly equal (1,1) and (2,2) line widths. For a subsample of sources with detected hyperfine satellites of the (2,2) lines, we fitted the hyperfine structure using the (2,2) optical depth derived from the (1,1) optical depth and the (2,2) to (1,1) main-beam brightness temperature ratio. This led to broader (2,2) line
widths that those from a hyperfine structure fit with measured (2,2) optical depth. Using the temperature ratio and the (1,1) optical depth instead of the measured (2,2) optical depth, we
found an underestimation of the rotational temperature and column density by a factor 0.64.
 \item We analysed trends of NH$_3$ line parameters vs galactocentric radius within the inner Galaxy. The rotational temperature, H$_2$ column density and NH$_3$ line widths show an approximately constant distribution. We obtained a decreasing NH$_3$ abundance with galactocentric radius, consistent with the trend presented in \cite{2011ApJ...741..110D}. We investigated their assumption to explain this distribution also with a decreasing nitrogen abundance: the N$_2$H$^+$/NH$_3$ column density ratio is constant within the inner Galaxy, in agreement with the modelling of a prestellar core \citep{2010A&A...513A..41H}, which indicates that an increase of the number of nitrogen atoms within a particular molecule might not lead to an enhanced decrease of its abundance.
\item NH$_3$ line parameters of ATLASGAL clumps were compared with those of cores within nearby molecular clouds. We obtained smaller mean velocity dispersions of cores in low-mass star-forming regions such as the Perseus molecular cloud and the Pipe Nebula than the mean NH$_3$ line width. This shows different dynamics between low- and high-mass star-forming clumps, while sources in the Ophiuchus molecular cloud associated with clustered star formation exhibit a velocity dispersion similar to the narrowest line widths of ATLASGAL sources. Lower mean values of the NH$_3$ column density in nearby molecular clouds than the mean of the ATLASGAL sample might show a lower peak column density of low-mass cores. Moreover, the mean kinetic temperature within a smaller beam width towards nearby molecular clouds is lower than the average kinetic temperature in a larger beam width around each ATLASGAL clump. This indicates a warmer surrounding of ATLASGAL sources than that of low-mass cores.
 \item Comparing the line widths of the NH$_3$ (1,1) inversion transition, tracing densities of $\sim 10^4$ cm$^{-3}$ \citep{1986A&A...157..207U}, with N$_2$H$^+$, probing higher densities of $\sim 2 \times 10^5$ cm$^{-3}$ \citep{2005ApJ...620..330A}, shows broader N$_2$H$^+$ than NH$_3$ line widths. Moreover, the NH$_3$/N$_2$H$^+$ line
width ratio is increasing with rising rotational temperature. Broader N$_2$H$^+$ than NH$_3$ lines lead to higher virial masses with a median virial parameter of 1.03, compared to 0.54 resulting from the NH$_3$ line width. Because the critical density and the radius derived from N$_2$H$^+$ agree with the density and the radius of a dust clump determined from the 870 $\mu$m emission, the dust is better represented by N$_2$H$^+$ than by NH$_3$.
\item We investigated deviations in the relative intensities of the NH$_3$ (1,1) satellite lines from LTE to distinguish between the processes that result in these hyperfine structure anomalies. As described in \cite{2007MNRAS.379..535L}, we calculated the ratio of the temperatures of the outer hyperfine components, $\alpha$, of the inner satellite lines, $\beta$, and of the sum of the outer and inner hyperfine structure lines on each side of the main line, $\gamma$. The median of $\alpha$ of $1.27 \pm 0.45$ differs significantly from 1 compared to a smaller deviation of the median values of $\beta$ of $0.9 \pm 0.3$ and of $\gamma$ of $1.06 \pm 0.37$ from 1. These parameters show that the anomalies of most ATLASGAL clumps are likely explained by a nonthermal excitation of the NH$_3$ (1,1) hyperfine levels. However, the intensity ratio of the satellite lines of two sources indicate systematic motion that results in the hyperfine structure anomalies.
\item In a second paper \cite{2015A&A...579A..91W}, we used the radial velocities derived in the first and fourth quadrant to resolve the kinematic distance ambiguity towards a large sample of high-mass star-forming regions discovered by ATLASGAL.
\end{enumerate}

\noindent
\textit{Acknowledgements.} M.W. participates in the CSIRO Astronomy and Space Science Student Program and acknowledges support from the ATNF staff within the course of the observations. We thank the referee, Erik Rosolowsky, for a careful reading of this article and the very useful comments and suggestions.

\bibliography{paperslibrary}

\begin{thebibliography}{127}
\expandafter\ifx\csname natexlab\endcsname\relax\def\natexlab#1{#1}\fi

\bibitem[{{Aguirre} {et~al.}(2011){Aguirre}, {Ginsburg}, {Dunham}, {Drosback},
  {Bally}, {Battersby}, {Bradley}, {Cyganowski}, {Dowell}, {Evans}, {Glenn},
  {Harvey}, {Rosolowsky}, {Stringfellow}, {Walawender}, \&
  {Williams}}]{2011ApJS..192....4A}
{Aguirre}, J.~E., {Ginsburg}, A.~G., {Dunham}, M.~K., {et~al.} 2011, \apjs,
  192, 4

\bibitem[{{Aikawa} {et~al.}(2005){Aikawa}, {Herbst}, {Roberts}, \&
  {Caselli}}]{2005ApJ...620..330A}
{Aikawa}, Y., {Herbst}, E., {Roberts}, H., \& {Caselli}, P. 2005, \apj, 620,
  330

\bibitem[{{Aikawa} {et~al.}(2012){Aikawa}, {Wakelam}, {Hersant}, {Garrod}, \&
  {Herbst}}]{2012ApJ...760...40A}
{Aikawa}, Y., {Wakelam}, V., {Hersant}, F., {Garrod}, R.~T., \& {Herbst}, E.
  2012, \apj, 760, 40

\bibitem[{{Anderson} {et~al.}(2011){Anderson}, {Bania}, {Balser}, \&
  {Rood}}]{2011ApJS..194...32A}
{Anderson}, L.~D., {Bania}, T.~M., {Balser}, D.~S., \& {Rood}, R.~T. 2011,
  \apjs, 194, 32

\bibitem[{{Andr\'e} {et~al.}(2000){Andr\'e}, {Ward-Thompson}, \&
  {Barsony}}]{2000prpl.conf...59A}
{Andr\'e}, P., {Ward-Thompson}, D., \& {Barsony}, M. 2000, Protostars and
  Planets IV, 59

\bibitem[{{Bania}(1977)}]{1977ApJ...216..381B}
{Bania}, T.~M. 1977, \apj, 216, 381

\bibitem[{{Battisti} \& {Heyer}(2014)}]{2014ApJ...780..173B}
{Battisti}, A.~J. \& {Heyer}, M.~H. 2014, \apj, 780, 173

\bibitem[{{Becker} {et~al.}(1990){Becker}, {White}, {McLean}, {Helfand}, \&
  {Zoonematkermani}}]{1990ApJ...358..485B}
{Becker}, R.~H., {White}, R.~L., {McLean}, B.~J., {Helfand}, D.~J., \&
  {Zoonematkermani}, S. 1990, \apj, 358, 485

\bibitem[{{Benjamin} {et~al.}(2003){Benjamin}, {Churchwell}, {Babler}, {Bania},
  {Clemens}, {Cohen}, {Dickey}, {Indebetouw}, {Jackson}, {Kobulnicky},
  {Lazarian}, {Marston}, {Mathis}, {Meade}, {Seager}, {Stolovy}, {Watson},
  {Whitney}, {Wolff}, \& {Wolfire}}]{2003PASP..115..953B}
{Benjamin}, R.~A., {Churchwell}, E., {Babler}, B.~L., {et~al.} 2003, \pasp,
  115, 953

\bibitem[{{Benson} {et~al.}(1998){Benson}, {Caselli}, \&
  {Myers}}]{1998ApJ...506..743B}
{Benson}, P.~J., {Caselli}, P., \& {Myers}, P.~C. 1998, \apj, 506, 743

\bibitem[{{Bergin} \& {Tafalla}(2007)}]{2007ARA&A..45..339B}
{Bergin}, E.~A. \& {Tafalla}, M. 2007, \araa, 45, 339

\bibitem[{{Bernard} {et~al.}(2010){Bernard}, {Paradis}, {Marshall}, {Montier},
  {Lagache}, {Paladini}, {Veneziani}, {Brunt}, {Mottram}, {Martin},
  {Ristorcelli}, {Noriega-Crespo}, {Compi{\`e}gne}, {Flagey}, {Anderson},
  {Popescu}, {Tuffs}, {Reach}, {White}, {Benedettini}, {Calzoletti},
  {Digiorgio}, {Faustini}, {Juvela}, {Joblin}, {Joncas}, {Mivilles-Deschenes},
  {Olmi}, {Traficante}, {Piacentini}, {Zavagno}, \&
  {Molinari}}]{2010A&A...518L..88B}
{Bernard}, J.-P., {Paradis}, D., {Marshall}, D.~J., {et~al.} 2010, \aap, 518,
  L88

\bibitem[{{Bertin} \& {Arnouts}(1996)}]{1996A&AS..117..393B}
{Bertin}, E. \& {Arnouts}, S. 1996, \aaps, 117, 393

\bibitem[{{Bertoldi} \& {McKee}(1992)}]{1992ApJ...395..140B}
{Bertoldi}, F. \& {McKee}, C.~F. 1992, \apj, 395, 140

\bibitem[{{Beuther} {et~al.}(2002){Beuther}, {Schilke}, {Menten}, {Motte},
  {Sridharan}, \& {Wyrowski}}]{2002ApJ...566..945B}
{Beuther}, H., {Schilke}, P., {Menten}, K.~M., {et~al.} 2002, \apj, 566, 945

\bibitem[{{Bontemps} {et~al.}(2010){Bontemps}, {Motte}, {Csengeri}, \&
  {Schneider}}]{2010A&A...524A..18B}
{Bontemps}, S., {Motte}, F., {Csengeri}, T., \& {Schneider}, N. 2010, \aap,
  524, A18

\bibitem[{{Brand} \& {Blitz}(1993)}]{1993A&A...275...67B}
{Brand}, J. \& {Blitz}, L. 1993, \aap, 275, 67

\bibitem[{{Brand} {et~al.}(2001){Brand}, {Cesaroni}, {Palla}, \&
  {Molinari}}]{2001A&A...370..230B}
{Brand}, J., {Cesaroni}, R., {Palla}, F., \& {Molinari}, S. 2001, \aap, 370,
  230

\bibitem[{{Brogan} {et~al.}(2011){Brogan}, {Hunter}, {Cyganowski}, {Friesen},
  {Chandler}, \& {Indebetouw}}]{2011ApJ...739L..16B}
{Brogan}, C.~L., {Hunter}, T.~R., {Cyganowski}, C.~J., {et~al.} 2011, \apjl,
  739, L16

\bibitem[{{Bronfman} {et~al.}(1988){Bronfman}, {Cohen}, {Alvarez}, {May}, \&
  {Thaddeus}}]{1988ApJ...324..248B}
{Bronfman}, L., {Cohen}, R.~S., {Alvarez}, H., {May}, J., \& {Thaddeus}, P.
  1988, \apj, 324, 248

\bibitem[{{Bronfman} {et~al.}(1996){Bronfman}, {Nyman}, \&
  {May}}]{1996A&AS..115...81B}
{Bronfman}, L., {Nyman}, L.-A., \& {May}, J. 1996, \aaps, 115, 81

\bibitem[{{Burton} {et~al.}(1975){Burton}, {Gordon}, {Bania}, \&
  {Lockman}}]{1975ApJ...202...30B}
{Burton}, W.~B., {Gordon}, M.~A., {Bania}, T.~M., \& {Lockman}, F.~J. 1975,
  \apj, 202, 30

\bibitem[{{Carey} {et~al.}(1998){Carey}, {Clark}, {Egan}, {Price}, {Shipman},
  \& {Kuchar}}]{1998ApJ...508..721C}
{Carey}, S.~J., {Clark}, F.~O., {Egan}, M.~P., {et~al.} 1998, \apj, 508, 721

\bibitem[{{Carey} {et~al.}(2000){Carey}, {Feldman}, {Redman}, {Egan},
  {MacLeod}, \& {Price}}]{2000ApJ...543L.157C}
{Carey}, S.~J., {Feldman}, P.~A., {Redman}, R.~O., {et~al.} 2000, \apjl, 543,
  L157

\bibitem[{{Carey} {et~al.}(2009){Carey}, {Noriega-Crespo}, {Mizuno}, {Shenoy},
  {Paladini}, {Kraemer}, {Price}, {Flagey}, {Ryan}, {Ingalls}, {Kuchar},
  {Pinheiro Gon{\c c}alves}, {Indebetouw}, {Billot}, {Marleau}, {Padgett},
  {Rebull}, {Bressert}, {Ali}, {Molinari}, {Martin}, {Berriman}, {Boulanger},
  {Latter}, {Miville-Deschenes}, {Shipman}, \& {Testi}}]{2009PASP..121...76C}
{Carey}, S.~J., {Noriega-Crespo}, A., {Mizuno}, D.~R., {et~al.} 2009, \pasp,
  121, 76

\bibitem[{{Caswell} {et~al.}(2010){Caswell}, {Fuller}, {Green}, {Avison},
  {Breen}, {Brooks}, {Burton}, {Chrysostomou}, {Cox}, {Diamond}, {Ellingsen},
  {Gray}, {Hoare}, {Masheder}, {McClure-Griffiths}, {Pestalozzi}, {Phillips},
  {Quinn}, {Thompson}, {Voronkov}, {Walsh}, {Ward-Thompson}, {Wong-McSweeney},
  {Yates}, \& {Cohen}}]{2010MNRAS.404.1029C}
{Caswell}, J.~L., {Fuller}, G.~A., {Green}, J.~A., {et~al.} 2010, \mnras, 404,
  1029

\bibitem[{{Caswell} {et~al.}(2011){Caswell}, {Fuller}, {Green}, {Avison},
  {Breen}, {Ellingsen}, {Gray}, {Pestalozzi}, {Quinn}, {Thompson}, \&
  {Voronkov}}]{2011MNRAS.417.1964C}
{Caswell}, J.~L., {Fuller}, G.~A., {Green}, J.~A., {et~al.} 2011, \mnras, 417,
  1964

\bibitem[{{Caswell} {et~al.}(1995){Caswell}, {Vaile}, {Ellingsen}, {Whiteoak},
  \& {Norris}}]{1995MNRAS.272...96C}
{Caswell}, J.~L., {Vaile}, R.~A., {Ellingsen}, S.~P., {Whiteoak}, J.~B., \&
  {Norris}, R.~P. 1995, \mnras, 272, 96

\bibitem[{{Cesaroni} {et~al.}(1994){Cesaroni}, {Churchwell}, {Hofner},
  {Walmsley}, \& {Kurtz}}]{1994A&A...288..903C}
{Cesaroni}, R., {Churchwell}, E., {Hofner}, P., {Walmsley}, C.~M., \& {Kurtz},
  S. 1994, \aap, 288, 903

\bibitem[{{Chira} {et~al.}(2013){Chira}, {Beuther}, {Linz}, {Schuller},
  {Walmsley}, {Menten}, \& {Bronfman}}]{2013A&A...552A..40C}
{Chira}, R.-A., {Beuther}, H., {Linz}, H., {et~al.} 2013, \aap, 552, A40

\bibitem[{{Clark} \& {Porter}(2004)}]{2004A&A...427..839C}
{Clark}, J.~S. \& {Porter}, J.~M. 2004, \aap, 427, 839

\bibitem[{{Cohen} \& {Thaddeus}(1977)}]{1977ApJ...217L.155C}
{Cohen}, R.~S. \& {Thaddeus}, P. 1977, \apjl, 217, L155

\bibitem[{{Contreras} {et~al.}(2013){Contreras}, {Schuller}, {Urquhart},
  {Csengeri}, {Wyrowski}, {Beuther}, {Bontemps}, {Bronfman}, {Henning},
  {Menten}, {Schilke}, {Walmsley}, {Wienen}, {Tackenberg}, \&
  {Linz}}]{2013A&A...549A..45C}
{Contreras}, Y., {Schuller}, F., {Urquhart}, J.~S., {et~al.} 2013, \aap, 549,
  A45

\bibitem[{{Csengeri} {et~al.}(2011){Csengeri}, {Bontemps}, {Schneider},
  {Motte}, {Gueth}, \& {Hora}}]{2011ApJ...740L...5C}
{Csengeri}, T., {Bontemps}, S., {Schneider}, N., {et~al.} 2011, \apjl, 740, L5

\bibitem[{{Csengeri} {et~al.}(2014){Csengeri}, {Urquhart}, {Schuller}, {Motte},
  {Bontemps}, {Wyrowski}, {Menten}, {Bronfman}, {Beuther}, {Henning}, {Testi},
  {Zavagno}, \& {Walmsley}}]{2014A&A...565A..75C}
{Csengeri}, T., {Urquhart}, J.~S., {Schuller}, F., {et~al.} 2014, \aap, 565,
  A75

\bibitem[{{Dame} {et~al.}(2001){Dame}, {Hartmann}, \&
  {Thaddeus}}]{2001ApJ...547..792D}
{Dame}, T.~M., {Hartmann}, D., \& {Thaddeus}, P. 2001, \apj, 547, 792

\bibitem[{{Duarte-Cabral} {et~al.}(2014){Duarte-Cabral}, {Bontemps}, {Motte},
  {Gusdorf}, {Csengeri}, {Schneider}, \& {Louvet}}]{2014A&A...570A...1D}
{Duarte-Cabral}, A., {Bontemps}, S., {Motte}, F., {et~al.} 2014, \aap, 570, A1

\bibitem[{{Duarte-Cabral} {et~al.}(2013){Duarte-Cabral}, {Bontemps}, {Motte},
  {Hennemann}, {Schneider}, \& {Andr{\'e}}}]{2013A&A...558A.125D}
{Duarte-Cabral}, A., {Bontemps}, S., {Motte}, F., {et~al.} 2013, \aap, 558,
  A125

\bibitem[{{Dunham} {et~al.}(2011){Dunham}, {Rosolowsky}, {Evans}, {Cyganowski},
  \& {Urquhart}}]{2011ApJ...741..110D}
{Dunham}, M.~K., {Rosolowsky}, E., {Evans}, II, N.~J., {Cyganowski}, C., \&
  {Urquhart}, J.~S. 2011, \apj, 741, 110

\bibitem[{{Dunham} {et~al.}(2010){Dunham}, {Rosolowsky}, {Evans}, {Cyganowski},
  {Aguirre}, {Bally}, {Battersby}, {Bradley}, {Dowell}, {Drosback}, {Ginsburg},
  {Glenn}, {Harvey}, {Merello}, {Schlingman}, {Shirley}, {Stringfellow},
  {Walawender}, \& {Williams}}]{2010ApJ...717.1157D}
{Dunham}, M.~K., {Rosolowsky}, E., {Evans}, II, N.~J., {et~al.} 2010, \apj,
  717, 1157

\bibitem[{{Eden} {et~al.}(2013){Eden}, {Moore}, {Morgan}, {Thompson}, \&
  {Urquhart}}]{2013MNRAS.431.1587E}
{Eden}, D.~J., {Moore}, T.~J.~T., {Morgan}, L.~K., {Thompson}, M.~A., \&
  {Urquhart}, J.~S. 2013, \mnras, 431, 1587

\bibitem[{{Egan} {et~al.}(1998){Egan}, {Shipman}, {Price}, {Carey}, {Clark}, \&
  {Cohen}}]{1998ApJ...494L.199E}
{Egan}, M.~P., {Shipman}, R.~F., {Price}, S.~D., {et~al.} 1998, \apjl, 494,
  L199

\bibitem[{{Evans}(1999)}]{1999ARA&A..37..311E}
{Evans}, II, N.~J. 1999, \araa, 37, 311

\bibitem[{{Fontani} {et~al.}(2004){Fontani}, {Cesaroni}, {Testi}, {Walmsley},
  {Molinari}, {Neri}, {Shepherd}, {Brand}, {Palla}, \&
  {Zhang}}]{2004A&A...414..299F}
{Fontani}, F., {Cesaroni}, R., {Testi}, L., {et~al.} 2004, \aap, 414, 299

\bibitem[{{Friesen} {et~al.}(2009){Friesen}, {Di Francesco}, {Shirley}, \&
  {Myers}}]{2009ApJ...697.1457F}
{Friesen}, R.~K., {Di Francesco}, J., {Shirley}, Y.~L., \& {Myers}, P.~C. 2009,
  \apj, 697, 1457

\bibitem[{{Giannetti} {et~al.}(2013){Giannetti}, {Brand}, {S{\'a}nchez-Monge},
  {Fontani}, {Cesaroni}, {Beltr{\'a}n}, {Molinari}, {Dodson}, \&
  {Rioja}}]{2013A&A...556A..16G}
{Giannetti}, A., {Brand}, J., {S{\'a}nchez-Monge}, {\'A}., {et~al.} 2013, \aap,
  556, A16

\bibitem[{{Giannetti} {et~al.}(2014){Giannetti}, {Wyrowski}, {Brand},
  {Csengeri}, {Fontani}, {Walmsley}, {Nguyen Luong}, {Beuther}, {Schuller},
  {G{\"u}sten}, \& {Menten}}]{2014A&A...570A..65G}
{Giannetti}, A., {Wyrowski}, F., {Brand}, J., {et~al.} 2014, \aap, 570, A65

\bibitem[{{Green} {et~al.}(2009){Green}, {Caswell}, {Fuller}, {Avison},
  {Breen}, {Brooks}, {Burton}, {Chrysostomou}, {Cox}, {Diamond}, {Ellingsen},
  {Gray}, {Hoare}, {Masheder}, {McClure-Griffiths}, {Pestalozzi}, {Phillips},
  {Quinn}, {Thompson}, {Voronkov}, {Walsh}, {Ward-Thompson}, {Wong-McSweeney},
  {Yates}, \& {Cohen}}]{2009MNRAS.392..783G}
{Green}, J.~A., {Caswell}, J.~L., {Fuller}, G.~A., {et~al.} 2009, \mnras, 392,
  783

\bibitem[{{Guilloteau} {et~al.}(1983){Guilloteau}, {Wilson}, {Batrla},
  {Martin}, \& {Pauls}}]{1983A&A...124..322G}
{Guilloteau}, S., {Wilson}, T.~L., {Batrla}, W., {Martin}, R.~N., \& {Pauls},
  T.~A. 1983, \aap, 124, 322

\bibitem[{{Hily-Blant} {et~al.}(2010){Hily-Blant}, {Walmsley}, {Pineau Des
  For{\^e}ts}, \& {Flower}}]{2010A&A...513A..41H}
{Hily-Blant}, P., {Walmsley}, M., {Pineau Des For{\^e}ts}, G., \& {Flower}, D.
  2010, \aap, 513, A41

\bibitem[{{Hindson} {et~al.}(2013){Hindson}, {Thompson}, {Urquhart}, {Faimali},
  {Johnston-Hollitt}, {Clark}, \& {Davies}}]{2013MNRAS.435.2003H}
{Hindson}, L., {Thompson}, M.~A., {Urquhart}, J.~S., {et~al.} 2013, \mnras,
  435, 2003

\bibitem[{{Ho} \& {Townes}(1983)}]{1983ARA&A..21..239H}
{Ho}, P.~T.~P. \& {Townes}, C.~H. 1983, \araa, 21, 239

\bibitem[{{Hoare} {et~al.}(2012){Hoare}, {Purcell}, {Churchwell}, {Diamond},
  {Cotton}, {Chandler}, {Smethurst}, {Kurtz}, {Mundy}, {Dougherty}, {Fender},
  {Fuller}, {Jackson}, {Garrington}, {Gledhill}, {Goldsmith}, {Lumsden},
  {Mart{\'{\i}}}, {Moore}, {Muxlow}, {Oudmaijer}, {Pandian}, {Paredes},
  {Shepherd}, {Spencer}, {Thompson}, {Umana}, {Urquhart}, \&
  {Zijlstra}}]{2012PASP..124..939H}
{Hoare}, M.~G., {Purcell}, C.~R., {Churchwell}, E.~B., {et~al.} 2012, \pasp,
  124, 939

\bibitem[{{Hunter} {et~al.}(2008){Hunter}, {Brogan}, {Indebetouw}, \&
  {Cyganowski}}]{2008ApJ...680.1271H}
{Hunter}, T.~R., {Brogan}, C.~L., {Indebetouw}, R., \& {Cyganowski}, C.~J.
  2008, \apj, 680, 1271

\bibitem[{{Hunter} {et~al.}(2000){Hunter}, {Churchwell}, {Watson}, {Cox},
  {Benford}, \& {Roelfsema}}]{2000AJ....119.2711H}
{Hunter}, T.~R., {Churchwell}, E., {Watson}, C., {et~al.} 2000, \aj, 119, 2711

\bibitem[{{Jackson} {et~al.}(2013){Jackson}, {Rathborne}, {Foster}, {Whitaker},
  {Sanhueza}, {Claysmith}, {Mascoop}, {Wienen}, {Breen}, {Herpin},
  {Duarte-Cabral}, {Csengeri}, {Longmore}, {Contreras}, {Indermuehle},
  {Barnes}, {Walsh}, {Cunningham}, {Brooks}, {Britton}, {Voronkov}, {Urquhart},
  {Alves}, {Jordan}, {Hill}, {Hoq}, {Finn}, {Bains}, {Bontemps}, {Bronfman},
  {Caswell}, {Deharveng}, {Ellingsen}, {Fuller}, {Garay}, {Green}, {Hindson},
  {Jones}, {Lenfestey}, {Lo}, {Lowe}, {Mardones}, {Menten}, {Minier}, {Morgan},
  {Motte}, {Muller}, {Peretto}, {Purcell}, {Schilke}, {Bontemps}, {Schuller},
  {Titmarsh}, {Wyrowski}, \& {Zavagno}}]{2013PASA...30...57J}
{Jackson}, J.~M., {Rathborne}, J.~M., {Foster}, J.~B., {et~al.} 2013, \pasa,
  30, 57

\bibitem[{{Johnstone} {et~al.}(2010){Johnstone}, {Rosolowsky}, {Tafalla}, \&
  {Kirk}}]{2010ApJ...711..655J}
{Johnstone}, D., {Rosolowsky}, E., {Tafalla}, M., \& {Kirk}, H. 2010, \apj,
  711, 655

\bibitem[{{Kauffmann} {et~al.}(2013){Kauffmann}, {Pillai}, \&
  {Goldsmith}}]{2013ApJ...779..185K}
{Kauffmann}, J., {Pillai}, T., \& {Goldsmith}, P.~F. 2013, \apj, 779, 185

\bibitem[{{Kraemer} \& {Jackson}(1995)}]{1995ApJ...439L...9K}
{Kraemer}, K.~E. \& {Jackson}, J.~M. 1995, \apjl, 439, L9

\bibitem[{{Kramer} {et~al.}(1998){Kramer}, {Stutzki}, {Rohrig}, \&
  {Corneliussen}}]{1998A&A...329..249K}
{Kramer}, C., {Stutzki}, J., {Rohrig}, R., \& {Corneliussen}, U. 1998, \aap,
  329, 249

\bibitem[{{Kroupa} {et~al.}(2011){Kroupa}, {Weidner}, {Pflamm-Altenburg},
  {Thies}, {Dabringhausen}, {Marks}, \& {Maschberger}}]{2011arXiv1112.3340K}
{Kroupa}, P., {Weidner}, C., {Pflamm-Altenburg}, J., {et~al.} 2011, ArXiv
  e-prints

\bibitem[{{Kurtz} {et~al.}(1994){Kurtz}, {Churchwell}, \&
  {Wood}}]{1994ApJS...91..659K}
{Kurtz}, S., {Churchwell}, E., \& {Wood}, D.~O.~S. 1994, \apjs, 91, 659

\bibitem[{{Lada} {et~al.}(2008){Lada}, {Muench}, {Rathborne}, {Alves}, \&
  {Lombardi}}]{2008ApJ...672..410L}
{Lada}, C.~J., {Muench}, A.~A., {Rathborne}, J., {Alves}, J.~F., \& {Lombardi},
  M. 2008, \apj, 672, 410

\bibitem[{{Ladd} {et~al.}(2005){Ladd}, {Purcell}, {Wong}, \&
  {Robertson}}]{2005PASA...22...62L}
{Ladd}, N., {Purcell}, C., {Wong}, T., \& {Robertson}, S. 2005, \pasa, 22, 62

\bibitem[{{Larson}(1981)}]{1981MNRAS.194..809L}
{Larson}, R.~B. 1981, \mnras, 194, 809

\bibitem[{{Le Bourlot}(1991)}]{1991A&A...242..235L}
{Le Bourlot}, J. 1991, \aap, 242, 235

\bibitem[{{Lee} {et~al.}(2012){Lee}, {Murray}, \&
  {Rahman}}]{2012ApJ...752..146L}
{Lee}, E.~J., {Murray}, N., \& {Rahman}, M. 2012, \apj, 752, 146

\bibitem[{{Longmore} {et~al.}(2007){Longmore}, {Burton}, {Barnes}, {Wong},
  {Purcell}, \& {Ott}}]{2007MNRAS.379..535L}
{Longmore}, S.~N., {Burton}, M.~G., {Barnes}, P.~J., {et~al.} 2007, \mnras,
  379, 535

\bibitem[{{Lu} {et~al.}(2014){Lu}, {Zhang}, {Liu}, {Wang}, \&
  {Gu}}]{2014ApJ...790...84L}
{Lu}, X., {Zhang}, Q., {Liu}, H.~B., {Wang}, J., \& {Gu}, Q. 2014, \apj, 790,
  84

\bibitem[{{Lumsden} {et~al.}(2013){Lumsden}, {Hoare}, {Urquhart}, {Oudmaijer},
  {Davies}, {Mottram}, {Cooper}, \& {Moore}}]{2013ApJS..208...11L}
{Lumsden}, S.~L., {Hoare}, M.~G., {Urquhart}, J.~S., {et~al.} 2013, \apjs, 208,
  11

\bibitem[{{Mangum} \& {Wootten}(1994)}]{1994ApJ...428L..33M}
{Mangum}, J.~G. \& {Wootten}, A. 1994, \apjl, 428, L33

\bibitem[{{Mangum} {et~al.}(1992){Mangum}, {Wootten}, \&
  {Mundy}}]{1992ApJ...388..467M}
{Mangum}, J.~G., {Wootten}, A., \& {Mundy}, L.~G. 1992, \apj, 388, 467

\bibitem[{{Matsakis} {et~al.}(1977){Matsakis}, {Brandshaft}, {Chui}, {Cheung},
  {Yngvesson}, {Cardiasmenos}, {Shanley}, \& {Ho}}]{1977ApJ...214L..67M}
{Matsakis}, D.~N., {Brandshaft}, D., {Chui}, M.~F., {et~al.} 1977, \apjl, 214,
  L67

\bibitem[{{Minier} {et~al.}(2009){Minier}, {Andr{\'e}}, {Bergman}, {Motte},
  {Wyrowski}, {Le Pennec}, {Rodriguez}, {Boulade}, {Doumayrou}, {Dubreuil},
  {Gallais}, {Hamon}, {Lagage}, {Lortholary}, {Martignac}, {Rev{\'e}ret},
  {Roussel}, {Talvard}, {Willmann}, \& {Olofsson}}]{2009A&A...501L...1M}
{Minier}, V., {Andr{\'e}}, P., {Bergman}, P., {et~al.} 2009, \aap, 501, L1

\bibitem[{{Molinari} {et~al.}(2010){Molinari}, {Swinyard}, {Bally}, {Barlow},
  {Bernard}, {Martin}, {Moore}, {Noriega-Crespo}, {Plume}, {Testi}, {Zavagno},
  {Abergel}, {Ali}, {Andr{\'e}}, {Baluteau}, {Benedettini}, {Bern{\'e}},
  {Billot}, {Blommaert}, {Bontemps}, {Boulanger}, {Brand}, {Brunt}, {Burton},
  {Campeggio}, {Carey}, {Caselli}, {Cesaroni}, {Cernicharo}, {Chakrabarti},
  {Chrysostomou}, {Codella}, {Cohen}, {Compiegne}, {Davis}, {de Bernardis}, {de
  Gasperis}, {Di Francesco}, {di Giorgio}, {Elia}, {Faustini}, {Fischera},
  {Fukui}, {Fuller}, {Ganga}, {Garcia-Lario}, {Giard}, {Giardino}, {Glenn},
  {Goldsmith}, {Griffin}, {Hoare}, {Huang}, {Jiang}, {Joblin}, {Joncas},
  {Juvela}, {Kirk}, {Lagache}, {Li}, {Lim}, {Lord}, {Lucas}, {Maiolo},
  {Marengo}, {Marshall}, {Masi}, {Massi}, {Matsuura}, {Meny}, {Minier},
  {Miville-Desch{\^e}nes}, {Montier}, {Motte}, {M{\"u}ller}, {Natoli}, {Neves},
  {Olmi}, {Paladini}, {Paradis}, {Pestalozzi}, {Pezzuto}, {Piacentini},
  {Pomar{\`e}s}, {Popescu}, {Reach}, {Richer}, {Ristorcelli}, {Roy}, {Royer},
  {Russeil}, {Saraceno}, {Sauvage}, {Schilke}, {Schneider-Bontemps},
  {Schuller}, {Schultz}, {Shepherd}, {Sibthorpe}, {Smith}, {Smith},
  {Spinoglio}, {Stamatellos}, {Strafella}, {Stringfellow}, {Sturm}, {Taylor},
  {Thompson}, {Tuffs}, {Umana}, {Valenziano}, {Vavrek}, {Viti}, {Waelkens},
  {Ward-Thompson}, {White}, {Wyrowski}, {Yorke}, \&
  {Zhang}}]{2010PASP..122..314M}
{Molinari}, S., {Swinyard}, B., {Bally}, J., {et~al.} 2010, \pasp, 122, 314

\bibitem[{{Moscadelli} \& {Goddi}(2014)}]{2014A&A...566A.150M}
{Moscadelli}, L. \& {Goddi}, C. 2014, \aap, 566, A150

\bibitem[{{Nejad} {et~al.}(1990){Nejad}, {Williams}, \&
  {Charnley}}]{1990MNRAS.246..183N}
{Nejad}, L.~A.~M., {Williams}, D.~A., \& {Charnley}, S.~B. 1990, \mnras, 246,
  183

\bibitem[{{Park}(2001)}]{2001A&A...376..348P}
{Park}, Y.-S. 2001, \aap, 376, 348

\bibitem[{{Perault} {et~al.}(1996){Perault}, {Omont}, {Simon}, {Seguin},
  {Ojha}, {Blommaert}, {Felli}, {Gilmore}, {Guglielmo}, {Habing}, {Price},
  {Robin}, {de Batz}, {Cesarsky}, {Elbaz}, {Epchtein}, {Fouque}, {Guest},
  {Levine}, {Pollock}, {Prusti}, {Siebenmorgen}, {Testi}, \&
  {Tiphene}}]{1996A&A...315L.165P}
{Perault}, M., {Omont}, A., {Simon}, G., {et~al.} 1996, \aap, 315, L165

\bibitem[{{Peretto} \& {Fuller}(2009)}]{2009A&A...505..405P}
{Peretto}, N. \& {Fuller}, G.~A. 2009, \aap, 505, 405

\bibitem[{{Pestalozzi} {et~al.}(2005){Pestalozzi}, {Minier}, \&
  {Booth}}]{2005A&A...432..737P}
{Pestalozzi}, M.~R., {Minier}, V., \& {Booth}, R.~S. 2005, \aap, 432, 737

\bibitem[{{Pillai} {et~al.}(2006){Pillai}, {Wyrowski}, {Carey}, \&
  {Menten}}]{2006A&A...450..569P}
{Pillai}, T., {Wyrowski}, F., {Carey}, S.~J., \& {Menten}, K.~M. 2006, \aap,
  450, 569

\bibitem[{{Plume} {et~al.}(1997){Plume}, {Jaffe}, {Evans},
  {Mart{\'{\i}}n-Pintado}, \& {G{\'o}mez-Gonz{\'a}lez}}]{1997ApJ...476..730P}
{Plume}, R., {Jaffe}, D.~T., {Evans}, II, N.~J., {Mart{\'{\i}}n-Pintado}, J.,
  \& {G{\'o}mez-Gonz{\'a}lez}, J. 1997, \apj, 476, 730

\bibitem[{{Purcell} {et~al.}(2012){Purcell}, {Longmore}, {Walsh}, {Whiting},
  {Breen}, {Britton}, {Brooks}, {Burton}, {Cunningham}, {Green},
  {Harvey-Smith}, {Hindson}, {Hoare}, {Indermuehle}, {Jones}, {Lo}, {Lowe},
  {Phillips}, {Thompson}, {Urquhart}, {Voronkov}, \&
  {White}}]{2012MNRAS.426.1972P}
{Purcell}, C.~R., {Longmore}, S.~N., {Walsh}, A.~J., {et~al.} 2012, \mnras,
  426, 1972

\bibitem[{{Ragan} {et~al.}(2011){Ragan}, {Bergin}, \&
  {Wilner}}]{2011ApJ...736..163R}
{Ragan}, S.~E., {Bergin}, E.~A., \& {Wilner}, D. 2011, \apj, 736, 163

\bibitem[{{Rathborne} {et~al.}(2008){Rathborne}, {Lada}, {Muench}, {Alves}, \&
  {Lombardi}}]{2008ApJS..174..396R}
{Rathborne}, J.~M., {Lada}, C.~J., {Muench}, A.~A., {Alves}, J.~F., \&
  {Lombardi}, M. 2008, \apjs, 174, 396

\bibitem[{{Rodriguez} {et~al.}(1982){Rodriguez}, {Canto}, \&
  {Moran}}]{1982ApJ...255..103R}
{Rodriguez}, L.~F., {Canto}, J., \& {Moran}, J.~M. 1982, \apj, 255, 103

\bibitem[{{Rohlfs} \& {Wilson}(2004)}]{2004tra..book.....R}
{Rohlfs}, K. \& {Wilson}, T.~L. 2004, {Tools of radio astronomy}

\bibitem[{{Rolleston} {et~al.}(2000){Rolleston}, {Smartt}, {Dufton}, \&
  {Ryans}}]{2000A&A...363..537R}
{Rolleston}, W.~R.~J., {Smartt}, S.~J., {Dufton}, P.~L., \& {Ryans}, R.~S.~I.
  2000, \aap, 363, 537

\bibitem[{{Roman-Duval} {et~al.}(2009){Roman-Duval}, {Jackson}, {Heyer},
  {Johnson}, {Rathborne}, {Shah}, \& {Simon}}]{2009ApJ...699.1153R}
{Roman-Duval}, J., {Jackson}, J.~M., {Heyer}, M., {et~al.} 2009, \apj, 699,
  1153

\bibitem[{{Rosolowsky} {et~al.}(2008){Rosolowsky}, {Pineda}, {Foster},
  {Borkin}, {Kauffmann}, {Caselli}, {Myers}, \&
  {Goodman}}]{2008ApJS..175..509R}
{Rosolowsky}, E.~W., {Pineda}, J.~E., {Foster}, J.~B., {et~al.} 2008, \apjs,
  175, 509

\bibitem[{{Scalise} {et~al.}(1989){Scalise}, {Rodriguez}, \&
  {Mendoza-Torres}}]{1989A&A...221..105S}
{Scalise}, Jr., E., {Rodriguez}, L.~F., \& {Mendoza-Torres}, E. 1989, \aap,
  221, 105

\bibitem[{{Schuller} {et~al.}(2009){Schuller}, {Menten}, {Contreras},
  {Wyrowski}, {Schilke}, {Bronfman}, {Henning}, {Walmsley}, {Beuther},
  {Bontemps}, {Cesaroni}, {Deharveng}, {Garay}, {Herpin}, {Lefloch}, {Linz},
  {Mardones}, {Minier}, {Molinari}, {Motte}, {Nyman}, {Reveret}, {Risacher},
  {Russeil}, {Schneider}, {Testi}, {Troost}, {Vasyunina}, {Wienen}, {Zavagno},
  {Kovacs}, {Kreysa}, {Siringo}, \& {Wei{\ss}}}]{2009A&A...504..415S}
{Schuller}, F., {Menten}, K.~M., {Contreras}, Y., {et~al.} 2009, \aap, 504, 415

\bibitem[{{Sevenster} {et~al.}(1997){Sevenster}, {Chapman}, {Habing},
  {Killeen}, \& {Lindqvist}}]{1997A&AS..124..509S}
{Sevenster}, M.~N., {Chapman}, J.~M., {Habing}, H.~J., {Killeen}, N.~E.~B., \&
  {Lindqvist}, M. 1997, \aaps, 124, 509

\bibitem[{{Shaver} {et~al.}(1983){Shaver}, {McGee}, {Newton}, {Danks}, \&
  {Pottasch}}]{1983MNRAS.204...53S}
{Shaver}, P.~A., {McGee}, R.~X., {Newton}, L.~M., {Danks}, A.~C., \&
  {Pottasch}, S.~R. 1983, \mnras, 204, 53

\bibitem[{{Simon} {et~al.}(2006){Simon}, {Rathborne}, {Shah}, {Jackson}, \&
  {Chambers}}]{2006ApJ...653.1325S}
{Simon}, R., {Rathborne}, J.~M., {Shah}, R.~Y., {Jackson}, J.~M., \&
  {Chambers}, E.~T. 2006, \apj, 653, 1325

\bibitem[{{Siringo} {et~al.}(2008){Siringo}, {Kreysa}, {Kovacs}, {Schuller},
  {Wei{\ss}}, {Esch}, {Gem{\"u}nd}, {Jethava}, {Lundershausen}, {G{\"u}sten},
  {Menten}, {Beelen}, {Bertoldi}, {Beeman}, {Haller}, \&
  {Colin}}]{2008SPIE.7020E..03S}
{Siringo}, G., {Kreysa}, E., {Kovacs}, A., {et~al.} 2008, in Society of
  Photo-Optical Instrumentation Engineers (SPIE) Conference Series, Vol. 7020,
  Society of Photo-Optical Instrumentation Engineers (SPIE) Conference Series,
  3

\bibitem[{{Siringo} {et~al.}(2007){Siringo}, {Weiss}, {Kreysa}, {Schuller},
  {Kovacs}, {Beelen}, {Esch}, {Gem{\"u}nd}, {Jethava}, {Lundershausen},
  {Menten}, {G{\"u}sten}, {Bertoldi}, {De Breuck}, {Nyman}, {Haller}, \&
  {Beeman}}]{2007Msngr.129....2S}
{Siringo}, G., {Weiss}, A., {Kreysa}, E., {et~al.} 2007, The Messenger, 129, 2

\bibitem[{{Sridharan} {et~al.}(2002){Sridharan}, {Beuther}, {Schilke},
  {Menten}, \& {Wyrowski}}]{2002ApJ...566..931S}
{Sridharan}, T.~K., {Beuther}, H., {Schilke}, P., {Menten}, K.~M., \&
  {Wyrowski}, F. 2002, \apj, 566, 931

\bibitem[{{Stutzki} \& {Guesten}(1990)}]{1990ApJ...356..513S}
{Stutzki}, J. \& {Guesten}, R. 1990, \apj, 356, 513

\bibitem[{{Stutzki} {et~al.}(1982){Stutzki}, {Ungerechts}, \&
  {Winnewisser}}]{1982A&A...111..201S}
{Stutzki}, J., {Ungerechts}, H., \& {Winnewisser}, G. 1982, \aap, 111, 201

\bibitem[{{Stutzki} \& {Winnewisser}(1985)}]{1985A&A...144...13S}
{Stutzki}, J. \& {Winnewisser}, G. 1985, \aap, 144, 13

\bibitem[{{Szymczak} {et~al.}(2002){Szymczak}, {Kus}, {Hrynek}, {K{\v e}pa}, \&
  {Pazderski}}]{2002A&A...392..277S}
{Szymczak}, M., {Kus}, A.~J., {Hrynek}, G., {K{\v e}pa}, A., \& {Pazderski}, E.
  2002, \aap, 392, 277

\bibitem[{{Tafalla} {et~al.}(2002){Tafalla}, {Myers}, {Caselli}, {Walmsley}, \&
  {Comito}}]{2002ApJ...569..815T}
{Tafalla}, M., {Myers}, P.~C., {Caselli}, P., {Walmsley}, C.~M., \& {Comito},
  C. 2002, \apj, 569, 815

\bibitem[{{Ungerechts} {et~al.}(1997){Ungerechts}, {Bergin}, {Goldsmith},
  {Irvine}, {Schloerb}, \& {Snell}}]{1997ApJ...482..245U}
{Ungerechts}, H., {Bergin}, E.~A., {Goldsmith}, P.~F., {et~al.} 1997, \apj,
  482, 245

\bibitem[{{Ungerechts} {et~al.}(1986){Ungerechts}, {Winnewisser}, \&
  {Walmsley}}]{1986A&A...157..207U}
{Ungerechts}, H., {Winnewisser}, G., \& {Walmsley}, C.~M. 1986, \aap, 157, 207

\bibitem[{{Urquhart} {et~al.}(2007){Urquhart}, {Busfield}, {Hoare}, {Lumsden},
  {Clarke}, {Moore}, {Mottram}, \& {Oudmaijer}}]{2007IAUS..237..482U}
{Urquhart}, J.~S., {Busfield}, A.~L., {Hoare}, M.~G., {et~al.} 2007, in IAU
  Symposium, Vol. 237, IAU Symposium, ed. B.~G. {Elmegreen} \& J.~{Palous},
  482--482

\bibitem[{{Urquhart} {et~al.}(2014{\natexlab{a}}){Urquhart}, {Csengeri},
  {Wyrowski}, {Schuller}, {Bontemps}, {Bronfman}, {Menten}, {Walmsley},
  {Contreras}, {Beuther}, {Wienen}, \& {Linz}}]{2014A&A...568A..41U}
{Urquhart}, J.~S., {Csengeri}, T., {Wyrowski}, F., {et~al.} 2014{\natexlab{a}},
  \aap, 568, A41

\bibitem[{{Urquhart} {et~al.}(2015){Urquhart}, {Figura}, {Moore}, {Csengeri},
  {Lumsden}, {Pillai}, {Thompson}, {Eden}, \& {Morgan}}]{2015MNRAS.452.4029U}
{Urquhart}, J.~S., {Figura}, C.~C., {Moore}, T.~J.~T., {et~al.} 2015, \mnras,
  452, 4029

\bibitem[{{Urquhart} {et~al.}(2014{\natexlab{b}}){Urquhart}, {Figura}, {Moore},
  {Hoare}, {Lumsden}, {Mottram}, {Thompson}, \&
  {Oudmaijer}}]{2014MNRAS.437.1791U}
{Urquhart}, J.~S., {Figura}, C.~C., {Moore}, T.~J.~T., {et~al.}
  2014{\natexlab{b}}, \mnras, 437, 1791

\bibitem[{{Urquhart} {et~al.}(2013{\natexlab{a}}){Urquhart}, {Moore},
  {Schuller}, {Wyrowski}, {Menten}, {Thompson}, {Csengeri}, {Walmsley},
  {Bronfman}, \& {K{\"o}nig}}]{2013MNRAS.431.1752U}
{Urquhart}, J.~S., {Moore}, T.~J.~T., {Schuller}, F., {et~al.}
  2013{\natexlab{a}}, \mnras, 431, 1752

\bibitem[{{Urquhart} {et~al.}(2011){Urquhart}, {Morgan}, {Figura}, {Moore},
  {Lumsden}, {Hoare}, {Oudmaijer}, {Mottram}, {Davies}, \&
  {Dunham}}]{2011MNRAS.418.1689U}
{Urquhart}, J.~S., {Morgan}, L.~K., {Figura}, C.~C., {et~al.} 2011, \mnras,
  418, 1689

\bibitem[{{Urquhart} {et~al.}(2013{\natexlab{b}}){Urquhart}, {Thompson},
  {Moore}, {Purcell}, {Hoare}, {Schuller}, {Wyrowski}, {Csengeri}, {Menten},
  {Lumsden}, {Kurtz}, {Walmsley}, {Bronfman}, {Morgan}, {Eden}, \&
  {Russeil}}]{2013MNRAS.435..400U}
{Urquhart}, J.~S., {Thompson}, M.~A., {Moore}, T.~J.~T., {et~al.}
  2013{\natexlab{b}}, \mnras, 435, 400

\bibitem[{{van der Tak} {et~al.}(2007){van der Tak}, {Black}, {Sch{\"o}ier},
  {Jansen}, \& {van Dishoeck}}]{2007A&A...468..627V}
{van der Tak}, F.~F.~S., {Black}, J.~H., {Sch{\"o}ier}, F.~L., {Jansen}, D.~J.,
  \& {van Dishoeck}, E.~F. 2007, \aap, 468, 627

\bibitem[{{Vasyunina} {et~al.}(2009){Vasyunina}, {Linz}, {Henning}, {Stecklum},
  {Klose}, \& {Nyman}}]{2009A&A...499..149V}
{Vasyunina}, T., {Linz}, H., {Henning}, T., {et~al.} 2009, \aap, 499, 149

\bibitem[{{Walmsley} \& {Ungerechts}(1983)}]{1983A&A...122..164W}
{Walmsley}, C.~M. \& {Ungerechts}, H. 1983, \aap, 122, 164

\bibitem[{{Walsh} {et~al.}(2011){Walsh}, {Breen}, {Britton}, {Brooks},
  {Burton}, {Cunningham}, {Green}, {Harvey-Smith}, {Hindson}, {Hoare},
  {Indermuehle}, {Jones}, {Lo}, {Longmore}, {Lowe}, {Phillips}, {Purcell},
  {Thompson}, {Urquhart}, {Voronkov}, {White}, \&
  {Whiting}}]{2011MNRAS.416.1764W}
{Walsh}, A.~J., {Breen}, S.~L., {Britton}, T., {et~al.} 2011, \mnras, 416, 1764

\bibitem[{{Walsh} {et~al.}(1998){Walsh}, {Burton}, {Hyland}, \&
  {Robinson}}]{1998MNRAS.301..640W}
{Walsh}, A.~J., {Burton}, M.~G., {Hyland}, A.~R., \& {Robinson}, G. 1998,
  \mnras, 301, 640

\bibitem[{{Wienen} {et~al.}(2015){Wienen}, {Wyrowski}, {Menten}, {Urquhart},
  {Csengeri}, {Walmsley}, {Bontemps}, {Russeil}, {Bronfman}, {Koribalski}, \&
  {Schuller}}]{2015A&A...579A..91W}
{Wienen}, M., {Wyrowski}, F., {Menten}, K.~M., {et~al.} 2015, \aap, 579, A91

\bibitem[{{Wienen} {et~al.}(2012){Wienen}, {Wyrowski}, {Schuller}, {Menten},
  {Walmsley}, {Bronfman}, \& {Motte}}]{2012A&A...544A.146W}
{Wienen}, M., {Wyrowski}, F., {Schuller}, F., {et~al.} 2012, \aap, 544, A146

\bibitem[{{Wilson} {et~al.}(1993){Wilson}, {Gaume}, \&
  {Johnston}}]{1993ApJ...402..230W}
{Wilson}, T.~L., {Gaume}, R.~A., \& {Johnston}, K.~J. 1993, \apj, 402, 230

\bibitem[{{Womack} {et~al.}(1992){Womack}, {Ziurys}, \&
  {Wyckoff}}]{1992ApJ...387..417W}
{Womack}, M., {Ziurys}, L.~M., \& {Wyckoff}, S. 1992, \apj, 387, 417

\bibitem[{{Wood} \& {Churchwell}(1989)}]{1989ApJS...69..831W}
{Wood}, D.~O.~S. \& {Churchwell}, E. 1989, \apjs, 69, 831

\bibitem[{{Wu} {et~al.}(2010{\natexlab{a}}){Wu}, {Evans}, {Shirley}, \&
  {Knez}}]{2010ApJS..188..313W}
{Wu}, J., {Evans}, II, N.~J., {Shirley}, Y.~L., \& {Knez}, C.
  2010{\natexlab{a}}, \apjs, 188, 313

\bibitem[{{Wu} {et~al.}(2010{\natexlab{b}}){Wu}, {Xu}, {Pandian}, {Yang},
  {Henkel}, {Menten}, \& {Zhang}}]{2010ApJ...720..392W}
{Wu}, Y.~W., {Xu}, Y., {Pandian}, J.~D., {et~al.} 2010{\natexlab{b}}, \apj,
  720, 392

\bibitem[{{Zhang} {et~al.}(2014){Zhang}, {Wang}, {Xu}, {Wyrowski}, \&
  {Menten}}]{2014ApJ...784..107Z}
{Zhang}, C.-P., {Wang}, J.-J., {Xu}, J.-L., {Wyrowski}, F., \& {Menten}, K.~M.
  2014, \apj, 784, 107

\bibitem[{{Zhang} {et~al.}(1999){Zhang}, {Hunter}, {Sridharan}, \&
  {Cesaroni}}]{1999ApJ...527L.117Z}
{Zhang}, Q., {Hunter}, T.~R., {Sridharan}, T.~K., \& {Cesaroni}, R. 1999,
  \apjl, 527, L117

\end{thebibliography}
\bibliographystyle{aa}

\end{document}